\PassOptionsToPackage{nopatch=footnote}{microtype}
\RequirePackage{anyfontsize}

\documentclass[pdfa,a4paper,UKenglish,numberwithinsect,autoref,cleveref]{lipics-v2021}
\nolinenumbers
\hideLIPIcs

\title{Complete Test Suites for Effectful Automata}

\author{B{\'a}lint Kocsis}
       {Institute for Computing and Information Sciences, Radboud University, Nijmegen, The Netherlands
          \and \url{https://balintkocsis.site}}
       {mail@balintkocsis.site}
       {https://orcid.org/0000-0003-0570-3031}
       {}
\author{Rick Koenders}
       {Department of Computer Science, Open University, Utrecht, The Netherlands
          \and \url{https://research.ou.nl/en/persons/rick-koenders}}
       {rick.koenders@ou.nl}
       {}
       {}
\author{Joshua Moerman}
       {Department of Computer Science, Open University, Utrecht, The Netherlands
          \and \url{https://joshuamoerman.nl}}
       {joshua.moerman@ou.nl}
       {https://orcid.org/0000-0001-9819-8374}
       {}
\author{Jurriaan Rot}
       {Institute for Computing and Information Sciences, Radboud University, Nijmegen, The Netherlands
          \and \url{http://jurriaan.me}}
       {jrot@cs.ru.nl}
       {https://orcid.org/0000-0002-1404-6232}
       {}
\authorrunning{B. Kocsis, R. Koenders, J. Moerman, and J. Rot}


\relatedversion{This is the full version of a conference paper of the same title to appear in the proceedings of {CSL} 2027.}
\funding{This research is supported by the NWO grant No.~VI.Vidi.223.096.}
\acknowledgements{We want to thank Henning Urbat and Fabian Lenke for discussions about join-semilattices, tensor products, and join-irreducible elements.
We are also very grateful to the anonymous reviewers for their time and useful comments, which improved the paper.}

\usepackage{mleftright}
\usepackage{refcount}
\usepackage{tikz,quiver}
\usepackage{xparse}
\usepackage{xspace}

\hypersetup{draft=false}
\usetikzlibrary{automata,positioning}

\creflabelformat{equation}{#2(#1)#3}
\creflabelformat{enumi}{#2(#1)#3}

\newaliascnt{assumption}{theorem}
\newtheorem{assumption}[assumption]{Assumption}
\aliascntresetthe{assumption}
\numberwithin{assumption}{section}
\crefname{assumption}{Assumption}{Assumptions}

\DeclareSymbolFont{bbold}{U}{bbold}{m}{n}
\DeclareSymbolFontAlphabet{\mathbbold}{bbold}

\NewDocumentCommand{\ie}{}{i.e.\ }
\NewDocumentCommand{\eg}{}{e.g.\ }
\NewDocumentCommand{\sth}{}{s.t.\ }
\NewDocumentCommand{\cf}{}{cf.~}
\NewDocumentCommand{\wrt}{}{w.r.t.~}
\NewDocumentCommand{\opcit}{s}{\textit{op.~cit.}\xspace}

\NewDocumentCommand{\itemref}{m m}{\namecref{#1}~\hyperref[#2]{\labelcref*{#1}\labelcref*{#2}}}

\makeatletter
\NewDocumentCommand{\an}{m}{a\@addn{#1}{\@an@letters}}
\NewDocumentCommand{\An}{m}{A\@addn{#1}{\@an@letters}}

\def\@an@letters{AaEeFfHhLlMmNnOoRrSsXx}
\def\@addn#1#2{%
  \begingroup%
    \edef\@arg{#1}%
    \ifx\@arg\@empty%
      \errmessage{Argument cannot be empty}%
    \fi%
    \edef\@first{\expandafter\@get@first\@arg&}%
    \edef\@list{#2}%
    \expandafter\@search\@list&%
  \endgroup%
}
\def\@gobbletoamp#1&{}
\def\@get@first#1#2&{#1}
\def\@search#1{\if#1&\else\expandafter\@search@step\expandafter#1\fi}
\def\@search@step#1{\if#1\@first n\@gobbletoamp\else\expandafter\@search\fi}
\makeatother

\renewcommand{\epsilon}{\varepsilon}
\renewcommand{\phi}{\varphi}
\NewDocumentCommand{\xto}{O{} m}{\xrightarrow[#1]{#2}}

\NewDocumentCommand{\leteq}{}{:=}
\NewDocumentCommand{\defeq}{}{\leteq}

\NewDocumentCommand{\set}{m o}{\{#1 \IfValueT{#2}{\mid #2}\}}
\NewDocumentCommand{\prodset}{}{\times}
\NewDocumentCommand{\sumset}{}{+}
\NewDocumentCommand{\funcset}{m m}{#2^{#1}}
\NewDocumentCommand{\powset}{}{\mathcal{P}_{\hspace{-1pt}f}}

\NewDocumentCommand{\dom}{m}{\operatorname{dom}\mleft(#1\mright)}
\NewDocumentCommand{\im}{m}{\operatorname{im}\mleft(#1\mright)}
\NewDocumentCommand{\card}{m}{|#1|}

\NewDocumentCommand{\N}{}{\mathbb{N}}
\NewDocumentCommand{\Z}{}{\mathbb{Z}}
\NewDocumentCommand{\1}{}{\mathbf{1}}
\NewDocumentCommand{\2}{}{\mathbf{2}}

\NewDocumentCommand{\defring}{}{\Z}
\NewDocumentCommand{\LinFunct}{O{\defring}}{\operatorname{Lin}_{#1}}
\NewDocumentCommand{\Lin}{O{\defring} m}{\LinFunct[#1]\mleft(#2\mright)}
\NewDocumentCommand{\Span}{m}{\operatorname{span}\mleft(#1\mright)}

\NewDocumentCommand{\latzero}{}{0}
\NewDocumentCommand{\latjoin}{}{\vee}

\NewDocumentCommand{\cat}{m}{\mathcal{#1}}
\NewDocumentCommand{\Ob}{m}{\operatorname{Ob}\mleft(#1\mright)}
\NewDocumentCommand{\Hom}{O{\operatorname{Hom}} m m}{#1\mleft(#2, #3\mright)}
\NewDocumentCommand{\com}{}{\circ}
\NewDocumentCommand{\id}{o}{1\IfValueT{#1}{_{#1}}}

\NewDocumentCommand{\ito}{}{\xto{\cong}}
\NewDocumentCommand{\inv}{s m}{\IfBooleanTF{#1}{\mleft(#2\mright)^{-1}}{#2^{-1}}}

\NewDocumentCommand{\Set}{}{\mathbf{Set}}

\NewDocumentCommand{\prodcat}{}{\times}

\NewDocumentCommand{\catsum}{}{\mathbin{+}}
\NewDocumentCommand{\coprojl}{}{\kappa_1}
\NewDocumentCommand{\coprojr}{}{\kappa_2}
\NewDocumentCommand{\casem}{m m}{\mleft[#1, #2\mright]}
\NewDocumentCommand{\coproj}{m}{\kappa_{#1}}

\NewDocumentCommand{\ladj}{m}{#1^\#}
\NewDocumentCommand{\radj}{m}{#1^\flat}

\NewDocumentCommand{\Kl}{m}{\operatorname{\mathbf{Kl}}\mleft(#1\mright)}
\newsavebox\sbklto
\savebox\sbklto{%
	\begin{tikzpicture}[baseline=-2.6pt]
		\filldraw[fill=white,draw=black,line width=0.4pt] circle (1.6pt);
	\end{tikzpicture}}
\NewDocumentCommand{\xklto}{m}{\mathrel{\ooalign{%
	$\xrightarrow{#1}$\cr%
	\hfil$\usebox\sbklto$\hfil}}}
\NewDocumentCommand{\klto}{}{\xklto{\phantom{m}}}

\NewDocumentCommand{\EM}{m}{\operatorname{\mathbf{EM}}\mleft(#1\mright)}
\NewDocumentCommand{\malg}{m}{\mathsf{#1}}
\NewDocumentCommand{\mext}{s m}{\IfBooleanTF{#1}{\mleft(#2\mright)}{#2}^\#}
\NewDocumentCommand{\algmap}{o}{\nu\IfValueT{#1}{_{#1}}}

\NewDocumentCommand{\klemb}{}{E}
\NewDocumentCommand{\klext}{}{V}
\NewDocumentCommand{\emfree}{}{S}
\NewDocumentCommand{\emforget}{}{U}
\NewDocumentCommand{\klcomp}{}{K}

\NewDocumentCommand{\tprod}{}{\otimes}
\NewDocumentCommand{\tunit}{}{\mathbbold{1}}
\NewDocumentCommand{\assoc}{o}{\alpha\IfValueT{#1}{_{#1}}}
\NewDocumentCommand{\lunit}{o}{\lambda\IfValueT{#1}{_{#1}}}
\NewDocumentCommand{\runit}{o}{\rho\IfValueT{#1}{_{#1}}}
\NewDocumentCommand{\braid}{o}{\beta\IfValueT{#1}{_{#1}}}

\NewDocumentCommand{\inthom}{m m}{\mleft[#1, #2\mright]}
\NewDocumentCommand{\intev}{o}{\epsilon\IfValueT{#1}{_{#1}}}

\NewDocumentCommand{\lstr}{o}{\sigma\IfValueT{#1}{_{#1}}}
\NewDocumentCommand{\effmf}{o}{\theta\IfValueT{#1}{_{#1}}}

\NewDocumentCommand{\emprod}{}{\mathbin{\scalebox{0.75}{$\boxtimes$}}}
\NewDocumentCommand{\emassoc}{o}{\assoc[#1]^\eff}
\NewDocumentCommand{\emlunit}{o}{\lunit[#1]^\eff}
\NewDocumentCommand{\emrunit}{o}{\runit[#1]^\eff}
\NewDocumentCommand{\emfreemf}{o}{\bar{\effmf}\IfValueT{#1}{_{#1}}}
\NewDocumentCommand{\emforgetmf}{o}{\omega\IfValueT{#1}{_{#1}}}

\NewDocumentCommand{\emhom}{m m}{\inthom{#1}{#2}_\eff}
\NewDocumentCommand{\emev}{o}{\intev^\eff\IfValueT{#1}{_{#1}}}
\NewDocumentCommand{\emcurry}{m}{\Lambda^\eff\mleft(#1\mright)}
\NewDocumentCommand{\emhominc}{o}{m\IfValueT{#1}{_{#1}}}
\NewDocumentCommand{\freehomiso}{o}{\psi\IfValueT{#1}{_{#1}}}

\NewDocumentCommand{\Alg}{o m}{\operatorname{\mathbf{Alg}}\IfValueT{#1}{_{#1}}\mleft(#2\mright)}

\NewDocumentCommand{\con}{o}{\zeta\IfValueT{#1}{_{#1}}}
\NewDocumentCommand{\emp}{o}{\eta\IfValueT{#1}{_{#1}}}
\NewDocumentCommand{\fext}{m}{\widetilde{#1}}
\NewDocumentCommand{\idext}{m}{#1^\ast}

\NewDocumentCommand{\alglift}{m}{\overline{#1}}

\NewDocumentCommand{\bcat}{}{\cat{C}}
\NewDocumentCommand{\dyn}{}{F}
\NewDocumentCommand{\inobj}{}{I}
\NewDocumentCommand{\outobj}{}{O}
\NewDocumentCommand{\aut}{m}{\mathcal{#1}}
\NewDocumentCommand{\Aut}{m m m}{\operatorname{\mathbf{Aut}}_{#2}^{#3}\mleft(#1\mright)}

\NewDocumentCommand{\fdyn}{}{\dyn^\ast}
\NewDocumentCommand{\flstr}{o}{\lstr[#1]^\ast}

\NewDocumentCommand{\rea}{m}{r_{#1}}
\NewDocumentCommand{\beh}{m}{B_{#1}}
\NewDocumentCommand{\run}{m}{\tau_{#1}}
\NewDocumentCommand{\obs}{m}{\ell_{#1}}

\NewDocumentCommand{\abcobj}{}{\Sigma}
\NewDocumentCommand{\words}{}{\abcobj^\ast}
\NewDocumentCommand{\empwd}{}{\epsilon}

\NewDocumentCommand{\specname}{}{S}
\NewDocumentCommand{\spec}{}{\aut{\specname}}
\NewDocumentCommand{\implname}{}{M}
\NewDocumentCommand{\impl}{}{\aut{\implname}}
\NewDocumentCommand{\fdom}{m}{\mathcal{#1}}

\NewDocumentCommand{\trconc}{}{\xi}
\NewDocumentCommand{\tsconc}{}{\cdot}
\NewDocumentCommand{\trext}{s m o}{\IfBooleanTF{#1}{\mleft(#2\mright)}{#2}^{\uparrow\IfValueT{#3}{#3}}}

\NewDocumentCommand{\wfdom}{m m}{\fdom{U}_{#1}^{#2}}

\NewDocumentCommand{\eff}{}{T}
\NewDocumentCommand{\effmult}{o}{\mu\IfValueT{#1}{_{#1}}}
\NewDocumentCommand{\effunit}{o O{\eff}}{\eta^{#2}\IfValueT{#1}{_{#1}}}

\NewDocumentCommand{\klcat}{}{\Kl{\eff}}
\NewDocumentCommand{\dlaw}{o}{\gamma\IfValueT{#1}{_{#1}}}
\NewDocumentCommand{\kldyn}{}{\dyn\ifmmode_{\operatorname{Kl}}\fi}

\NewDocumentCommand{\emcat}{}{\EM{\eff}}
\NewDocumentCommand{\emdyn}{}{\dyn\ifmmode_{\operatorname{EM}}\fi}
\NewDocumentCommand{\emfreedyn}{o}{\phi\IfValueT{#1}{_{#1}}}
\NewDocumentCommand{\klcompdyn}{o}{%
  \IfValueTF{#1}{\mleft(\emfreedyn_{\operatorname{Kl}}\mright)_{#1}}%
                {\emfreedyn_{\operatorname{Kl}}}}
\NewDocumentCommand{\emforgetdyn}{o}{\psi\IfValueT{#1}{_{#1}}}

\NewDocumentCommand{\emfdyn}{}{\emdyn^\ast}
\NewDocumentCommand{\emcon}{o}{%
  \IfValueTF{#1}{\mleft(\con_{\operatorname{EM}}\mright)_{#1}}
                {\con_{\operatorname{EM}}}}
\NewDocumentCommand{\ememp}{o}{%
  \IfValueTF{#1}{\mleft(\emp_{\operatorname{EM}}\mright)_{#1}}
                {\emp_{\operatorname{EM}}}}
\NewDocumentCommand{\emfreefdyn}{o}{\emfreedyn[#1]^\ast}

\NewDocumentCommand{\emlstr}{o}{%
  \IfValueTF{#1}{\mleft(\lstr_{\operatorname{EM}}\mright)_{#1}}
                {\lstr_{\operatorname{EM}}}}
\NewDocumentCommand{\emflstr}{o}{%
  \IfValueTF{#1}{\mleft(\flstr_{\operatorname{EM}}\mright)_{#1}}
                {\flstr_{\operatorname{EM}}}}

\NewDocumentCommand{\emtrconc}{}{\trconc_{\operatorname{EM}}}
\NewDocumentCommand{\emfreets}{s m}{%
  \mathchoice{\widetilde{\emfree}}{\widetilde{\emfree}}{\tilde{\emfree}}{\tilde{\emfree}}%
  \IfBooleanTF{#1}{\mleft(#2\mright)}{#2}}

\NewDocumentCommand{\emwfdom}{m m}{\fdom{U}_{#1,\operatorname{EM}}^{#2}}
\NewDocumentCommand{\klwfdom}{m m}{\fdom{U}_{#1,\operatorname{Kl}}^{#2}}

\tikzset{
  automaton/.style={
    node distance=1.2cm,
    shorten >=1pt,
    auto,
    initial text={},
    initial distance=0.6cm,
    every loop/.append style={xscale=2},
    every edge/.append style={bend angle=15},
    every node/.append style={align=center},
  },
  weighted automaton/.style={
    automaton,
    accepting distance=0.6cm,
    inweight/.style={initial by arrow,initial text=$##1$},
    outweight/.style={accepting by arrow,accepting text=$##1$},
  },
}

\NewDocumentCommand{\exNFA}{o}{\aut{A}\IfValueT{#1}{_{#1}}}
\NewDocumentCommand{\exlang}{o}{L\IfValueT{#1}{_{#1}}}
\NewDocumentCommand{\exWFA}{o}{\aut{B}\IfValueT{#1}{_{#1}}}
\NewDocumentCommand{\exwlang}{o}{\widetilde{L}\IfValueT{#1}{_{#1}}}

\begin{document}

\maketitle

\begin{abstract}
  Complete test suites for automata are (finite) sets of words
  that enable testing an implementation against a specification:
  when a (black-box) automaton passes the test suite, completeness guarantees
  that it is equivalent to the specification on all (infinitely many) words.
  A well-known method for constructing complete test suites for deterministic automata
  is the W-method, used in conformance testing and automata learning.

  Based on a recent categorical generalisation of the W-method,
  we propose a framework for~the construction and analysis of test suites
  for a certain class of effectful automata defined by commutative monads.
  By working in the Eilenberg-Moore category of the associated monad at hand,
  we obtain a generic notion of \emph{effectful W-method}, together with a completeness theorem.
  We show how to apply the framework and derive complete test suites
  for non-deterministic and weighted automata.
\end{abstract}

\section{Introduction}

The \emph{W-method} is a celebrated result in conformance testing of
deterministic finite state machines (FSMs) by Chow~\cite{Cho:78} and Vasilevskii~\cite{Vas:73}.
The W-method generates a finite test suite based on a specification FSM.
Crucially, this test suite is \emph{complete},
meaning that any (black-box) implementation that passes the test suite
and has no more states than the specification is necessarily behaviourally equivalent.
Behavioural equivalence is a strong property:
it means that the two systems give equal outputs on any of the infinitely many inputs.
On top of that, the test suite generated by the W-method is only polynomial in size.

One area where conformance testing is used is \emph{active automata learning}.
Algorithms such as L*~\cite{Ang:87} and its descendants
assume access to an equivalence oracle, that is, a way of deciding
whether a hypothesis is correct with respect to the system under learning (SUL).
In active automata learning, the SUL is typically assumed to be black-box.
A common way to realise the equivalence oracle is through conformance testing,
where the hypothesis is viewed as the specification and the SUL as the implementation.
Libraries such as LearnLib~\cite{FHS:25} and AALpy~\cite{MAP+:21}
have implemented the W-method and its extensions.

Active automata learning has been generalised to various types of automata,
such as non-deterministic automata~\cite{BHK+:09},
weighted automata~\cite{BV:96},
stochastic systems with observable non-determinism~\cite{TAB+:21},
and many others, including categorical frameworks that generalise and extend
some of these algorithms~\cite{BKR:19,US:20,vHSS:20,CPS:21,AP:26}.
This motivates us to also extend the W-method to a broader class of automata,
since these algorithms all rely on equivalence queries.
In particular, the main challenge that we address in this paper is to develop
a framework for the W-method and its completeness that applies to types of automata
beyond deterministic ones, such as non-deterministic automata and weighted automata.

The aim of constructing test suites for non-deterministic FSMs is not new.
Petrenko and Yevtushenko describe a complete test suite in~\cite{PY:14}
with many references to prior work; we discuss more related work in~\cref{sec:related-work}.
In this line of research, the fault domain, \ie the class of models for which
the test suite is complete, is fixed: it is the class of automata of a certain size.
Our aim instead is to generalise the W-method specifically
without fixing our fault domain \emph{a priori}.
This idea of taking a more liberal fault domain follows~\cite{KR:25} and is strongly connected to
the notion of \emph{$k$-$A$-completeness} by Vaandrager and Melse~\cite{VM:25}.

To this end, we build on the generalised W-method developed in~\cite{KR:25}
at the level of automata in monoidal closed categories,
which includes instances such as deterministic automata, Mealy and Moore machines,
and deterministic nominal automata~\cite{BKL:11}.
Weighted automata over a field are also described in~\cite{KR:25},
but the examples there require additional analysis beyond what the framework offers,
and a systematic theory is missing.
In particular, non-deterministic automata and weighted automata over~$\Z$ are not covered in \opcit

Our approach is to develop a categorical framework for conformance testing
based on \emph{effectful automata}, \ie automata that have
an effectful transition structure modelled using a monad.
Formally, these are automata in the Kleisli category of a monad,
as already studied by Arbib and Manes~\cite{AM:75b},
and they naturally cover automata with
non-deterministic, weighted, or stochastic transition structure.
To obtain a general framework, we first refine the theory in~\cite{KR:25}
by assuming an arbitrary strong endofunctor for describing the transition structure of automata.
We then aim to instantiate this theory to effectful automata.

However, this leads to a technical challenge: the assumptions are not met
for automata in Kleisli categories, as these categories do not have enough structure
to construct the test suites required by the W-method.
Therefore, we exploit the algebraic structure of effectful automata
by working in Eilenberg-Moore categories of commutative monads.
Showing that the framework applies to these categories is non-trivial,
and it is one of the main contributions of this paper.

Based on this development, we reach our key contributions:
the notions of \emph{effectful state cover} and \emph{effectful state characteriser}
that are used to construct the \emph{effectful W-method}.
We derive the completeness of the effectful W-method
by instantiating the general theory to Eilenberg-Moore categories.
Furthermore, we illustrate the effectful W-method by deriving complete test suites
for non-deterministic automata and weighted automata over the ring~$\defring$.

The paper is structured as follows.
\Cref{sec:automata} provides some background on categorical automata theory.
In~\cref{sec:w-method}, we present our generalised W-method and its completeness theorem.
In~\cref{sec:effectful-automata}, we describe our setting for
studying effectful automata in Kleisli and Eilenberg-Moore categories of commutative monads.
We then introduce the effectful W-method
and prove its completeness in~\cref{sec:effectful-testing}.
\Cref{sec:applications} covers applications of the theory to
non-deterministic automata and weighted automata over~$\defring$.
Finally, we conclude with related work in~\cref{sec:related-work}
and a discussion in~\cref{sec:conclusion}.

We assume familiarity with basic category theory,
such as categories, functors, natural transformations,
limits, colimits, adjunctions, and monads.
These notions are covered by standard textbooks (\eg \cite{Mac:78,AHS:06,Awo:10,Rie:16}).
We also rely heavily on symmetric monoidal closed categories and some further categorical notions,
which are listed in~\cref{sec:prelim}.

\section{Automata in categories}\label{sec:automata}

We recall the basics of
automata in a category (see \eg \cite{AM:75a,AT:90,Gog:75}).
Throughout the paper, we fix a \emph{base category}~$\bcat$, a functor~$\dyn \colon \bcat \to \bcat$,
an \emph{initial state object}~$\inobj$, and an \emph{output object}~$\outobj$.

\begin{definition}\label{def:aut}
  \An{\dyn} \emph{$\dyn$-automaton} in~$\bcat$ consists of
  a \emph{state space}~$Q \in \bcat$,
  a \emph{transition morphism}~$\delta \colon \dyn Q \to Q$,
  an \emph{input morphism}~$i \colon \inobj \to Q$,
  and an \emph{output morphism}~$o \colon Q \to \outobj$.
\end{definition}
If the functor~$\dyn$ is clear from context, we simply use the term \emph{automaton}.

\begin{example}\label{ex:dfa}
  Let~$\bcat$ be the category~$\Set$ of sets and functions,
  let~$\dyn$ be defined as $\dyn A \leteq A \prodset \abcobj$
  for some finite alphabet~$\abcobj$,
  let~$\inobj$ be the singleton set~$\1 \defeq \set{*}$,
  and let~$\outobj$ be the two-element set~$\2 \defeq \set{0,1}$.
  Then \an{\dyn} $\dyn$-automaton is given by a set of states~$Q$,
  a transition function~$\delta \colon Q \prodset \abcobj \to Q$,
  an initial state~$i \colon \1 \to Q$,
  and an acceptance predicate~$o \colon Q \to \2$.
  Thus, \an{\dyn} $\dyn$-automaton with a finite state space
  is a \emph{deterministic finite automaton} (DFA)~\cite{RS:59}.
\end{example}

\begin{example}\label{ex:seq-aut}
  The previous example can be generalised by replacing the base category~$\Set$
  with an arbitrary \emph{monoidal category}~$(\bcat, \tprod, \tunit)$ (see \cref{def:mon-cat}).
  For such a category~$\bcat$ and any alphabet object~$\abcobj \in \bcat$,
  we can set $\dyn A \leteq A \tprod \abcobj$.
  Letting $\inobj \leteq \tunit$ while leaving the output object~$\outobj$ arbitrary,
  we obtain the notion of \emph{sequential automata} as defined by
  Ad{\'a}mek and Trnkov{\'a}~\cite{AT:90}, also called \emph{machines} by Goguen~\cite{Gog:75}.
  This notion encompasses many instances such as weighted automata~\cite{Sch:61},
  semilattice automata~\cite{MAM+:14}, and nominal automata~\cite{BKL:11}.
\end{example}

The transition morphism of \an{\dyn} $\dyn$-automaton
is an \emph{algebra} for the functor~$\dyn$.
In general, \an{\dyn} \emph{$\dyn$-algebra} (or simply \emph{algebra})
is a pair~$(A, a)$ with $A \in \bcat$ and $a \colon \dyn A \to A$.
\An{\dyn} $\dyn$-algebra \emph{homomorphism}~$f \colon (A, a) \to (B, b)$
is a map~$f \colon A \to B$ such that $f \com a = b \com \dyn f$.
Together, these form a category~$\Alg{\dyn}$.
Moreover, the notion of algebra homomorphism extends to automata.
A \emph{homomorphism} of $\dyn$-automata
from~$(Q_1, \delta_1, i_1, o_1)$ to~$(Q_2, \delta_2, i_2, o_2)$
is \an{\dyn} $\dyn$-algebra map~$f \colon (Q_1, \delta_1) \to (Q_2, \delta_2)$
such that $f \com i_1 = i_2$ and $o_2 \com f = o_1$.
$\dyn$-automata and their homomorphisms form a category~$\Aut{\dyn}{\inobj}{\outobj}$.

The behaviour of automata can be expressed using free algebras.
Given an object~$A \in \bcat$, a \emph{free $\dyn$-algebra over~$A$}
consists of an object~$\fdyn A \in \bcat$ and morphisms
$\con[A] \colon \dyn\fdyn A \to \fdyn A$ and~$\emp[A] \colon A \to \fdyn A$
such that for every~$B \in \bcat$, $b \colon \dyn B \to B$, and~$f \colon A \to B$,
there exists a unique $\dyn$-algebra morphism~$\fext{f} \colon (\fdyn A, \con[A]) \to (B, b)$
with $\fext{f} \com \emp[A] = f$.
We refer to~$\fext{f}$ as the \emph{free $\dyn$-extension} of~$f$.
For an algebra~$(A, a)$,
we write~$\idext{a}$ for~$\fext{\id[A]} \colon (\fdyn A, \con[A]) \to (A, a)$.
If free $\dyn$-algebras exist over all objects of~$\bcat$,
we say that $\dyn$~is a \emph{varietor}~\cite{AT:90}.
The following result provides a useful sufficient condition
for verifying that a functor is a varietor.
\begin{lemma}[\cite{Ada:74,Kel:80}]\label{lem:varietor}
  If $\bcat$~has and $\dyn$~preserves colimits of $\omega$-chains,
  then $\dyn$~is a varietor.
\end{lemma}

In the context of automata theory, $\fdyn\inobj$~generalises the set of words:
it is an object describing all possible inputs to an automaton.
This motivates the following definitions.
\begin{definition}\label{def:rea-beh}
  Let~$\dyn$ be a varietor and let~$\aut{A} = (Q, \delta, i, o)$ be \an{\dyn} $\dyn$-automaton.
  \begin{itemize}
	  \item A \emph{behaviour} (or \emph{language}) is a morphism~$\fdyn\inobj \to \outobj$.
    \item The \emph{reachability map}~$\rea{\aut{A}} \colon \fdyn\inobj \to Q$ of~$\aut{A}$
    is the free $\dyn$-extension of the input morphism~$i$.
    \item The \emph{behaviour}~$\beh{\aut{A}} \colon \fdyn\inobj \to \outobj$ of~$\aut{A}$
    is the composite~$\fdyn\inobj \xto{\rea{\aut{A}}} Q \xto{o} \outobj$.
    \end{itemize}
\end{definition}
Intuitively, the reachability map
sends an input word to the state reached after processing that word.
The behaviour, then, is defined by observing the output of the corresponding~state.
\begin{example}\label{ex:dfa-rea-beh}
  Since $\Set$~is cocomplete and Cartesian closed,
  the functor~$\dyn \leteq (-) \prodset \abcobj$ corresponding to DFAs (\cref{ex:dfa})
  preserves colimits, so it is a varietor by~\cref{lem:varietor}.
  The set~$\fdyn\1$ is isomorphic to~$\words$, the set of words over~$\abcobj$.
  Therefore, a behaviour~$\words \to \2$ is precisely
  a language in the classical sense, \ie a set of words.

  The reachability and behaviour maps in~\cref{def:rea-beh} instantiate to the standard notions in the case of DFAs.
  Recall that, for a given DFA~$\aut{A} = (Q, \delta, q_0, \Gamma)$,
  the \emph{generalised transition function}~$\delta^* \colon Q \prodset \words \to Q$
  is defined inductively by $\delta^*(q, \empwd) \defeq q$
  and $\delta^*(q, va) \defeq \delta(\delta^*(q, v), a)$.
  Then the reachability map of~$\aut{A}$ is given by $\rea{\aut{A}}(v) = \delta^*(q_0, v)$,
  and the behaviour of~$\aut{A}$ coincides with the usual \emph{accepted language} of~$\aut{A}$:
  $\beh{\aut{A}}(v) = 1$ iff $\rea{\aut{A}}(v) \in \Gamma$.
\end{example}

An important automata-theoretic notion that is missing from~\cref{def:rea-beh}
is the \emph{observability map}.
For a DFA~$(Q, \delta, q_0, \Gamma)$,
the observability map~$\obs{\aut{A}} \colon Q \to \funcset{\words}{\2}$
sends a state~$q \in Q$ to the accepted language of~$q$,
defined as $\obs{\aut{A}}(q) \defeq \set{v \in \words}[\delta^*(q, v) \in \Gamma]$.

To define this generally,
we use a categorical analogue of the set of languages~$\funcset{\words}{\2}$.
We assume that our base category~$\bcat$ is \emph{monoidal closed} (\cref{def:mon-cat});
we use the notation~$\inthom{A}{B}$ for the internal hom.
Then we may construct the object~$\inthom{\fdyn\tunit}{\outobj}$,
thought of as the \emph{behaviour object},
and define the observability map as a morphism~$Q \to \inthom{\fdyn\tunit}{\outobj}$.
This approach to behaviours was introduced by Goguen~\cite{Gog:75}
in the case of sequential automata (\cref{ex:seq-aut}).

To be able to define the observability map this way for an arbitrary functor~$\dyn$,
we need a \emph{left strength} for~$\fdyn$ (see \cref{def:strength}).
If we assume, besides closedness,
that $\bcat$~is a \emph{symmetric monoidal category} (\cref{def:mon-cat}),
then any left strength~$\lstr$ for~$\dyn$ induces a
canonical left strength~$\flstr$ for~$\fdyn$~\cite[Proposition~6.7]{MU:22}.

\begin{definition}\label{def:run-obs}
  Suppose $\bcat$~is symmetric monoidal closed and
  $\dyn$~is a varietor with left strength~$\lstr$.
  Denote the induced strength for~$\fdyn$ by~$\flstr$,
  and let~$\aut{A} = (Q, \delta, i, o)$ be an automaton.
  \begin{itemize}
    \item The \emph{run map}~$\run{\aut{A}} \colon Q \tprod \fdyn\tunit \to Q$ of~$\aut{A}$
    is the composite
    \[
      Q \tprod \fdyn\tunit \xto{\flstr[Q,\tunit]} \fdyn(Q \tprod \tunit)
        \xto{\fdyn\runit[Q]} \fdyn Q \xto{\idext{\delta}} Q,
    \]
    where $\runit$~is the right unitor of~$\bcat$.
    \item The \emph{observability map}~$\obs{\aut{A}} \colon Q \to \inthom{\fdyn\tunit}{\outobj}$
    of~$\aut{A}$ is the currying of~$Q \tprod \fdyn\tunit \xto{\run{\aut{A}}} Q \xto{o} \outobj$.
  \end{itemize}
\end{definition}

\section{Generalised W-method}\label{sec:w-method}

In this section, we present a categorical version of the W-method for $\dyn$-automata,
following closely the testing framework introduced in~\cite{KR:25}.
The categorical W-method and the proof of its completeness presented in~\opcit
is a direct generalisation of the completeness proof in~\cite{Moe:20}
(also the appendix of~\cite{KJR:24}), which ultimately goes back to~\cite{Vas:73,Cho:78}.
We first give an overview of the W-method for DFAs, then present our categorical generalisation.

Conformance testing aims to decide the equivalence of two automata,
a known \emph{specification} and a black-box \emph{implementation},
by checking their behaviour on a well-chosen, finite set of inputs, called a \emph{test suite}.
The W-method~\cite{Vas:73,Cho:78} is a classical way
to construct such test suites for deterministic finite (Mealy-)automata.
It relies on two special sets of words:
a state cover is used to reach all (reachable) states, and
a characterisation set is used to distinguish all pairs of (inequivalent) states.

More precisely,
given a specification DFA~$\spec = (Q, \delta, q_0, \Gamma)$ over an alphabet~$\abcobj$,
a set of words~$C \subseteq \words$ is a called a \emph{state cover} for~$\spec$
if $\empwd \in C$ and for every reachable state~$q \in Q$, there exists a word~$v \in C$
(called an \emph{access sequence} for~$q$) such that $\delta^*(q_0, v) = q$.
Moreover, a set of words~$W \subseteq \words$ is called
a \emph{characterisation set} for~$\spec$ if $\empwd \in W$
and for any pair of states~$p, q \in Q$ with $\obs{\spec}(p) \ne \obs{\spec}(q)$,
there exists a word~$v \in W$ (called a \emph{distinguishing sequence} for $p$~and~$q$)
such that $\delta^*(p, v) \in \Gamma$ and $\delta^*(q, v) \notin \Gamma$ or vice versa.

Given such sets $C$~and~$W$ and a natural number~$k$,
we define a test suite~$T = C \cdot \abcobj^{\le k+1} \cdot W$,
where $\cdot$~denotes the concatenation of languages
and $\abcobj^{\le m}$~is the set of words of length at~most~$m$.
That is, $T$~is obtained by concatenating words from the state cover
with words from the characerisation set, with at~most~$k+1$~letters inserted in the middle.
This test suite, which we call the \emph{W-suite}, has the property that it is \emph{$k$-complete},
meaning that for any inequivalent automaton with at~most~$k$~more states than~$\spec$,
we can find a test case~$v \in T$ that witnesses the difference
(\ie one accepts~$v$ and the other rejects it).

\begin{figure}
  \centering
  \begin{tikzpicture}[automaton]
    \node[state,initial]   (q0)               {$q_0$};
    \node[state]           (q1) [right=of q0] {$q_1$};
    \node[state,accepting] (q2) [right=of q1] {$q_2$};
    \node[state,accepting] (q3) [right=of q2] {$q_3$};
    \path[->]
      (q0) edge                        node       {$a$} (q1)
      (q0) edge[loop above]            node       {$b$} ()
      (q1) edge[bend left=70,xscale=2] node       {$a$} (q3)
      (q1) edge[bend left]             node       {$b$} (q2)
      (q2) edge[bend left]             node       {$a$} (q1)
      (q2) edge[bend left=70,xscale=2] node       {$b$} (q0)
      (q3) edge[loop above]            node       {$a$} ()
      (q3) edge                        node[swap] {$b$} (q2);
  \end{tikzpicture}
  \caption{Deterministic automaton for the language~$L_1 = \words a \abcobj$ with $\abcobj = \set{a,b}$.}
  \label{fig:dfa-example}
\end{figure}
\begin{example}\label{ex:w-method-og}
  Consider the automaton~$\spec$ over the alphabet~$\abcobj = \set{a,b}$
  depicted in~\cref{fig:dfa-example}.
  In~$\spec$, the states~$q_0$, $q_1$, $q_2$, and~$q_3$
  are reached by the words $\empwd$, $a$, $ab$, and~$aa$, respectively.
  Hence, $C \leteq \set{\empwd, a, aa, ab}$~is a state cover for~$\spec$.
  Furthermore, $W \leteq \set{\empwd, a}$~is a characterisation set for~$\spec$:
  the word~$\empwd$ distinguishes the rejecting states ($q_0$~and~$q_1$)
  from the accepting states ($q_2$~and~$q_3$),
  and the word~$a$ distinguishes both $q_0$~from~$q_1$ and $q_2$~from~$q_3$.

  The W-suite with~$k = 0$ is then given by $T = C \cdot (\set{\empwd} \cup \abcobj) \cdot W$.
  Performing the calculations, we see that
  $T = \set{\empwd, a, aa, aaa, aaaa, aab, aaba, ab, aba, abaa, abb, abba, b, ba}$.
  Since the test suite~$T$ is $0$-complete, every automaton with at~most~$4$~states
  which agrees with~$\spec$ on the acceptance of the $14$~words in~$T$
  accepts exactly the language~$L_1 = \words a \abcobj$ of~$\spec$.
\end{example}

We now turn to the categorical generalisation of the W-method.
\begin{assumption}\label{ass:w-method}
  Throughout this section, we assume that
  $\bcat$~is a symmetric monoidal closed category with binary coproducts, that
  $\dyn$~is a strong functor, and that
  $\dyn$~is a varietor.
\end{assumption}

Categorically speaking, a subset of a set~$A$ corresponds to
an equivalence class of monomorphisms (injective functions) with codomain~$A$.
To simplify the theory, we instead work with \emph{generalised elements},
\ie arbitrary morphisms into~$A$.
Intuitively, the subset represented by a morphism~$f \colon B \to A$
is the image~$\im{f} \subseteq A$ of~$f$.
In particular, collections of inputs, such as test suites,
can be represented as morphisms into~$\fdyn\inobj$.
Although finiteness of test suites is crucial for testing in practice,
it is not required for the theory.
\begin{definition}\label{def:test-suite}
  A \emph{test suite} is a morphism in~$\bcat$ with codomain~$\fdyn\inobj$.
\end{definition}

Let us fix
a specification automaton~$\spec = (\specname, \delta_\specname, i_\specname, o_\specname)$.
An intuitive way to categorify state covers is
as a morphism~$c \colon C \to \fdyn\inobj$ such that $\rea{\spec} \com c$~is an epimorphism.
However, to prove completeness, we need an `internal' way to assign states to access sequences.
This suggests to demand that $\rea{\spec} \com c$~be a \emph{split} epi,
\ie that it have a right inverse~$\specname \to C$.
Unfortunately, as explained in~\cite{KR:25}, this condition is too strong in many categories.
Following~\opcit, we use an alternative definition that equips~$C$ with an algebra structure.

To state the definition, we need some notation.
Given maps~$f,g$ with common codomain~$X$, a morphism~$f \to g$
is a map~$h$ such that $f = g \com h$ (\ie a map in the slice category~$\bcat/X$).
Regarding $f$~and~$g$ as subsets of~$X$, such a morphism expresses that $f$~is a subset of~$g$.
In particular, for DFAs (\cref{ex:dfa}), where $\fdyn\inobj \cong \words$ (\cref{ex:dfa-rea-beh}),
if $c \colon A \to \fdyn\inobj$~is thought of as a collection of words, then
the existence of a map~$\emp[\inobj] \to c$ intuitively means that `$c$~contains the empty word'.
\begin{definition}\label{def:st-cov}
  A \emph{state cover} for~$\spec$ is a morphism~$c \colon C \to \fdyn\inobj$
  together with maps $\delta_C \colon \dyn C \to C$~and~$i_C \colon \emp[\inobj] \to c$
  such that $\rea{\spec} \com c$~is
  \an{\dyn} $\dyn$-algebra morphism~$(C, \delta_C) \to (\specname, \delta_\specname)$.
\end{definition}
The condition that $\rea{\spec} \com c$~be an algebra map
is equivalent to the following commuting diagram:%
\begin{equation}\label{eq:st-cov}
  \begin{tikzcd}
    {\dyn C} & {\dyn\fdyn\inobj} & {\fdyn\inobj} \\
    C & {\fdyn\inobj} & \specname
    \arrow["{\dyn c}", from=1-1, to=1-2]
    \arrow["{\delta_C}"', from=1-1, to=2-1]
    \arrow["{\con[\inobj]}", from=1-2, to=1-3]
    \arrow["{\rea{\spec}}", from=1-3, to=2-3]
    \arrow["c"', from=2-1, to=2-2]
    \arrow["{\rea{\spec}}"', from=2-2, to=2-3]
  \end{tikzcd}
\end{equation}

The idea is that $C$~is a set of representative access sequences for the
reachable states of~$\spec$, and $\delta_C$~imitates the transition structure of~$\spec$ on~$C$.
For~$\dyn \leteq (-) \prodset \abcobj$, the intuitive reading of
diagram~\labelcref{eq:st-cov} is the following: for each $v \in C$~and~$s \in \abcobj$,
the word~$\delta_C(v,s) \in C$ is an access sequence for the state~$\rea{\spec}(vs)$.
The map~$\delta_C$ is part of the structure because it is required to be a morphism in~$\bcat$.
In general, there may be many such maps,
but the choice does not matter as far as completeness is concerned.
In~$\Set$, some~$\delta_C$ can always be defined if $C$~is a state cover
in the original sense of the word (\cf \cref{prop:effectful-st-cov-set}).

The condition for a set~$W \subseteq \words$ containing~$\empwd$
to be a characterisation set can be stated in the following form: for all $p,q \in \specname$,
$\obs{\spec}(p) \cap W = \obs{\spec}(q) \cap W$ implies $\obs{\spec}(p) = \obs{\spec}(q)$.
That is, if two states are equivalent on words in~$W$, then they are equivalent on all words.
The definition of abstract characterisation sets (which we call \emph{state characterisers})
is obtained from this by replacing~$p,q \in \specname$
with generalised elements~$p,q \colon A \to \specname$.
\begin{definition}\label{def:state-characteriser}
  A \emph{state characteriser} for~$\spec$ is a pair of maps
  $w \colon W \to \fdyn\tunit$~and~$i_W \colon \emp[\tunit] \to w$
  \sth for any~$p,q \colon A \to \specname$, if
  \(
    \specname \xto{\obs{\spec}} \inthom{\fdyn\tunit}{\outobj}
              \xto{\inthom{w}{\id}} \inthom{W}{\outobj}
  \)
  equalises $p$~and~$q$, then so does~$\obs{\spec}$.
\end{definition}
We often omit the $\delta_C$, $i_C$, and $i_W$~components of state covers and state characterisers,
in which case they are understood to be existentially quantified.

Before defining the W-suite, we introduce auxiliary operations related to concatenation.
Intuitively, the map~$\trconc$ below is a generalisation of word concatenation,
the operation~$\tsconc$ extends concatenation to sets of words,
and $\trext{a}[k]$~consists of words that are obtained by appending
an arbitrary sequence of at~most~$k$~letters to the end of some word in~$a$.
\begin{definition}\label{def:word-extension}\hfill
  \begin{itemize}
    \item The morphism~$\trconc \colon \fdyn\inobj \tprod \fdyn\tunit \to \fdyn\inobj$
    is defined as the composite
    \[
      \fdyn\inobj \tprod \fdyn\tunit
        \xto{\flstr[\fdyn\inobj,\tunit]} \fdyn(\fdyn\inobj \tprod \tunit)
        \xto{\fdyn\runit[\fdyn\inobj]} \fdyn\fdyn\inobj
        \xto{\idext{\con[\inobj]}} \fdyn\inobj.
    \]
    \item Given $a \colon A \to \fdyn\inobj$~and~$b \colon B \to \fdyn\tunit$,
    we define the map~$a \tsconc b \colon A \tprod B \to \fdyn\inobj$
    as~$\trconc \com (a \tprod b)$.
    \item Given~$a \colon A \to \fdyn\inobj$,
    we define the map~$\trext{a} \colon A \catsum \dyn A \to \fdyn\inobj$
    as~$\casem{a}{\con[\inobj] \com \dyn a}$.
    Furthermore, we write~$\trext*{-}[k]$ for the $k$-fold iteration of~$\trext*{-}$
    and denote the domain of~$\trext{a}[k]$ by~$\trext{A}[k]$.
  \end{itemize}
\end{definition}

\noindent The W-suite can now be constructed by concatenating words from the state cover with
words from the state characteriser, with at~most~$k+1$ arbitrary letters inserted in between.
\begin{definition}\label{def:w-suite}
  Given a state cover~$c$ for~$\spec$, a state characteriser~$w$ for~$\spec$, and a
  natural number~$k$, we define the \emph{W-suite} as~$\trext{c}[k+1] \tsconc w$.
\end{definition}

Test coverage may be expressed using a collection of automata called a \emph{fault domain}.
A test suite is then said to be \emph{complete} if for any inequivalent implementation
from the given fault domain, we can find a failing test case witnessing the error.
This is formally expressed as follows,
using the behaviour maps $\beh{\spec}$~and~$\beh{\impl}$ from~\cref{def:rea-beh}.
\begin{definition}\hfill
  \begin{itemize}
    \item A \emph{fault domain} is a class of automata,
    \ie a subclass of~$\Ob{\Aut{\dyn}{\inobj}{\outobj}}$.
    \item Let~$t$ be a test suite and let~$\fdom{U}$ be a fault domain.
    We say that $t$~is \emph{complete for~$\spec$ \wrt $\fdom{U}$}
    if for all~$\impl \in \fdom{U}$,
    $\beh{\spec} \com t = \beh{\impl} \com t$ implies $\beh{\spec} = \beh{\impl}$.
  \end{itemize}
\end{definition}

\noindent The fault domain corresponding to $k$-completeness for DFAs is
$\fdom{U}_k = \set{\impl}[\card{\impl} \le \card{\spec} + k]$,
where $\card{\aut{A}}$~denotes the number of states of~$\aut{A}$.
We instead consider the fault domain~$\wfdom{c}{k}$ that, intuitively, consists of those automata
each of whose states can be reached by a word from~$c$ followed by at~most~$k$ letters.
In the case of Mealy machines, completeness with respect to a similar fault domain
has been studied by Vaandrager and Melse~\cite{VM:25} (see also~\cite{Maa:20}).
\begin{definition}\label{def:w-fdom}
  Let~$c \colon C \to \fdyn\inobj$ and let~$k \in \N$.
  The fault domain~$\wfdom{c}{k}$ is defined as
  \(
    \wfdom{c}{k} \defeq
      \set{\impl \in \Aut{\dyn}{\inobj}{\outobj}}
          [\trext{c}[k]~\text{is a state cover for~$\impl$}]
  \).
\end{definition}
The theorem below states the completeness of the W-method.
An automaton~$\aut{A}$ is called \emph{observable} if $\obs{\aut{A}}$~is a monomorphism.
Observability is a sort of minimality condition: it means all states have different behaviour.
\begin{theorem}\label{thm:w-comp}
  Let~$c$ be a state cover for~$\spec$,
  let~$w$ be a state characteriser for~$\spec$,
  let~$k \in \N$,
  and suppose $\spec$~is observable.
  Then the W-suite~$\trext{c}[k+1] \tsconc w$ is complete for~$\spec$ \wrt $\wfdom{c}{k}$.
\end{theorem}

Instantiating the theory to DFAs, we recover the original W-method.
\begin{example}\label{ex:w-method-dfa}
  The setting of deterministic automata (\cref{ex:dfa}) satisfies~\cref{ass:w-method}:
  the category~$\Set$ is cocomplete and Cartesian closed.
  We saw in~\cref{ex:dfa-rea-beh} that
  the functor~$\dyn \leteq (-) \prodset \abcobj$ is a varietor.
  Its strength~$\lstr$ is given by $\lstr[A,B](a,(b,s)) = ((a,b),s)$.

  As explained above, the notions of state cover and state characteriser
  reduce to the classical notions for DFAs (after taking images).
  In particular, given a (traditional) state cover~$C$, we can define~$\delta_C(v,s)$
  simply by choosing any~$u \in C$ satisfying $\delta^*(q_0,u) = \delta^*(q_0,vs)$.

  Returning to the automaton~$\spec$ in~\cref{fig:dfa-example},
  we have the state cover~$C$ and the state characteriser~$W$ constructed in~\cref{ex:w-method-og}.
  The maps $c$~and~$w$ are simply the inclusions into~$\words$.
  Instantiating~\cref{def:w-suite} with~$k = 0$, the W-suite is then~$t = \trext{c} \tsconc w$.
  Concretely, we have
  \(
    T \defeq \dom{t}
      = (C \sumset \dyn C) \prodset W
      \cong C \prodset (\set{\empwd} \cup \abcobj) \prodset W
  \),
  and $t \colon T \to \words$~is given by $t(u, a, v) = uav$.
  The corresponding set of words~$\im{t}$ is precisely the set~$T$ from~\cref{ex:w-method-og}.

  Completeness of the test suite~$t$ now follows from~\cref{thm:w-comp}.
  Note that compared to~\cref{ex:w-method-og}, where we used $k$-completeness,
  the fault domain~$\wfdom{c}{0}$ in this case is different: it consists of automata
  for which $C$~is a state cover, which does not include all automata with at~most~$4$~states.
  However, as $k$~increases, $\wfdom{c}{k}$~can contain DFAs
  with exponentially more states than~$\spec$.
  (See~\cite{VM:25} for a comparison.)
\end{example}

\section{Effectful automata}\label{sec:effectful-automata}

The automaton model defined in~\cref{sec:automata} is an abstraction of deterministic automata.
However, as observed by Arbib and Manes~\cite{AM:75b},
it also covers instances such as non-deterministic, weighted, or stochastic automata
if we choose appropriate Kleisli categories as base categories.
Towards developing a theory of conformance testing covering such examples,
we therefore consider automata in Kleisli categories of commutative monads.
Since monads are often associated with effects in computer science,
we call these automata \emph{effectful}.

We wish to apply the abstract framework from~\cref{sec:w-method} to effectful automata.
Unfortunately, Kleisli categories in general are not closed,
so they do not satisfy~\cref{ass:w-method},
preventing us to construct the W-suite (\cref{def:w-suite}) and apply
the associated completeness result (\cref{thm:w-comp}) directly.
We therefore move
from the Kleisli category to the richer Eilenberg-Moore category (as in~\cite{AM:75b}),
which \emph{is} monoidal closed under certain conditions~\cite{Sea:13}.
Thus, automata in the Kleisli category provide a `syntactic' presentation,
whereas the actual computations happen in the Eilenberg-Moore category.

This section aims to provide the technical apparatus for
studying conformance testing of effectful automata in this way.
We first illustrate our high-level approach in~\cref{sec:effectful-overview}.
Then, in~\cref{sec:effectful-setting}, we describe the technical ingredients precisely
and verify that the assumptions of the testing framework (\cref{ass:w-method})
hold in the Eilenberg-Moore category.

\subsection{Overview: automata in Kleisli and Eilenberg-Moore categories}\label{sec:effectful-overview}

Let~$\eff$ be a monad describing an effect type,
and consider the left triangle in the diagram~below:%
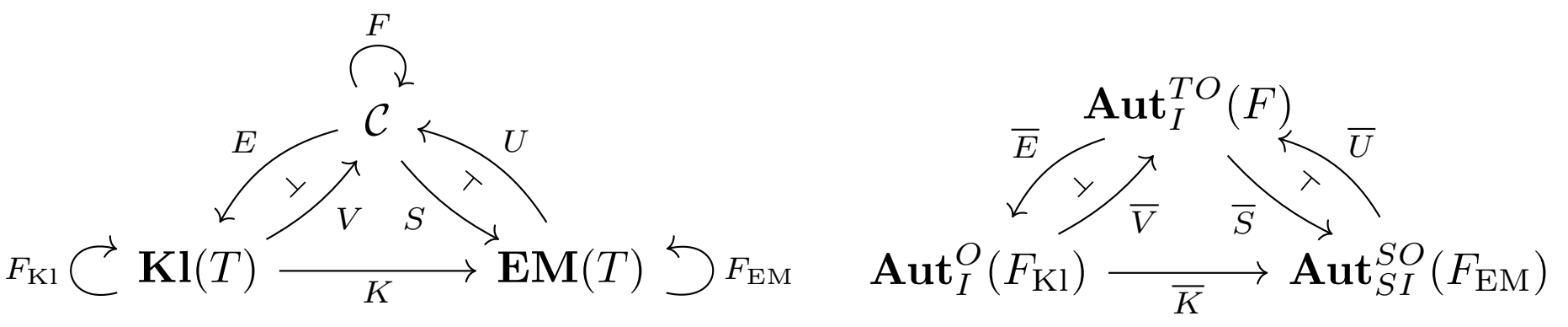
\begin{equation}\label{eq:aut-triangle}
  \begin{tikzcd}[baseline=(align.center),column sep=1.5em]
    & \bcat & \\
    \klcat && \emcat
    \arrow["\dyn", from=1-2, to=1-2, loop, in=60, out=120, distance=5mm]
    \arrow[""{name=0, anchor=center, inner sep=0}, "\klemb"', shift right, curve={height=5pt}, from=1-2, to=2-1]
    \arrow[""{name=1, anchor=center, inner sep=0}, "\emfree"'{pos=0.4}, shift right, curve={height=2pt}, from=1-2, to=2-3]
    \arrow[""{name=2, anchor=center, inner sep=0}, "\klext"'{pos=0.6}, shift right, curve={height=2pt}, from=2-1, to=1-2]
    \arrow["\kldyn", from=2-1, to=2-1, loop, in=165, out=-165, distance=5mm]
    \arrow["\klcomp"', from=2-1, to=2-3]
    \arrow[""{name=3, anchor=center, inner sep=0}, "\emforget"', shift right, curve={height=5pt}, from=2-3, to=1-2]
    \arrow["\emdyn"', from=2-3, to=2-3, loop, in=13, out=-13, distance=5mm]
    \arrow["\dashv"{anchor=center, rotate=-45}, draw=none, from=0, to=2]
    \arrow["\dashv"{anchor=center, rotate=50, name=align}, draw=none, from=1, to=3]
  \end{tikzcd}
  \hspace{1em}
  \begin{tikzcd}[baseline=(align.center),column sep=-1em]
    & {\Aut{\dyn}{\inobj}{\eff\outobj}} & \\
    {\Aut{\kldyn}{\inobj}{\outobj}} && {\Aut{\emdyn}{\emfree\inobj}{\emfree\outobj}}
    \arrow[""{name=0, anchor=center, inner sep=0}, "{\alglift{\klemb}}"', shift right, curve={height=4pt}, from=1-2, to=2-1]
    \arrow[""{name=1, anchor=center, inner sep=0}, "{\alglift{\emfree}}"'{pos=0.4}, shift right, curve={height=2pt}, from=1-2, to=2-3]
    \arrow[""{name=2, anchor=center, inner sep=0}, "{\alglift{\klext}}"'{pos=0.6}, shift right, curve={height=2pt}, from=2-1, to=1-2]
    \arrow["{\alglift{\klcomp}}"', from=2-1, to=2-3]
    \arrow[""{name=3, anchor=center, inner sep=0}, "{\alglift{\emforget}}"', shift right, curve={height=4pt}, from=2-3, to=1-2]
    \arrow["\dashv"{anchor=center, rotate=-45}, draw=none, from=0, to=2]
    \arrow["\dashv"{anchor=center, rotate=50, name=align}, draw=none, from=1, to=3]
  \end{tikzcd}
\end{equation}
Here, $\klemb \dashv \klext$~is the Kleisli adjunction,
$\emfree \dashv \emforget$~is the Eilenberg-Moore adjunction,
and $\klcomp$~is the comparison functor.
Our idea is to start with a base functor~$\dyn \colon \bcat \to \bcat$
that describes the transition structure of `plain', non-effectful automata,
and then `extend' it to functors $\kldyn \colon \klcat \to \klcat$
and~$\emdyn \colon \emcat \to \emcat$ that commute with the left adjoints.
The functors $\kldyn$~and~$\emdyn$ can then be used
for modelling effectful generalisations of $\dyn$-automata.
(This idea for Kleisli categories occurred already in~\cite{AM:75b} and also in~\cite{vHKR+:19}.)

The original situation can then be lifted to the induced categories of automata,
obtaining the triangle on the right of~\labelcref{eq:aut-triangle}.
Here, the functors that are overlined are induced by the respective functors in the first diagram.
In particular, the functor~$\alglift{\klcomp}$ provides a sort of
\emph{generalised determinisation} construction that also endows the resulting automaton
with algebraic structure, similarly to coalgebraic determinisation~\cite{JSS:15}.

\begin{example}\label{ex:running-example-nfa}
  We recall the notion of a \emph{non-deterministic finite automaton} (NFA)~\cite{RS:59}
  to be used as a running example.
  Fixing some finite alphabet~$\abcobj$, an NFA consists of
  a finite set~$Q$ of \emph{states},
  a \emph{transition function}~$\delta \colon Q \prodset \abcobj \to \mathcal{P}Q$,
  a set~$Q_0 \subseteq Q$ of \emph{initial} states, and
  a set~$\Gamma \subseteq Q$ of \emph{accepting} states.
  In the next section, we will show that this data is exactly captured by a (finite) object
  of~$\Aut{\kldyn}{\inobj}{\outobj}$ for suitable choices of $\kldyn$, $\inobj$, and~$\outobj$.
  For the effect~$\eff$, we take the (finite) powerset monad,
  which is generally associated with non-determinism.
  Then $\emdyn$-automata will turn out to be equivalent to JSL-automata~\cite{MAM+:14},
  and the functor~$\alglift{\klcomp}$ will coincide with the
  determinisation of NFAs via the powerset construction.
\end{example}

\subsection{From deterministic to effectful automata}\label{sec:effectful-setting}

In the remainder of the section, we give the necessary technical ingredients
for carrying out the idea outlined above.
The main goal is to verify the assumptions of the testing theory of~\cref{sec:w-method}
for automata in~$\emcat$ (\cref{thm:em-sat-ass}).
To achieve this, we make stronger assumptions on the base category compared to those in~\cref{ass:w-method},
which are however sufficient for the purposes of the present paper.
\begin{assumption}\label{ass:effectful-base}
  For the rest of the section, we assume that
  $\bcat$~is a complete and cocomplete symmetric monoidal closed category, and that
  $\dyn$~is strong and preserves colimits of $\omega$-chains.
\end{assumption}

Throughout this section, we fix a monad~$(\eff, \effmult, \effunit)$
and use the notation introduced in diagram~\labelcref{eq:aut-triangle}
for the associated categories and functors.
To make~$\emcat$ symmetric monoidal closed,
we need to assume that $\eff$~is a \emph{commutative monad} (\cref{def:comm-monad});
we denote its structure map by~$\effmf[A,B] \colon \eff A \tprod \eff B \to \eff(A \tprod B)$.
Moreover, we need some additional technical assumptions on~$\eff$
to guarantee the existence of colimits and free algebras in~$\emcat$.
\begin{assumption}\label{ass:monad}
  We assume that the monad~$\eff$ is commutative,
  that $\emcat$~has reflexive coequalisers,
  and that $\eff$~preserves colimits of $\omega$-chains.
\end{assumption}

We first consider the problem of extending~$\dyn$ to~$\klcat$.
As discussed above, we need a functor~$\kldyn \colon \klcat \to \klcat$
such that $\kldyn$-automata are effectful generalisations of $\dyn$-automata.
This can be achieved by demanding that $\dyn$~and~$\kldyn$ commute with the embedding~$\klemb$
of~$\bcat$ into the Kleisli category~$\klcat$, that is, $\kldyn\klemb = \klemb\dyn$.
We call \an{\kldyn}~$\kldyn$ satisfying this equation a \emph{Kleisli extension} of~$\dyn$.
Kleisli extensions are easily obtained using \emph{distributive laws}
(also called \emph{$\mathcal{K}\ell$-laws} in~\cite{JSS:15}; see \cref{def:dist-law}).
\begin{theorem}[\cite{AM:75b,Mul:94}]\label{thm:dlaw-kldyn-bi}
  There is a bijective correspondence between Kleisli extensions of~$\dyn$
  and distributive laws~$\dlaw \colon \dyn\eff \to \eff\dyn$ of~$\dyn$ over~$\eff$.
  Given~$\dlaw$, the Kleisli extension~$\kldyn$ is defined as
  $\kldyn A \defeq \dyn A$ for~$A \in \bcat$ and
  $\kldyn f \defeq \dlaw[B] \com \dyn f$ for~$f \colon A \klto B$.
\end{theorem}
\begin{assumption}\label{ass:dist-law}
  We assume a distributive law~$\dlaw \colon \dyn\eff \to \eff\dyn$
  of~$\dyn$ over~$\eff$, inducing a Kleisli extension~$\kldyn \colon \klcat \to \klcat$ of~$\dyn$.
\end{assumption}

Thus, $\Aut{\kldyn}{\inobj}{\outobj}$ is the category of effectful automata.
\An{\kldyn} $\kldyn$-automaton is like an ordinary $\dyn$-automaton,
but whose transition map returns $\eff$-structured collections of states.
\begin{example}[continuing~\cref{ex:running-example-nfa}]\label{ex:running-example-nfa-kl}
  The \emph{finitary powerset} monad~$\powset$ is defined on sets as
  $\powset A \defeq \set{X \subseteq A}[X\text{ is finite}]$.
  The unit~$\effunit[A][\powset] \colon A \to \powset A$ is given by $a \mapsto \set{a}$
  and the multiplication~$\effmult[A] \colon \powset\powset A \to \powset A$ is given by
  \(
    \mathcal{X} \mapsto \bigcup_{X \in \mathcal{X}} X
      = \set{a \in A}[\exists X \in \mathcal{X}.\,a \in X]
  \).
  The Kleisli category~$\Kl{\powset}$ models finitely branching non-determinism:
  a Kleisli map~$A \klto B$ maps an input in~$A$ to a finite set of possible outputs in~$B$.

  Since NFAs are the non-deterministic extension of DFAs,
  we continue~\cref{ex:dfa} and add the monad~$\powset$.
  We set $\bcat \leteq \Set$,
  $\dyn A \leteq A \prodset \abcobj$ for some finite alphabet~$\abcobj$,
  $\eff \leteq \powset$, and $\inobj \leteq \1$.
  However, instead of~$\2$, we take~$\outobj \leteq \1$ (since $\powset\1 \cong \2$).
  The arguments given in~\cref{ex:dfa-rea-beh,ex:w-method-dfa}, combined with the
  completeness of~$\Set$, show that \cref{ass:effectful-base} holds in this example.
  \Cref{ass:monad} is also satisfied: commutativity is shown by
  the maps $\effmf[A,B] \colon \powset A \prodset \powset B \to \powset(A \prodset B)$
  given by $(X,Y) \mapsto X \prodset Y$.
  Lastly, for~\cref{ass:dist-law}, we use
  the distributive law~$\dlaw[A] \colon \powset A \prodset \abcobj \to \powset(A \prodset \abcobj)$
  given by $(X, a) \mapsto X \!\prodset\! \set{a}$ (as in~\cite{AM:75b,HJS:07}).

  Hence, we obtain a Kleisli extension~$\kldyn$ with $\kldyn A = A \prodset \abcobj$.
  Spelling out~\cref{def:aut}, we get that a finite $\kldyn$-automaton consists of:
  a finite set~$Q$ of states;
  a non-deterministic transition function~$\delta \colon Q \prodset \abcobj \klto Q$;
  a function~$i \colon \1 \klto Q$, corresponding to a set~$i(*)$ of initial states; and
  a function~$o \colon Q \klto \1$, corresponding to a set~$\inv{o}(\1)$ of accepting states.
  Thus, we see that finite $\kldyn$-automata are equivalent to NFAs
  as defined in~\cref{ex:running-example-nfa}.
\end{example}

Similarly to the Kleisli extension, we would also like some sort of extension of~$\dyn$ to~$\emcat$.
Since the objects of~$\emcat$ are not (necessarily) in bijection with those of~$\bcat$,
it is unrealistic to expect a \emph{strict} extension,
\ie one that commutes with~$\emfree$ up to equality.
Instead, we are after a functor~$\emdyn \colon \emcat \to \emcat$ such that
there is a natural isomorphism~$\emdyn\emfree \ito \emfree\dyn$.
We call such \an{\emdyn}~$\emdyn$ an \emph{Eilenberg-Moore extension} of~$\dyn$.

Mulry~\cite{Mul:02} provided a sufficient condition for the existence of Eilenberg-Moore extensions.
Below, in~\cref{thm:dlaw-emdyn-equiv},
we recall this result and extend it with a uniqueness argument.
For this, we use that any $\eff$-algebra~$(A, a)$ arises as the coequaliser~\(
  \begin{tikzcd}[cramped]
    {\emfree\eff A} & {\emfree A} & {(A, a)}
    \arrow["{\emfree a}"', shift right, from=1-1, to=1-2]
    \arrow["{\effmult[A]}", shift left, from=1-1, to=1-2]
    \arrow["a", from=1-2, to=1-3]
  \end{tikzcd}
\),
called the \emph{canonical presentation} of~$(A, a)$~\cite{Mac:78}.

\begin{theorem}\label{thm:dlaw-emdyn-equiv}
  Given a distributive law~$\dlaw \colon \dyn\eff \to \eff\dyn$ of~$\dyn$ over~$\eff$,
  there exist a functor $\emdyn \colon \emcat \to \emcat$
  and a natural isomorphism~$\emfreedyn \colon \emdyn\emfree \ito \emfree\dyn$ such that
  \begin{enumerate}[(1)]
    \item\label{prop:emdyn-pres-can} $\emdyn$~preserves canonical presentations, and
    \item\label{prop:emfreedyn-effmult} the following diagram commutes:
  \end{enumerate}
    \begin{equation}\label{eq:emfreedyn-effmult}
      \begin{tikzcd}
        {\emdyn\emfree\eff} & {\emfree\dyn\eff} & {\emfree\eff\dyn} \\
        {\emdyn\emfree} && {\emfree\dyn}
        \arrow["{\emfreedyn\eff}", from=1-1, to=1-2]
        \arrow["{\emdyn\effmult}"', from=1-1, to=2-1]
        \arrow["{\emfree\dlaw}", from=1-2, to=1-3]
        \arrow["{\effmult\dyn}", from=1-3, to=2-3]
        \arrow["\emfreedyn"', from=2-1, to=2-3]
      \end{tikzcd}
    \end{equation}
    Further, for any functor~$\emdyn' \colon \emcat \to \emcat$ preserving canonical presentations
    and any natural isomorphism~$\emfreedyn' \colon \emdyn'\emfree \ito \emfree\dyn$
    making~\labelcref{eq:emfreedyn-effmult} commute,
    there is a unique natural transformation~$\iota \colon \emdyn' \to \emdyn$
    with $\emfreedyn \circ \iota\emfree = \emfreedyn'$,
    and this $\iota$~is a natural isomorphism.
\end{theorem}

Since we already assumed a distributive law~$\dlaw$,
we can apply~\cref{thm:dlaw-emdyn-equiv} to obtain a functor~$\emdyn \colon \emcat \to \emcat$
and a natural isomorphism $\emfreedyn \colon \emdyn\emfree \ito \emfree\dyn$ as above.
Then $\emdyn$-automata can be seen as algebraic versions of $\dyn$-automata
whose state spaces are equipped with algebraic structure preserved by the transition maps.
\begin{example}[continuing~\cref{ex:running-example-nfa-kl}]\label{ex:running-example-nfa-em}
  We describe the Eilenberg-Moore extension~$\emdyn$ for NFAs.
  An algebra for~$\powset$ is precisely a \emph{join-semilattice} (JSL),
  \ie a set~$A$ equipped with an associative, commutative, and idempotent
  binary operation~$\latjoin \colon A \prodset A \to A$ and a unit~$\latzero \in A$ for~$\latjoin$.
  A \emph{JSL-homomorphism} is a function that preserves $\latjoin$~and~$\latzero$.
  The \emph{direct sum}~$A \oplus B$ of JSLs $A$~and~$B$
  is the Cartesian product~$A \prodset B$ equipped with pointwise join and unit operations.
  It is the categorical coproduct in the category of JSLs and JSL-homomorphisms,
  with coprojections~$\coprojl(a) = (a, \latzero)$ and~$\coprojr(b) = (\latzero, b)$,
  and cotupling~$\casem{f}{g}(a, b) = f(a) \latjoin g(b)$.
  
  The Eilenberg-Moore extension~$\emdyn$ is then given by
  $\emdyn A = \bigoplus_{s \in \abcobj} A = A \oplus \dotsb \oplus A$.
  We write~$\coproj{s} \colon A \to \bigoplus_{s \in \abcobj} A$
  for the coprojection at~$s$.
  The map~\(
    \emfreedyn[A] \colon \bigoplus_{s \in \abcobj} \powset A \to \powset(A \prodset \abcobj)
  \)
  is the cotupling of the maps~$\emfreedyn[A,s] \colon \powset A \to \powset(A \prodset \abcobj)$
  given by $\emfreedyn[A,s](X) = X \prodset \set{s}$, and its inverse~$\inv{\emfreedyn[A]}$
  sends a set~$X \subseteq A \prodset \abcobj$ to the tuple~$(X_s)_{s \in \abcobj}$
  where $X_s = \set{a \in A}[(a,s) \in X]$.

  Spelling out~\cref{def:aut} with $\inobj \leteq \outobj \leteq \powset\1 \cong \2$
  (ordered by $0 < 1$), \an{\emdyn} $\emdyn$-automaton is given by:
  a join-semilattice~$Q$;
  a JSL-homomorphism~$\delta \colon \bigoplus_{s \in \abcobj} Q \to Q$, corresponding to a family
  of JSL-homomorphisms~$\delta_s \defeq \delta \com \coproj{s} \colon Q \to Q$ ($s \in \abcobj$);
  a JSL-homomorphism~$i \colon \2 \to Q$, corresponding to
  an initial state~$q_0 \defeq i(1) \in Q$;
  and a JSL-homomorphism~$o \colon Q \to \2$, corresponding to
  a set~$\Gamma \defeq \inv{o}(1) \subseteq Q$ of accepting states
  satisfying $\latzero \notin \Gamma$ and
  $p \latjoin q \in \Gamma$ iff $p \in \Gamma$ or $q \in \Gamma$.
  Thus, $\emdyn$-automata are equivalent to \emph{JSL-automata}~\cite{MAM+:14}.
\end{example}

\begin{remark}
  $\emdyn$-algebras are equivalent to the $\lambda$-algebras of Arbib and Manes~\cite{AM:75b}.
  The latter are defined as objects equipped with both
  \an{\dyn} $\dyn$-algebra (qua functor) and \an{\eff} $\eff$-algebra (qua monad) structure
  satisfying a compatibility axiom involving the distributive law~$\dlaw$.
\end{remark}

The following lemma expresses that the family~$\emfreedyn[A]$ is natural \wrt Kleisli maps.
\begin{lemma}\label{lem:klcompdyn}
  The family~$\emfreedyn$ extends to a
  natural isomorphism~$\klcompdyn \colon \emdyn\klcomp \ito \klcomp\kldyn$.
\end{lemma}

Putting everything so far together, we are now in the situation
on the left of diagram~\labelcref{eq:aut-triangle}.
The next step is to lift all these functors and adjunctions
to the respective categories of automata.
Hermida and Jacobs~\cite[Theorem~2.14]{HJ:98} showed
how to lift adjunctions to categories of algebras.
A straightforward adaptation of their proof to include initial state and output morphisms
gives us the adjunctions in the right triangle of diagram~\labelcref{eq:aut-triangle}.
\begin{theorem}\label{thm:triangle-lift}
	The triangle on the left in \labelcref{eq:aut-triangle} lifts to the triangle on the right in \labelcref{eq:aut-triangle}.
\end{theorem}
Here, the adjunction~$\alglift{\klemb} \dashv \alglift{\klext}$ is
induced by the identity natural transformation on~$\kldyn\klemb = \klemb\dyn$,
the adjunction~$\alglift{\emfree} \dashv \alglift{\emforget}$ is induced by~$\emfreedyn$,
and the functor~$\alglift{\klcomp}$ is induced by~$\klcompdyn$ from~\cref{lem:klcompdyn}.

Intuitively, the map~$\alglift{\klemb}$ turns a plain automaton into
an effectful automaton without any real effects.
It is analogous to how every DFA is also an NFA.
Conversely, the map~$\alglift{\klext}$ provides
a generalisation of the usual powerset construction for NFAs:
it turns an effectful automaton with state space~$Q$
into an equivalent plain automaton with state space~$\eff Q$.
The map~$\alglift{\klcomp}$ acts like~$\alglift{\klext}$,
but it additionally endows the resulting automaton with algebraic structure.
Finally, $\alglift{\emfree}$~is simply the composition $\alglift{\klcomp} \circ \alglift{\klemb}$,
and the map~$\alglift{\emforget}$ forgets the $\eff$-algebra structure.

\begin{example}[continuing~\cref{ex:running-example-nfa-em}]
  Using the definitions given so far, we see that in our running example,
  $\alglift{\klcomp}$~sends an NFA~$\aut{A} = (Q, \delta, i, o)$
  to the JSL-automaton~$(\powset Q, \bar{\delta}, \klcomp i, \klcomp o)$,
  where $\bar{\delta}_s \defeq \bar{\delta} \com \coproj{s}$~is the composite~\(
    \powset Q
      \xto{\emfreedyn[Q,s]} \powset(Q \prodset \abcobj)
      \xto{\klcomp\delta} \powset Q
  \).
  Unfolding definitions, we get that $\bar{\delta}_s$~is given by
  $\bar{\delta}_s(X) = \bigcup_{q \in X} \delta(q, s)$.
  Moreover, identifying $i$~and~$o$ with sets of states $Q_0$~and~$\Gamma$, respectively,
  $\klcomp i \colon \2 \to \powset Q$~corresponds to
  the state~$(\klcomp i)(1) = Q_0$ of~$\alglift{\klcomp}\aut{A}$,
  and $\klcomp o \colon \powset Q \to \2$~corresponds to the set~\(
    \overline{\Gamma} \defeq \inv*{\klcomp o}(1)
      = \set{X \in \powset Q}[X \cap\, \Gamma \ne \emptyset]
  \).
  Hence, the automaton~$\alglift{\klcomp}\aut{A}$ is precisely
  the determinisation of the NFA~$\aut{A}$ in the classical sense~\cite{RS:59}.
\end{example}

As our final step, we need to extend the given strength~$\lstr$ for~$\dyn$ to~$\emdyn$.
Since $\emdyn$~is defined in terms of the distributive law~$\dlaw$,
and the monoidal structure on~$\emcat$ relies on
the map~$\effmf$ of the commutative monad~$\eff$,
it is natural that we need a compatibility condition
between $\lstr$, $\dlaw$, and~$\effmf$ to achieve this.
\begin{assumption}\label{ass:dlaw-effmf-lstr}
  We assume that the following diagram commutes:
  \begin{equation}
    \begin{tikzcd}[column sep=3.5em]
      {\eff A \tprod \dyn\eff B} & {\dyn(\eff A \tprod \eff B)} & {\dyn\eff(A \tprod B)} \\
      {\eff A \tprod \eff\dyn B} & {\eff(A \tprod \dyn B)} & {\eff\dyn(A \tprod B)}
      \arrow["{\lstr[\eff A,\eff B]}", from=1-1, to=1-2]
      \arrow["{\id[\eff A] \tprod \dlaw[B]}"', from=1-1, to=2-1]
      \arrow["{\dyn\effmf[A,B]}", from=1-2, to=1-3]
      \arrow["{\dlaw[A \tprod B]}", from=1-3, to=2-3]
      \arrow["{\effmf[A,\dyn B]}"', from=2-1, to=2-2]
      \arrow["{\eff\lstr[A,B]}"', from=2-2, to=2-3]
    \end{tikzcd}
  \end{equation}
\end{assumption}
\begin{example}
  One can check that \cref{ass:dlaw-effmf-lstr} holds for NFAs (\cref{ex:running-example-nfa-kl}).
  More generally, if $\dyn = (-) \prodset \abcobj$ and $\lstr[A,B](a,(b,s)) = ((a,b),s)$,
  then the assumption holds for any commutative monad~$\eff$ if $\dlaw$~is defined as
  $\dlaw[A] = \effmf[A,\abcobj] \com (\id[\eff A] \prodset \effunit[\abcobj])$
  (as in~\cref{ex:running-example-nfa-kl}).
\end{example}

Under~\cref{ass:dlaw-effmf-lstr}, it can be shown that $\emdyn$~is a strong functor.
Furthermore, our assumptions on $\dyn$~and~$\eff$ ensure that
$\emdyn$~preserves colimits of $\omega$-chains; hence, it is a varietor (\cref{lem:varietor}).
The following theorem summarises these results.
\begin{theorem}\label{thm:em-sat-ass}
  If \cref{ass:effectful-base,ass:monad,ass:dist-law,ass:dlaw-effmf-lstr} hold,
  then \cref{ass:w-method} holds for~$\bcat \leteq \emcat$ and~$\dyn \leteq \emdyn$.
\end{theorem}

We conclude this section with a closer look at behaviours in $\klcat$~and~$\emcat$.
Recall from~\cref{sec:automata} that
the free $\dyn$-algebra over~$A \in \bcat$ is denoted by~$(\fdyn A, \con[A], \emp[A])$.
\begin{proposition}\label{prop:kldyn-free-alg}
  For all~$A \in \bcat$,
  the triple~$(\fdyn A, \klemb\con[A], \klemb\emp[A])$
  is a free $\kldyn$-algebra over~$A$.
  Moreover, for any automaton~$\aut{A} \in \Aut{\kldyn}{\inobj}{\outobj}$,
  we have $\rea{\aut{A}} = \rea{\alglift{\klext}\aut{A}}$
  and $\beh{\aut{A}} = \beh{\alglift{\klext}\aut{A}}$.
\end{proposition}
The first part of the above proposition shows that $\kldyn$~is a varietor;
hence, it makes sense to talk about the behaviour of $\kldyn$-automata.
In particular, a behaviour in~$\klcat$ is a morphism~$\fdyn\inobj \to \eff\outobj$.
The second part shows how to compute the reachability map and the behaviour
of an $\kldyn$-automaton: they coincide with those of its determinisation.

We already argued that $\emdyn$~is a varietor;
we denote the free $\emdyn$-algebra functor by~$\emfdyn$.
Since $\klcat$~is essentially the full subcategory of~$\emcat$ on the free $\eff$-algebras,
it is not surprising that we have an analogous proposition for~$\emcat$.
Let us write~$\mext{f} \colon \emfree A \to \malg{B}$
for the free $\eff$-extension of a map~$f \colon A \to B$
into \an{\eff} $\eff$-algebra~$\malg{B} = (B, b)$, given by $\mext{f} \defeq b \com \emfree f$.
\begin{proposition}\label{prop:emdyn-free-alg}
  For all~$A \in \bcat$,
  the triple~$(\emfree\fdyn A, \emfree\con[A] \com \emfreedyn[A], \emfree\emp[A])$
  is a free $\emdyn$-algebra over~$\emfree A$.
  Hence, there is a natural isomorphism~$\emfreefdyn \colon \emfdyn\emfree \ito \emfree\fdyn$
  such that for any automaton~$\aut{A} \in \Aut{\emdyn}{\emfree\inobj}{\emfree\outobj}$,
  we have $\rea{\aut{A}} = \mext{\rea{\alglift{\emforget}\aut{A}}} \com \emfreefdyn[\inobj]$
  and $\beh{\aut{A}} = \mext{\beh{\alglift{\emforget}\aut{A}}} \com \emfreefdyn[\inobj]$.
\end{proposition}

\section{Effectful test suites}\label{sec:effectful-testing}

As explained in~\cref{sec:effectful-automata},
we wish to test effectful automata in~$\klcat$ as automata in~$\emcat$.
In this section, we accomplish this program
and derive a special form of the W-method for $\emdyn$-automata,
which we call the \emph{effectful W-method}.
The effectful W-suite is constructed from
effectful versions of state covers and state characterisers.
We obtain the completeness of the effectful W-method
by applying the completeness theorem of the
`ordinary' categorical W-method (\cref{thm:w-comp}) to~$\emcat$.

We continue to work under~\cref{ass:effectful-base,ass:monad,ass:dist-law,ass:dlaw-effmf-lstr}
and employ the notations used so far.
We also fix a specification~$\spec \in \Aut{\emdyn}{\emfree\inobj}{\emfree\outobj}$
and write~$Q$ for its state space (which is \an{\eff} $\eff$-algebra).

\subparagraph*{The effectful W-method}

Whereas the general categorical definitions of state covers and state characterisers
in~\cref{sec:w-method} are stated in the category in which the specification lives,
the notions of \emph{effectful state cover} and \emph{effectul state characteriser}
introduced below are presented as structure in the base category~$\bcat$
instead of the category~$\emcat$ of~$\spec$.
Accordingly, the specification~$\spec$ is transported to~$\bcat$
using the functor~$\alglift{\emforget}$.

\begin{definition}\label{def:effectful-st-cov}
  An \emph{effectful state cover} for~$\spec$ is a map~$c \colon C \to \fdyn\inobj$ in~$\bcat$
  together with maps $\delta_C \colon \dyn C \to \eff C$~and~$i_C \colon \emp[\inobj] \to c$
  such that the following diagram in~$\bcat$ commutes:
  \begin{equation}\label{eq:effectful-st-cov}
    \begin{tikzcd}
      {\dyn C} & {\dyn\fdyn\inobj} & {\fdyn\inobj} \\
      {\eff C} & {\eff\fdyn\inobj} & Q
      \arrow["{\dyn c}", from=1-1, to=1-2]
      \arrow["{\delta_C}"', from=1-1, to=2-1]
      \arrow["{\con[\inobj]}", from=1-2, to=1-3]
      \arrow["{\rea{\alglift{\emforget}\spec}}", from=1-3, to=2-3]
      \arrow["{\eff c}"', from=2-1, to=2-2]
      \arrow["{\mext{\rea{\alglift{\emforget}\spec}}}"', from=2-2, to=2-3]
    \end{tikzcd}
  \end{equation}
\end{definition}
\Cref{def:effectful-st-cov} is formally quite similar
to~\cref{def:st-cov} stated for~$\alglift{\emforget}\spec$.
The main difference is that~$\delta_C$ now returns
\an{\eff} $\eff$-structured collection of elements of~$C$ instead of a single element.
In other words, we are allowed to use the algebraic structure of~$Q$ to reach states.

As explained in~\cref{sec:w-method}, the map~$\delta_C$ is included in the definition
because it is required to be a morphism in~$\bcat$.
In~$\Set$, however, there is an an equivalent characterisation of state covers without~$\delta_C$
that is analogous to the classical definition of state covers in terms of access sequences
(albeit in the more general context of effectful $\dyn$-automata).
\begin{proposition}\label{prop:effectful-st-cov-set}
  Suppose $\bcat = \Set$ and write~$R = \im{\rea{\alglift{\emforget}\spec}}$
  for the set of reachable states of~$\alglift{\emforget}\spec$.
  A set~$C \subseteq \fdyn\inobj$ is an effectful state cover for~$\spec$ iff
  $\emp[\inobj](a) \in C$ for all~$a \in \inobj$ and for all~$q \in R$,
  there exists~$z \in \eff C$ such that $\mext{\rea{\alglift{\emforget}\spec}}(\eff c(z)) = q$.
\end{proposition}

In contrast to effectful state covers, the algebraic structure of~$\spec$
is not used for the definition of effectful state characterisers.
Instead, we simply return to the base category via the functor~$\alglift{\emforget}$
and treat~$\spec$ as a plain deterministic automaton.
\begin{definition}\label{def:effectful-st-char}
  An \emph{effectful state characteriser} for~$\spec$ is
  a state characteriser for~$\alglift{\emforget}\spec$.
\end{definition}

\begin{definition}
  Given an effectful state cover~$c$ for~$\spec$,
  an effectful state characteriser~$w$ for~$\spec$, and a natural number~$k$,
  we define the \emph{effectful W-suite} as~$\trext{c}[k+1] \tsconc w$.
\end{definition}
Note that the construction in the definition above is the same as in~\cref{def:w-suite}.
However, the inputs are different; in particular, an effectful state cover for~$\spec$
is \emph{not} the same as a state cover for~$\alglift{\emforget}\spec$.
(The analogous statement for state characterisers is true by definition.)

Effectful state covers and state characterisers are equivalent to
\emph{freely generated} state covers and state characterisers in~$\emcat$,
which justifies the correctness of our definitions.
For~$a \colon A \to \fdyn B$, let us write~\(
  \emfreets{a} \defeq \inv*{\emfreefdyn[B]} \com \emfree a \colon \emfree A \to \emfdyn\emfree B
\),
where $\emfreefdyn$~is the map from~\cref{prop:emdyn-free-alg}.
We then call a morphism freely generated if it is of the form~$\emfreets{a}$.
Intuitively, if $t \colon T \to \fdyn\inobj$~is a test suite, viewed as a subset of~$\fdyn\inobj$,
then $\emfreets{t}$~is the subalgebra of~$\emfdyn\emfree\inobj \cong \emfree\fdyn\inobj$
freely generated by the elements of~$t$.
\begin{theorem}\label{thm:effectful-iff-free}
  Let~$C \xto{c} \fdyn\inobj$, $\dyn C \xto{\delta_C} \eff C$, $\emp[\inobj] \xto{i_C} c$,
  $W \xto{w} \fdyn\tunit$, and~$\emp[\tunit] \xto{i_W} w$.
  Then:
  \begin{enumerate}[(1)]
    \item\label{prop:effectful-iff-free:st-cov}
    $(c, \delta_C, i_C)$~is an effectful state cover for~$\spec$ iff
    $(\emfreets{c}, \klcomp\delta_C \com \emfreedyn[C], \emfree i_C)$~is a state cover for~$\spec$.
    \item\label{prop:effectful-iff-free:st-char}
    $(w, i_W)$~is an effectful state characteriser for~$\spec$ iff
    $(\emfreets{w}, \emfree i_W)$~is a state characteriser~for~$\spec$.
  \end{enumerate}
\end{theorem}

The following result states that under the correspondences of~\cref{thm:effectful-iff-free},
the W-suite (\cref{def:w-suite}) constructed in~$\emcat$ is essentially the same as
the effectful W-suite.
\begin{lemma}\label{lem:emfree-w-suite}
  Let~$c \colon C \to \fdyn\inobj$ be an effectful state cover for~$\spec$,
  let~$w \colon W \to \fdyn\tunit$ be an effectful state characteriser for~$\spec$,
  and let us write~$\emprod$ for the monoidal product in~$\emcat$.
  Then for any~$k \in \N$, there is an isomorphism~\(
    \trext*{\emfree C}[k+1] \emprod \emfree W
      \cong \emfree\mleft(\trext{C}[k+1] \tprod W\mright)
  \)
  making the following diagram commute:
  \[\begin{tikzcd}[column sep=-1.8em]
    {\trext*{\emfree C}[k+1] \emprod \emfree W} & \cong & {\emfree\mleft(\trext{C}[k+1] \tprod W\mright)} \\
    & {\emfdyn\emfree\inobj}
    \arrow["{\trext*{\emfreets{c}}[k+1] \hspace{0.5pt}\tsconc\hspace{1pt} \emfreets{w}}"'{pos=0.4}, from=1-1, to=2-2]
    \arrow["{\emfreets*{\trext{c}[k+1] \hspace{0.5pt}\tsconc\hspace{1pt} w}}"{pos=0.4}, from=1-3, to=2-2]
  \end{tikzcd}\]
\end{lemma}

\subparagraph*{Completeness of the effectful W-method}

\Cref{thm:em-sat-ass} shows that
the testing theory of~\cref{sec:w-method} is applicable to $\emdyn$-automata.
Combining our results on the effectful W-method with~\cref{thm:w-comp},
we obtain the following completeness theorem for $\emdyn$-automata.
To state the theorem, we introduce the effectful version
of the fault domain~$\wfdom{c}{k}$ from~\cref{def:w-fdom}:
for~$C \xto{c} \fdyn\inobj$, we write~\(
  \emwfdom{c}{k} \defeq
    \set{\impl \in \Aut{\emdyn}{\emfree\inobj}{\emfree\outobj}}
        [\trext{c}[k]~\text{is an effectful state cover for~$\impl$}]
\).
\begin{theorem}\label{thm:effectful-w-comp}
  Let~$c$ be an effectful state cover for~$\spec$,
  let~$w$ be an effectful state characteriser for~$\spec$,
  let~$k \in \N$,
  and suppose $\spec$~is observable.
  Then the test suite~$\emfreets*{\trext{c}[k+1] \tsconc w}$
  is complete for~$\spec$ \wrt $\emwfdom{c}{k}$.
\end{theorem}

Finally, we consider the special case when the specification~$\spec$
originates from the Kleisli category~$\klcat$.
For~$\spec \in \Aut{\kldyn}{\inobj}{\outobj}$,
we overload the term `effectful state cover for~$\spec$'
to mean `effectful state cover for~$\alglift{\klcomp}\spec$',
and similarly for `effectful state characteriser'.
Note that in this case, the automaton~$\alglift{\emforget}\hspace{1pt}\alglift{\klcomp}\spec$
used in~\cref{def:effectful-st-cov,def:effectful-st-char} coincides with~$\alglift{\klext}\spec$.
Furthermore, we write~\(
  \klwfdom{c}{k} \defeq
    \set{\impl \in \Aut{\kldyn}{\inobj}{\outobj}}
        [\trext{c}[k]~\text{is an effectful state cover for~$\impl$}]
\).
\begin{corollary}\label{cor:effectful-w-comp}
  Let~$\spec \in \Aut{\kldyn}{\inobj}{\outobj}$ be an automaton.
  Let~$c$ be an effectful state cover for~$\spec$,
  let~$w$ be an effectful state characteriser for~$\spec$,
  let~$k \in \N$,
  and suppose $\alglift{\klcomp}\spec$~is observable.
  Then the test suite~$\klemb\mleft(\trext{c}[k+1] \tsconc w\mright)$
  is complete for~$\spec$ \wrt $\klwfdom{c}{k}$.
\end{corollary}

\section{Applications}\label{sec:applications}

We instantiate the effectful W-method to
non-deterministic automata and weighted automata over the integers.
In both cases, we give example automata and derive a complete test suite.
Additional details for the examples can be found in~\cref{sec:app-details}.

\subsection{Nondeterministic automata}\label{sec:nfa-testing}

\begin{figure}
  \centering
  \begin{tikzpicture}[automaton]
    \node[state,initial]   (q0)               {$q_0$};
    \node[state]           (q1) [right=of q0] {$q_1$};
    \node[state]           (q2) [right=of q1] {$q_2$};
    \node[state,accepting] (q3) [right=of q2] {$q_3$};
    \path[->] (q0) edge[loop above] node {$a,b$} ()
              (q0) edge             node {$a$}   (q1)
              (q1) edge             node {$a,b$} (q2)
              (q2) edge             node {$a,b$} (q3);
  \end{tikzpicture}
  \caption{The non-deterministic automaton~$\exNFA[2]$ for the language~$\exlang[2] = \words a \abcobj^2$ with $\abcobj = \set{a,b}$.}
  \label{fig:nfa}
\end{figure}

Consider the family of regular languages~$\exlang[n] \defeq \words a \abcobj^n$
over~$\abcobj \leteq \set{a,b}$ for~$n \in \N$.
These are typical examples of languages accepted by a small NFA (with $n+2$~states)
whose determinisation is exponentially large (with $2^{n+1}$~states).
The NFA~$\exNFA[2]$ accepting~$\exlang[2]$ is depicted in~\cref{fig:nfa}.
The automata~$\exNFA[n]$ for arbitrary~$n$ look similar but with a longer chain of states.
We construct the effectful W-suite for~$\exNFA[2]$ as follows.

\subparagraph*{State cover}

Specialising~\cref{prop:effectful-st-cov-set} to NFAs (\cref{ex:running-example-nfa-kl}), we get:
\begin{proposition}
  Let~$\spec$ be an NFA and write~$R = \im{\rea{\alglift{\klext}\spec}}$
  for the set of reachable states of~$\alglift{\klext}\spec$.
  A set~$C \subseteq \words$ of words is an effectful state cover for~$\spec$ iff
  $\empwd \in C$ and for all~$q \in R$, there exist words~$v_1,\dotsc,v_n \in C$
  such that $\bigcup_{i=1}^n \rea{\alglift{\klext}\spec}(v_i) = q$.
\end{proposition}
Note that we can use unions (the algebraic structure of~$\powset Q$) to cover states,
meaning that the size of an effectful state cover for an NFA may be smaller
than a state cover for its determinised DFA.
In this example, one can check that the set~$C \leteq \set{\empwd, a, ab, abb}$
with just four elements is an effectful state cover for~$\exNFA[2]$, even though $\card{R} = 8$.

The above effectful state cover~$C$ is minimal \wrt the number of states.
In general, constructing a minimal effectful state cover is equivalent to
finding an access sequence for each reachable state that
cannot be expressed as a union of other reachable states.
Such states are precisely the \emph{join-irreducible} elements
of the subsemilattice of~$\powset Q$ generated by~$R$.
This is similar to the construction of the \emph{canonical residual finite state automaton}
of a language~$L$~\cite{DLT:02} (also called \emph{jiromaton} in~\cite{MAM+:14}).

A (not necessarily minimal) effectful state cover~$C$ for an NFA~$\spec$
can be computed incrementally, as follows.
Start with~$C \leteq \set{\empwd}$,
and perform a traversal on the determinisation of~$\spec$.
For each new state, check if it can be expressed as a union of states reached by~$C$.
If not, add its access sequence to~$C$; otherwise, discard the current branch.
Incidentally, this procedure also produces the map~$\delta_C$ (\cref{def:effectful-st-cov}).
Here, $\delta_C \colon C \prodset \abcobj \to \powset C$~is given by:
\begin{align*}
  \empwd, a &\mapsto \empwd \vee a  &  a, a &\mapsto \empwd \vee a \vee ab  &  ab, a &\mapsto \empwd \vee a \vee abb  &  abb, a &\mapsto \empwd \vee a \\
  \empwd, b &\mapsto \empwd         &  a, b &\mapsto \empwd \vee ab         &  ab, b &\mapsto \empwd \vee abb         &  abb, b &\mapsto \empwd
\end{align*}

\subparagraph*{State characteriser}

For an effectful state characteriser, we need to distinguish \emph{all}
pairs of inequivalent states in~$\powset Q$ (and not just the reachable ones).
We can take \eg the set~$W \leteq \set{\empwd, a, aa, aaa}$:
the element~$\empwd$ distinguishes all sets containing~$q_3$ from those not containing~$q_3$,
the element~$a$ does the same for~$q_2$, and so on.

\subparagraph*{Test suite}

We conclude the example by constructing the effectful W-suite for $k = 0$.
As in~\cref{ex:w-method-dfa}, it is the map~$t \colon T \to \words$
with $T = C \prodset (\set{\empwd} \cup \abcobj) \prodset W$ and $t(u, a, v) = uav$.
The resulting test suite has $24$~words (after removing duplicates).

When considering the effectful W-suite for the entire family~$\exNFA[n]$,
we observe that the sizes of effectful state covers and state characterisers
are linear in~$n$: both contain exactly $n+2$~elements.
This makes the resulting test suite \emph{quadratic} in~$n$ (for a fixed~$k$).
This is more efficient than constructing a W-suite for the determinised automata.
In particular, the automaton~$\alglift{\klext}\exNFA[n]$ has $2^{n+1}$~states,
which is thus the size of the minimal state cover.
This leads to a test suite of \emph{exponential} size.
(The state characteriser is still linear with $n+1$~elements.)

We should remark that \emph{completeness} means something different in these two testing scenarios.
In the non-deterministic case, the test suite is complete for the class of NFAs
for which $C = \set{\empwd, a, ab, abb}$~is an effectful state cover.
This also includes DFAs, but only those with at most~$\card{C} = 4$ states.
The deterministic W-suite is similar but the state cover contains more elements,
and hence the associated fault domain contains larger DFAs (but no NFAs).

\subsection{Weighted automata}\label{sec:wfa-testing}

We instantiate the automata-theoretic definitions of~\cref{sec:effectful-automata}
to obtain weighted finite automata (WFA), using weights from the ring~$\defring$.
As semirings, rings are quite special since they have additive inverses;
this greatly simplifies the computation of state covers and state characterisers.
However, even though we work with~$\defring$ in this example,
our theory in~\cref{sec:effectful-automata} applies in principle to
arbitrary commutative semirings, such as the tropical semiring.

Like for NFAs (\cref{ex:running-example-nfa-kl}),
we use the data~$\dyn \leteq (-) \prodset \abcobj$, $\inobj \leteq \outobj \leteq \1$
and choose the monad~\(
  \eff X \leteq \Lin{X} \defeq
    \set{v \colon X \to \defring}[v(x) \ne 0\text{ for finitely many~$x$}]
\)
to capture weighted transitions.
The set~$\Lin{X}$ consists of all \emph{formal linear combinations} of elements in~$X$.

Morphisms~$A \klto B$ in the Kleisli category~$\Kl{\LinFunct}$ are functions~$A \to \Lin{B}$;
these assign every input in~$A$ to a linear combination of elements in~$B$.
The Eilenberg-Moore category~$\EM{\LinFunct}$ is
the category of $\defring$-modules (\ie Abelian groups) and linear maps.

We already saw that \cref{ass:effectful-base} holds,
and one can also verify that \cref{ass:monad,ass:dist-law,ass:dlaw-effmf-lstr} are satisfied.
By~\cref{thm:dlaw-kldyn-bi,thm:dlaw-emdyn-equiv}, we obtain
both a Kleisli extension~$\kldyn$ and an Eilenberg-Moore extension~$\emdyn$.
Unfolding definitions, a finite $\kldyn$-automaton consists of:
a finite set~$Q$ of states;
a weighted transition function~$\delta \colon Q \prodset \abcobj \klto Q$, assigning
each state~$q \in Q$ and letter~$s \in \abcobj$ to a linear combination of successor states;
a morphism~$i \colon \1 \klto Q$, given by an initial vector in~$\defring^Q$; and
a morphism~$o \colon Q \klto \1$, given by a final vector in~$\defring^Q$.

Many variations of weighted automata appear in the literature
(see \eg \cite{Pag:66,DKV:09,BBB+:12,vHee:20,DK:21}),
mostly differing in the way transitions are defined (\eg using matrices).
These are equivalent to our definition (provided $Q$~and~$\abcobj$ are finite sets),
which is in line with the work of Arbib and Manes~\cite{AM:75a,AM:75b}.
We note that $\emdyn$-automata closely resemble
the \emph{linear weighted automata} of~\cite{BBB+:12},
and the functor~$\alglift{\klcomp}$ coincides with their `weighted powerset construction'
that maps a WFA to a deterministic automaton
whose states are formal linear combinations of states of the original WFA.
We refer to such formal linear combinations of states as \emph{configurations}.

\subparagraph*{Example WFA}

\begin{figure}
  \centering
  \begin{tikzpicture}[weighted automaton]
    \node[state,inweight=1]  (q0)               {$q_0$};
    \node[state]             (q1) [right=of q0] {$q_1$};
    \node[state]             (q2) [right=of q1] {$q_2$};
    \node[state,outweight=1] (q3) [right=of q2] {$q_3$};
    \path[->] (q0) edge[loop above] node {$a/2$ \\ $b/1$} ()
              (q0) edge             node {$a/1$}          (q1)
              (q1) edge             node {$a/1$ \\ $b/1$} (q2)
              (q2) edge             node {$a/1$ \\ $b/1$} (q3);
  \end{tikzpicture}
  \caption{The weighted automaton~$\exWFA[2]$.}
  \label{fig:wfa}
\end{figure}

We consider the weighted automata~$\exWFA[n]$ obtained from the NFAs~$\exNFA[n]$
defined in~\cref{sec:nfa-testing} by assigning weight~$2$
to the $a$-loop on~$q_0$ and weight~$1$ to all other transitions.
The initial weight is~$1$ for initial states and $0$~for the others,
and similarly, the final weight is~$1$ for accepting states and $0$~for the rest.
The case~$\exWFA[2]$ is depicted in~\cref{fig:wfa},
for which we now construct the effectful W-suite.

\subparagraph*{State cover}

In the case of WFAs, \cref{prop:effectful-st-cov-set} specialises to the following:
\begin{proposition}
  Let~$\spec$ be a WFA and write~$R = \im{\rea{\alglift{\klext}\spec}}$
  for the set of reachable configurations of~$\alglift{\klext}\spec$.
  A set~$C \subseteq \words$ of words is an effectful state cover for~$\spec$ iff
  $\empwd \in C$ and for all~$q \in R$,
  there exist~$v_1,\dotsc,v_n \in C$ and~$a_1,\dotsc,a_n \in \defring$
  such that $\sum_{i=1}^n a_i \rea{\alglift{\klext}\spec}(v_i) = q$.
\end{proposition}
Instead of unions, we are now allowed to use linear combinations of configurations
to cover additional configurations.
In particular, we can use subtraction and multiplication by a scalar,
which allows for much greater flexibility and thus leads to smaller state covers.

Analogously to the case of NFAs, an effectful state cover can be constructed incrementally
by performing a traversal on the configuration space and extending the state cover any time
we encounter a configuration that is linearly independent of previous ones.
In this example, $C \leteq \set{\empwd, a, aa, aaa}$~is an effectful state cover.
Again, this gives a map~$\delta_C \colon C \prodset \abcobj \to \Lin{C}$,
and we highlight one equation to show the linear structure:
$\delta_C(a, b) = 2 \cdot \empwd - 2 \cdot a + 1 \cdot aa$.

\subparagraph*{State characteriser}

We can show that, for a WFA~$\spec$ with state space~$Q$,
the observability map of the automaton~$\alglift{\klext}\spec$
is a linear map~$\Lin{Q} \to \funcset{\words}{\defring}$ from the configuration space
to the space of weighted languages (equipped with the pointwise structure induced by~$\defring$).
Thus, to distinguish configurations, we can use the module structure
of the configuration space: two states are equivalent on a set~$T$ if and only if
the language of their difference assigns~$0$ to all words in~$T$.
Hence, for the state characteriser, we need to pick a set of words~$W$ such that,
for all configurations~$v$, if $\obs{\alglift{\klext}\spec}(v)(u) = 0$ for all~$u \in W$,
then $\obs{\alglift{\klext}\spec}(v) = 0$.

As an example,
given the configuration~$v = x_0 \cdot q_0 + x_1 \cdot q_1 + x_2 \cdot q_2 + x_3 \cdot q_3$,
we have $\obs{\alglift{\klext}\spec}(v)(\empwd) = x_3$.
Therefore, the condition that the output be~$0$ yields the requirement~$x_3 = 0$.
Similarly, the input~$a$ yields the requirement~$x_2 = 0$,
$aa$~yields~$x_1 = 0$, and $aaa$~yields~$x_0 = 0$.
Hence, we can take~$W \leteq \set{\empwd, a, aa, aaa}$:
if a configuration~$v$ outputs zero on all these words,
then we know that $v = 0$ (and $\obs{\alglift{\klext}\spec}(v) = 0$ in particular).

\subparagraph*{Test suite}

Like before, the effectful W-suite is given by
$T = C \prodset (\set{\empwd} \cup \abcobj) \prodset W$, and it again has $24$~words.
For arbitrary~$n$, the minimal state cover and state characteriser for~$\exWFA[n]$
will both contain $n+2$~words, and the resulting test suite will be quadratic in~$n$.

\section{Related work}\label{sec:related-work}

Conformance testing of non-deterministic systems
has been considered before by Petrenko and Yevtushenko~\cite{PY:14}.
A big difference between our work and theirs is how non-determinism is observed.
In their setting, non-determinism means that each run may produce a different output,
even for a single test.
Hence, it may be necessary to run the same test multiple times.
In our case, a non-deterministic automaton always accepts a word
if there is \emph{some} accepting run.
In short, we assume the observable behaviour is a deterministic function~$\words \to \2$.

Another difference is the overall approach.
We try to generalise the known W-method and then describe the fault domain;
they fix the fault domain upfront (namely, all models of a certain size)
and then construct a complete test suite.
The work of Petrenko and Yevtushenko has lead to complete test suites
for other conformance relations, such as ioco~\cite{BJM:19}.
Moreover, it has been formalised by Sachtleben in Isabelle/HOL~\cite{Sac:22}.

Our framework in \cref{sec:w-method} is a mild generalisation of the one in~\cite{KR:25}:
the present version works with an arbitrary strong endofunctor
for describing the transition structure.
This level of abstraction simplifies many definitions and proofs.
Moreover, the framework covers deterministic tree automata;
the investigation of such an application is left for future work.
The main extension of the current paper is a systematic framework for effectful automata.

In van Heerdt's thesis~\cite{vHee:20},
the L* learning algorithm is generalised to a categorical framework.
An earlier article~\cite[Section~6]{vHSS:17} also includes an outline of a general approach
to conformance testing, based on cofree algebras and factorisation systems
instead of the monoidal closed structure in the current paper and~\cite{KR:25}.
However, the completeness theorem of~\cite{vHSS:17} has different conditions on the (implicit) fault domain:
notably, the implementations are assumed to be observable
and to have no more states than the specification.
Again, another key difference with the current paper is that
we focus explicitly on effectful automata.

Distributive laws of functors over monads (and monads over functors)
also play a key role in coalgebraic determinisation and trace semantics (\eg \cite{JSS:15,HJS:07}).
We follow an algebraic approach to automata in the style of Arbib and Manes
instead of a coalgebraic one.
Determinisation and minimisation for these types of automata was studied in~\cite{vHKR+:19}.

\section{Conclusion}\label{sec:conclusion}

We introduced a general categorical framework for testing effectful automata,
which are Arbib--Manes-style algebraic automata in the Kleisli category of a monad.
The key step underlying our approach was developing the theory of
the W-method and its completeness in the Eilenberg-Moore category of the given monad,
instead of its Kleisli category.
This analysis resulted in the \emph{effectful W-method} and its completeness.
We instantiated the theory to obtain test suites for non-deterministic automata
and weighted automata over the integers.

\subparagraph*{Future work}

Besides sequential (word-based) automata, our categorical testing framework in~\cref{sec:w-method}
also covers \emph{deterministic bottom-up tree automata}~\cite{CDG+:08}
by choosing a polynomial functor for~$\dyn$.
However, we did not pursue tree automata in the present paper because
\cref{ass:dlaw-effmf-lstr} is \emph{not} satisfied for
our main examples of interest ($\eff \leteq \powset$~and~$\eff \leteq \LinFunct$).
One possible approach to solve this issue is to involve \emph{contexts},
as done also \eg in~\cite{vHKR+:22}.

Probabilistic automata can be modelled as effectful automata
for the finite-support distribution monad; hence, our W-method applies here as well.
However, we think such stochastic models need additional analysis:
our test suite expects an exact output distribution for each test,
but probabilistic systems in practice often cannot guarantee exact measurements
and instead need to approximate the output distribution via sampling.

Our notion of state characteriser is a map~$W \to \fdyn\tunit$ satisfying a condition
on the induced restriction map~$\inthom{\fdyn\tunit}{\outobj} \to \inthom{W}{\outobj}$.
This means that more refined test suites such as
the Wp-method~\cite{FvBK+91} and the HSI-method~\cite{PvBY:96} cannot be captured,
as they may require this set to depend on the state (or rather, the behaviour of the state).
A possible generalisation is to consider a map~$\inthom{\fdyn\tunit}{\outobj} \to Z$
as the required data, similar to the notion of \emph{wrapper} in~\cite{vHee:20}.

In the examples that we considered, we did not have to minimise
when moving from the Kleisli category to the Eilenberg-Moore category.
However, this might not always be the case, and our completeness result
(\cref{thm:effectful-w-comp}) requires at least an observable specification.
In general, minimising can also be useful to obtain smaller test suites.
We leave an integration of the current framework
with algorithms for minimising automata for future work.

\bibliographystyle{plainurl}
\bibliography{refs}

\begin{thebibliography}{10}

\bibitem{Ada:74}
Ji{\v r}{\'i} Ad{\'a}mek.
\newblock Free algebras and automata realizations in the language of categories.
\newblock {\em Commentationes Mathematicae Universitatis Carolinae}, 015(4):589--602, 1974.
\newblock URL: \url{https://dml.cz/handle/10338.dmlcz/105583}.

\bibitem{AT:90}
Ji{\v r}{\'i} Ad{\'a}mek and V{\v e}ra Trnkov{\'a}.
\newblock {\em Automata and Algebras in Categories}, volume~37 of {\em Mathematics and its Applications}.
\newblock Springer Dordrecht, 1990.

\bibitem{AHS:06}
Jir{\v r}{\'i} Ad{\'a}mek, Horst Herrlich, and George~E. Strecker.
\newblock {\em Abstract and Concrete Categories: {The} Joy of Cats}.
\newblock TAC Reprints, 2006.
\newblock URL: \url{https://tac.mta.ca/tac/reprints/articles/17/tr17abs.html}.

\bibitem{Ang:87}
Dana Angluin.
\newblock Learning regular sets from queries and counterexamples.
\newblock {\em Information and Computation}, 75(2):87--106, 1987.
\newblock \href {https://doi.org/10.1016/0890-5401(87)90052-6} {\path{doi:10.1016/0890-5401(87)90052-6}}.

\bibitem{AM:75a}
Michael~A. Arbib and Ernest~G. Manes.
\newblock Adjoint machines, state-behavior machines, and duality.
\newblock {\em Journal of Pure and Applied Algebra}, 6(3):313--344, 1975.
\newblock \href {https://doi.org/10.1016/0022-4049(75)90028-6} {\path{doi:10.1016/0022-4049(75)90028-6}}.

\bibitem{AM:75b}
Michael~A. Arbib and Ernest~G. Manes.
\newblock Fuzzy machines in a category.
\newblock {\em Bulletin of the Australian Mathematical Society}, 13(2):169--210, 1975.
\newblock \href {https://doi.org/10.1017/S0004972700024412} {\path{doi:10.1017/S0004972700024412}}.

\bibitem{AP:26}
Quentin Aristote and Daniela Petri{\c s}an.
\newblock Learning bottom-up tree automata valued in monoidal categories.
\newblock In Nathalie Bertrand and Stefan Milius, editors, {\em 29th International Conference on Foundations of Software Science and Computation Structures ({FoSSaCS} 2026)}, volume 16503 of {\em Lecture Notes in Computer Science}, pages 21--42, 2026.
\newblock \href {https://doi.org/10.1007/978-3-032-22730-0_2} {\path{doi:10.1007/978-3-032-22730-0_2}}.

\bibitem{Awo:10}
Steve Awodey.
\newblock {\em Category Theory}.
\newblock Oxford Logic Guides. Oxford University Press, 2010.
\newblock \href {https://doi.org/10.1093/acprof:oso/9780198568612.001.0001} {\path{doi:10.1093/acprof:oso/9780198568612.001.0001}}.

\bibitem{BKR:19}
Simone Barlocco, Clemens Kupke, and Jurriaan Rot.
\newblock Coalgebra learning via duality.
\newblock In Miko{\l}aj Boja{\'n}czyk and Alex Simpson, editors, {\em 22nd International Conference on Foundations of Software Science and Computation Structures ({FoSSaCS} 2019)}, volume 11425 of {\em Lecture Notes in Computer Science}, pages 62--79, 2019.
\newblock \href {https://doi.org/10.1007/978-3-030-17127-8_4} {\path{doi:10.1007/978-3-030-17127-8_4}}.

\bibitem{Bec:69}
Jon Beck.
\newblock Distributive laws.
\newblock In H.~Appelgate, M.~Barr, J.~Beck, F.~W. Lawvere, F.~E.~J. Linton, E.~Manes, M.~Tierney, F.~Ulmer, and B.~Eckmann, editors, {\em Seminar on Triples and Categorical Homology Theory}, volume~80 of {\em Lecture Notes in Mathematics}, pages 119--140, 1969.
\newblock \href {https://doi.org/10.1007/BFb0083084} {\path{doi:10.1007/BFb0083084}}.

\bibitem{BV:96}
Francesco Bergadano and Stefano Varricchio.
\newblock Learning behaviors of automata from multiplicity and equivalence queries.
\newblock {\em {SIAM} Journal on Computing}, 25(6):1268--1280, 1996.
\newblock \href {https://doi.org/10.1137/S009753979326091X} {\path{doi:10.1137/S009753979326091X}}.

\bibitem{BKL:11}
Miko{\l}aj Boja{\'n}czyk, Bartek Klin, and S{\l}awomir Lasota.
\newblock Automata with group actions.
\newblock In {\em 26th Annual Symposium on Logic in Computer Science ({LICS} 2011)}, pages 355--364, 2011.
\newblock \href {https://doi.org/10.1109/LICS.2011.48} {\path{doi:10.1109/LICS.2011.48}}.

\bibitem{BHK+:09}
Benedikt Bollig, Peter Habermehl, Carsten Kern, and Martin Leucker.
\newblock {Angluin}-style learning of {NFA}.
\newblock In Craig Boutilier, editor, {\em 21st International Joint Conference on Artificial Intelligence ({IJCAI} 2009)}, pages 1004--1009, 2009.
\newblock URL: \url{https://www.ijcai.org/proceedings/2009}.

\bibitem{BBB+:12}
Filippo Bonchi, Marcello~M. Bonsangue, Michele Boreale, Jan J. M.~M. Rutten, and Alexandra Silva.
\newblock A coalgebraic perspective on linear weighted automata.
\newblock {\em Information and Computation}, 211:77--105, 2012.
\newblock \href {https://doi.org/10.1016/J.IC.2011.12.002} {\path{doi:10.1016/J.IC.2011.12.002}}.

\bibitem{BJM:19}
Petra~{van den} Bos, Ramon Janssen, and Joshua Moerman.
\newblock n-complete test suites for {IOCO}.
\newblock {\em Software Quality Journal}, 27(2):563--588, 2019.
\newblock \href {https://doi.org/10.1007/S11219-018-9422-X} {\path{doi:10.1007/S11219-018-9422-X}}.

\bibitem{Cho:78}
Tsun~S. Chow.
\newblock Testing software design modeled by finite-state machines.
\newblock {\em {IEEE} Transactions on Software Engineering}, 4(3):178--187, 1978.
\newblock \href {https://doi.org/10.1109/TSE.1978.231496} {\path{doi:10.1109/TSE.1978.231496}}.

\bibitem{CPS:21}
Thomas Colcombet, Daniela Petri{\c s}an, and Riccardo Stabile.
\newblock Learning automata and transducers: A categorical approach.
\newblock In Christel Baier and Jean Goubault{-}Larrecq, editors, {\em 29th Annual Conference on Computer Science Logic ({CSL} 2021)}, volume 183 of {\em Leibniz International Proceedings in Informatics}, pages 15:1--15:17, 2021.
\newblock \href {https://doi.org/10.4230/LIPIcs.CSL.2021.15} {\path{doi:10.4230/LIPIcs.CSL.2021.15}}.

\bibitem{CDG+:08}
Hubert Comon, Max Dauchet, R{\'e}mi Gilleron, Florent Jacquemard, Denis Lugiez, Christof L{\"o}ding, Sophie Tison, and Marc Tommasi.
\newblock Tree automata techniques and applications, 2008.
\newblock URL: \url{https://inria.hal.science/hal-03367725}.

\bibitem{DLT:02}
Fran{\c c}ois Denis, Aur{\'e}lien Lemay, and Alain Terlutte.
\newblock Residual finite state automata.
\newblock {\em Fundamenta Informaticae}, 51(4):339--368, 2002.
\newblock \href {https://doi.org/10.1007/3-540-44693-1_13} {\path{doi:10.1007/3-540-44693-1_13}}.

\bibitem{DKV:09}
Manfred Droste, Werner Kuich, and Heiko Vogler, editors.
\newblock {\em Handbook of Weighted Automata}.
\newblock Monographs in Theoretical Computer Science. {An} {EATCS} Series. Springer, 2009.
\newblock \href {https://doi.org/10.1007/978-3-642-01492-5} {\path{doi:10.1007/978-3-642-01492-5}}.

\bibitem{DK:21}
Manfred Droste and Dietrich Kuske.
\newblock Weighted automata.
\newblock In Jean{-}{\'E}ric Pin, editor, {\em Handbook of Automata Theory}, pages 113--150. EMS Press, 2021.
\newblock \href {https://doi.org/10.4171/automata-1/4} {\path{doi:10.4171/automata-1/4}}.

\bibitem{FHS:25}
Markus Frohme, Falk Howar, and Bernhard Steffen.
\newblock {LearnLib}: 10 years later.
\newblock In Ruzica Piskac and Zvonimir Rakamari{\'c}, editors, {\em 37th International Conference on Computer Aided Verification ({CAV} 2025)}, volume 15934 of {\em Lecture Notes in Computer Science}, pages 141--160, 2025.
\newblock \href {https://doi.org/10.1007/978-3-031-98685-7_7} {\path{doi:10.1007/978-3-031-98685-7_7}}.

\bibitem{FvBK+91}
Susumu Fujiwara, Gregor~{von} Bochmann, Ferhat Khendek, Mokhtar Amalou, and Abderrazak Ghedamsi.
\newblock Test selection based on finite state models.
\newblock {\em {IEEE} Transactions on Software Engineering}, 17(6):591--603, 1991.
\newblock \href {https://doi.org/10.1109/32.87284} {\path{doi:10.1109/32.87284}}.

\bibitem{Gog:75}
Joseph~A. Goguen.
\newblock Discrete-time machines in closed monoidal categories {I}.
\newblock {\em Journal of Computer and System Sciences}, 10(1):1--43, 1975.
\newblock \href {https://doi.org/10.1016/S0022-0000(75)80012-2} {\path{doi:10.1016/S0022-0000(75)80012-2}}.

\bibitem{HJS:07}
Ichiro Hasuo, Bart Jacobs, and Ana Sokolova.
\newblock Generic trace semantics via coinduction.
\newblock {\em Logical Methods in Computer Science}, 3(4:11):1--36, 2007.
\newblock \href {https://doi.org/10.2168/LMCS-3(4:11)2007} {\path{doi:10.2168/LMCS-3(4:11)2007}}.

\bibitem{vHee:20}
Gerco~{van} Heerdt.
\newblock {\em {CALF}: Categorical Automata Learning Framework}.
\newblock PhD thesis, University College London, 2020.
\newblock URL: \url{https://discovery.ucl.ac.uk/id/eprint/10110356}.

\bibitem{vHKR+:19}
Gerco~{van} Heerdt, Tobias Kapp{\'e}, Jurriaan Rot, Matteo Sammartino, and Alexandra Silva.
\newblock Tree automata as algebras: Minimisation and determinisation.
\newblock In Markus Roggenbach and Ana Sokolova, editors, {\em 8th Conference on Algebra and Coalgebra in Computer Science ({CALCO} 2019)}, volume 139 of {\em Leibniz International Proceedings in Informatics}, pages 6:1--6:22, 2019.
\newblock \href {https://doi.org/10.4230/LIPIcs.CALCO.2019.6} {\path{doi:10.4230/LIPIcs.CALCO.2019.6}}.

\bibitem{vHKR+:22}
Gerco~{van} Heerdt, Tobias Kapp{\'e}, Jurriaan Rot, Matteo Sammartino, and Alexandra Silva.
\newblock A categorical framework for learning generalised tree automata.
\newblock In Helle~Hvid Hansen and Fabio Zanasi, editors, {\em 16th International Workshop on Coalgebraic Methods in Computer Science ({CMCS} 2022)}, volume 13225 of {\em Lecture Notes in Computer Science}, pages 67--87, 2022.
\newblock \href {https://doi.org/10.1007/978-3-031-10736-8_4} {\path{doi:10.1007/978-3-031-10736-8_4}}.

\bibitem{vHSS:17}
Gerco~{van} Heerdt, Matteo Sammartino, and Alexandra Silva.
\newblock {CALF}: Categorical automata learning framework.
\newblock In Valentin Goranko and Mads Dam, editors, {\em 26th Annual Conference on Computer Science Logic ({CSL} 2017)}, volume~82 of {\em Leibniz International Proceedings in Informatics}, pages 29:1--29:24, 2017.
\newblock \href {https://doi.org/10.4230/LIPIcs.CSL.2017.29} {\path{doi:10.4230/LIPIcs.CSL.2017.29}}.

\bibitem{vHSS:20}
Gerco~{van} Heerdt, Matteo Sammartino, and Alexandra Silva.
\newblock Learning automata with side-effects.
\newblock In Daniela Petri{\c s}an and Jurriaan Rot, editors, {\em 15th International Workshop on Coalgebraic Methods in Computer Science ({CMCS} 2020)}, volume 12094 of {\em Lecture Notes in Computer Science}, pages 68--89, 2020.
\newblock \href {https://doi.org/10.1007/978-3-030-57201-3_5} {\path{doi:10.1007/978-3-030-57201-3_5}}.

\bibitem{HJ:98}
Claudio Hermida and Bart Jacobs.
\newblock Structural induction and coinduction in a fibrational setting.
\newblock {\em Information and Computation}, 145(2):107--152, 1998.
\newblock \href {https://doi.org/10.1006/inco.1998.2725} {\path{doi:10.1006/inco.1998.2725}}.

\bibitem{JSS:15}
Bart Jacobs, Alexandra Silva, and Ana Sokolova.
\newblock Trace semantics via determinization.
\newblock {\em Journal of Computer and System Sciences}, 81(5):859--879, 2015.
\newblock \href {https://doi.org/10.1016/j.jcss.2014.12.005} {\path{doi:10.1016/j.jcss.2014.12.005}}.

\bibitem{Joh:75}
Peter~T. Johnstone.
\newblock Adjoint lifting theorems for categories of algebras.
\newblock {\em Bulletin of the London Mathematical Society}, 7(3):294--297, 1975.
\newblock \href {https://doi.org/10.1112/blms/7.3.294} {\path{doi:10.1112/blms/7.3.294}}.

\bibitem{Joh:02}
Peter~T. Johnstone.
\newblock {\em Sketches of an Elephant: A Topos Theory Compendium, Vol. 1}.
\newblock Oxford University Press, 2002.
\newblock \href {https://doi.org/10.1093/oso/9780198534259.001.0001} {\path{doi:10.1093/oso/9780198534259.001.0001}}.

\bibitem{Kel:80}
G.~Max Kelly.
\newblock A unified treatment of transfinite constructions for free algebras, free monoids, colimits, associated sheaves, and so on.
\newblock {\em Bulletin of the Australian Mathematical Society}, 22(1):1--83, 1980.
\newblock \href {https://doi.org/10.1017/S0004972700006353} {\path{doi:10.1017/S0004972700006353}}.

\bibitem{Koc:70}
Anders Kock.
\newblock Monads on symmetric monoidal closed categories.
\newblock {\em Archiv der Mathematik}, 21(1):1--10, 1970.
\newblock \href {https://doi.org/10.1007/BF01220868} {\path{doi:10.1007/BF01220868}}.

\bibitem{Koc:71}
Anders Kock.
\newblock Closed categories generated by commutative monads.
\newblock {\em Journal of the Australian Mathematical Society}, 12(4):405--424, 1971.
\newblock \href {https://doi.org/10.1017/S1446788700010272} {\path{doi:10.1017/S1446788700010272}}.

\bibitem{Koc:72}
Anders Kock.
\newblock Strong functors and monoidal monads.
\newblock {\em Archiv der Mathematik}, 23(1):113--120, 1972.
\newblock \href {https://doi.org/10.1007/BF01304852} {\path{doi:10.1007/BF01304852}}.

\bibitem{KR:25}
B{\'a}lint Kocsis and Jurriaan Rot.
\newblock Complete test suites for automata in monoidal closed categories.
\newblock In Parosh~Aziz Abdulla and Delia Kesner, editors, {\em 28th International Conference on Foundations of Software Science and Computation Structures ({FoSSaCS} 2025)}, volume 15691 of {\em Lecture Notes in Computer Science}, pages 198--219, 2025.
\newblock \href {https://doi.org/10.1007/978-3-031-90897-2_10} {\path{doi:10.1007/978-3-031-90897-2_10}}.

\bibitem{KJR:24}
Loes Kruger, Sebastian Junges, and Jurriaan Rot.
\newblock Small test suites for active automata learning.
\newblock In Bernd Finkbeiner and Laura Kov{\'a}cs, editors, {\em 30th International Conference on Tools and Algorithms for the Construction and Analysis of Systems ({TACAS} 2024)}, volume 14571 of {\em Lecture Notes in Computer Science}, pages 109--129, 2024.
\newblock \href {https://doi.org/10.1007/978-3-031-57249-4_6} {\path{doi:10.1007/978-3-031-57249-4_6}}.

\bibitem{Lin:69}
Fred E.~J. Linton.
\newblock Coequalizers in categories of algebras.
\newblock In H.~Appelgate, M.~Barr, J.~Beck, F.~W. Lawvere, F.~E.~J. Linton, E.~Manes, M.~Tierney, F.~Ulmer, and B.~Eckmann, editors, {\em Seminar on Triples and Categorical Homology Theory}, volume~80 of {\em Lecture Notes in Mathematics}, pages 75--90, 1969.
\newblock \href {https://doi.org/10.1007/BFb0083082} {\path{doi:10.1007/BFb0083082}}.

\bibitem{Maa:20}
Timo Maarse.
\newblock Active {Mealy} machine learning using action refinements.
\newblock Master's thesis, Radboud University Nijmegen, The Netherlands, 2020.
\newblock URL: \url{https://www.cs.ru.nl/F.Vaandrager/thesis-2019_NWI-IMC048_s4416295.pdf}.

\bibitem{Mac:78}
Saunders {Mac Lane}.
\newblock {\em Categories for the Working Mathematician}, volume~5 of {\em Graduate Texts in Mathematics}.
\newblock Springer, 1978.
\newblock \href {https://doi.org/10.1007/978-1-4757-4721-8} {\path{doi:10.1007/978-1-4757-4721-8}}.

\bibitem{MU:22}
Dylan McDermott and Tarmo Uustalu.
\newblock What makes a strong monad?
\newblock {\em Electronic Proceedings in Theoretical Computer Science}, 360:113--133, 2022.
\newblock \href {https://doi.org/10.4204/EPTCS.360.6} {\path{doi:10.4204/EPTCS.360.6}}.

\bibitem{Moe:20}
Joshua Moerman.
\newblock {\em Nominal Techniques and Black Box Testing for Automata Learning}.
\newblock PhD thesis, Radboud University, 2019.
\newblock URL: \url{https://joshuamoerman.nl/thesis}.

\bibitem{Mul:94}
Philip~S. Mulry.
\newblock Lifting theorems for {Kleisli} categories.
\newblock In Stephen Brookes, Michael Main, Austin Melton, Michael Mislove, and David Schmidt, editors, {\em 9th International Conference on Mathematical Foundations of Programming Semantics ({MFPS} 1993)}, volume 802 of {\em Lecture Notes in Computer Science}, pages 304--319, 1994.
\newblock \href {https://doi.org/10.1007/3-540-58027-1_15} {\path{doi:10.1007/3-540-58027-1_15}}.

\bibitem{Mul:02}
Philip~S. Mulry.
\newblock Lifting results for categories of algebras.
\newblock {\em Theoretical Computer Science}, 278(1):257--269, 2002.
\newblock \href {https://doi.org/10.1016/S0304-3975(00)00338-8} {\path{doi:10.1016/S0304-3975(00)00338-8}}.

\bibitem{MAP+:21}
Edi Muskardin, Bernhard~K. Aichernig, Ingo Pill, Andrea Pferscher, and Martin Tappler.
\newblock {AALpy}: An active automata learning library.
\newblock {\em Innovations in Systems and Software Engineering}, 18(3):417--426, 2022.
\newblock \href {https://doi.org/10.1007/S11334-022-00449-3} {\path{doi:10.1007/S11334-022-00449-3}}.

\bibitem{MAM+:14}
Robert S.~R. Myers, Ji{\v r}{\'i} Ad{\'a}mek, Stefan Milius, and Henning Urbat.
\newblock Canonical nondeterministic automata.
\newblock In Marcello~M. Bonsangue, editor, {\em 12th International Workshop on Coalgebraic Methods in Computer Science ({CMCS} 2014)}, volume 8446 of {\em Lecture Notes in Computer Science}, pages 189--210, 2014.
\newblock \href {https://doi.org/10.1007/978-3-662-44124-4_11} {\path{doi:10.1007/978-3-662-44124-4_11}}.

\bibitem{Pag:66}
Carl~V. Page.
\newblock Equivalences between probabilistic and deterministic sequential machines.
\newblock {\em Information and Control}, 9(5):469--520, 1966.
\newblock \href {https://doi.org/10.1016/S0019-9958(66)80012-8} {\path{doi:10.1016/S0019-9958(66)80012-8}}.

\bibitem{PvBY:96}
Alexandre Petrenko, Gregor~{von} Bochmann, and Ming~Yu Yao.
\newblock On fault coverage of tests for finite state specifications.
\newblock {\em Computer Networks and {ISDN} Systems}, 29(1):81--106, 1996.
\newblock \href {https://doi.org/10.1016/S0169-7552(96)00019-0} {\path{doi:10.1016/S0169-7552(96)00019-0}}.

\bibitem{PY:14}
Alexandre Petrenko and Nina Yevtushenko.
\newblock Adaptive testing of nondeterministic systems with {FSM}.
\newblock In {\em 15th International Symposium on High-Assurance Systems Engineering ({HASE} 2014)}, pages 224--228, 2014.
\newblock \href {https://doi.org/10.1109/HASE.2014.39} {\path{doi:10.1109/HASE.2014.39}}.

\bibitem{RS:59}
Michael~O. Rabin and Dana~S. Scott.
\newblock Finite automata and their decision problems.
\newblock {\em {IBM} Journal of Research and Development}, 3(2):114--125, 1959.
\newblock \href {https://doi.org/10.1147/rd.32.0114} {\path{doi:10.1147/rd.32.0114}}.

\bibitem{Rie:16}
Emily Riehl.
\newblock {\em Category theory in context}.
\newblock Aurora: {Dover} modern math originals. Dover Publications, 2016.
\newblock URL: \url{https://math.jhu.edu/~eriehl/context}.

\bibitem{Sac:22}
Robert Sachtleben.
\newblock {\em An Approach for the Verification and Synthesis of Complete Test Generation Algorithms for Finite State Machines}.
\newblock PhD thesis, Bremen University, Germany, 2022.
\newblock \href {https://doi.org/10.26092/ELIB/1665} {\path{doi:10.26092/ELIB/1665}}.

\bibitem{Sch:61}
Marcel{-}Paul Sch{\"u}tzenberger.
\newblock On the definition of a family of automata.
\newblock {\em Information and Control}, 4(2-3):245--270, 1961.
\newblock \href {https://doi.org/10.1016/S0019-9958(61)80020-X} {\path{doi:10.1016/S0019-9958(61)80020-X}}.

\bibitem{Sea:13}
Gavin~J. Seal.
\newblock Tensors, monads and actions.
\newblock {\em Theory and Applications of Categories}, 28(15):403--433, 2013.
\newblock \href {https://doi.org/10.70930/tac/euyoegw2} {\path{doi:10.70930/tac/euyoegw2}}.

\bibitem{TAB+:21}
Martin Tappler, Bernhard~K. Aichernig, Giovanni Bacci, Maria Eichlseder, and Kim~G. Larsen.
\newblock L*-based learning of {Markov} decision processes (extended version).
\newblock {\em Formal Aspects of Computing}, 33(4--5):575--615, 2021.
\newblock \href {https://doi.org/10.1007/S00165-021-00536-5} {\path{doi:10.1007/S00165-021-00536-5}}.

\bibitem{US:20}
Henning Urbat and Lutz Schr{\"o}der.
\newblock Automata learning: An algebraic approach.
\newblock In Holger Hermanns, Lijun Zhang, Naoki Kobayashi, and Dale Miller, editors, {\em 35th Annual Symposium on Logic in Computer Science ({LICS} 2020)}, pages 900--914, 2020.
\newblock \href {https://doi.org/10.1145/3373718.3394775} {\path{doi:10.1145/3373718.3394775}}.

\bibitem{VM:25}
Frits~W. Vaandrager and Ivo Melse.
\newblock New fault domains for conformance testing of finite state machines.
\newblock In Patricia Bouyer and Jaco~{van de} Pol, editors, {\em 36th International Conference on Concurrency Theory ({CONCUR} 2025)}, volume 348 of {\em Leibniz International Proceedings in Informatics}, pages 34:1--34:22, 2025.
\newblock \href {https://doi.org/10.4230/LIPIcs.CONCUR.2025.34} {\path{doi:10.4230/LIPIcs.CONCUR.2025.34}}.

\bibitem{Vas:73}
M.~P. Vasilevskii.
\newblock Failure diagnosis of automata.
\newblock {\em Cybernetics and Systems Analysis}, 9(4):653--665, 1973.
\newblock \href {https://doi.org/10.1007/BF01068590} {\path{doi:10.1007/BF01068590}}.

\end{thebibliography}
\clearpage

\appendix

\section{Categorical preliminaries}\label[appendix]{sec:prelim}

We recall some standard categorical definitions that are used throughout the paper.

\begin{definition}[\cite{Mac:78}]\label{def:mon-cat}\hfill
  \begin{enumerate}[(1)]
    \item
    A \emph{monoidal category} is a category~$\cat{C}$ equipped with
    \begin{itemize}
      \item a functor~$\tprod\, \colon \cat{C} \prodcat \cat{C} \to \cat{C}$
      (called the \emph{tensor product}),
      \item an object~$\tunit \in \cat{C}$ (called the \emph{tensor unit}),
      \item a natural isomorphism~\(
        \assoc[A,B,C] \colon (A \tprod B) \tprod C \ito A \tprod (B \tprod C)
      \)
      (called the \emph{associator}),
      \item a natural isomorphism~$\lunit[A] \colon \tunit \tprod A \ito A$
      (called the \emph{left unitor}), and
      \item a natural isomorphism~$\runit[A] \colon A \tprod \tunit \ito A$
      (called the \emph{right unitor})
    \end{itemize}
    such that the following diagrams commute:
    \[
      \begin{tikzcd}[column sep=-2.3em]
        & {(A \tprod B) \tprod (C \tprod D)} & \\
        {((A \tprod B) \tprod C) \tprod D} && {A \tprod (B \tprod (C \tprod D))} \\
        {(A \tprod (B \tprod C)) \tprod D} && {A \tprod ((B \tprod C) \tprod D)}
        \arrow["{\assoc[A,B,C \tprod D]}", from=1-2, to=2-3]
        \arrow["{\assoc[A \tprod B,C,D]}", from=2-1, to=1-2]
        \arrow["{\assoc[A,B,C] \tprod \id[D]}"', from=2-1, to=3-1]
        \arrow["{\assoc[A,B \tprod C,D]}"', from=3-1, to=3-3]
        \arrow["{\id[A] \tprod \assoc[B,C,D]}"', from=3-3, to=2-3]
      \end{tikzcd}
      \begin{tikzcd}[column sep=-0.2em]
        {(A \tprod \tunit) \tprod B} && {A \tprod (\tunit \tprod B)} \\
        & {A \tprod B}
        \arrow["{\assoc[A,\tunit,B]}", from=1-1, to=1-3]
        \arrow["{\runit[A] \tprod \id[B]}"', from=1-1, to=2-2]
        \arrow["{\id[A] \tprod \lunit[B]}", from=1-3, to=2-2]
      \end{tikzcd}
    \]

    \item
    A \emph{symmetric monoidal category} is a monoidal category equipped with an
    additional natural isomorphism~$\braid[A,B] \colon A \tprod B \ito B \tprod A$
    (called the \emph{braiding}) such that $\braid[B,A] \com \braid[A,B] = \id[A \tprod B]$
    and the following diagram commutes:
    \[\begin{tikzcd}[column sep=4em]
      {(A \tprod B) \tprod C} & {A \tprod (B \tprod C)} & {(B \tprod C) \tprod A} \\
      {(B \tprod A) \tprod C} & {B \tprod (A \tprod C)} & {B \tprod (C \tprod A)}
      \arrow["{\assoc[A,B,C]}", from=1-1, to=1-2]
      \arrow["{\braid[A,B] \tprod \id[C]}"', from=1-1, to=2-1]
      \arrow["{\braid[A,B \tprod C]}", from=1-2, to=1-3]
      \arrow["{\assoc[B,C,A]}", from=1-3, to=2-3]
      \arrow["{\assoc[B,A,C]}"', from=2-1, to=2-2]
      \arrow["{\id[B] \tprod \braid[A,C]}"', from=2-2, to=2-3]
    \end{tikzcd}\]

    \item
    A monoidal category~$\cat{C}$ is said to be \emph{closed} if for all~$A \in \bcat$,
    the functor~$(-) \tprod A$ has a right adjoint~$\inthom{A}{-}$, called the \emph{internal hom}.
    A \emph{(symmetric) monoidal closed category} is
    a (symmetric) monoidal category that is closed.
  \end{enumerate}
\end{definition}

\begin{definition}[\cite{Koc:72}]\label{def:strength}
  Let~$\cat{C}$ be a monoidal category and let~$F \colon \cat{C} \to \cat{C}$ be an endofunctor.
  A \emph{left strength} for~$F$ is a
  natural transformation~$\lstr[A,B] \colon A \tprod FB \to F(A \tprod B)$
  making the following diagrams commute:
  \[
    \begin{tikzcd}[column sep=4em]
      {(A \tprod B) \tprod FC} & {A \tprod (B \tprod FC)} \\
      & {A \tprod F(B \tprod C)} \\
      {F((A \tprod B) \tprod C)} & {F(A \tprod (B \tprod C))}
      \arrow["{\assoc[A,B,FC]}", from=1-1, to=1-2]
      \arrow["{\lstr[A \tprod B,C]}"', from=1-1, to=3-1]
      \arrow["{\id[A] \tprod \lstr[B,C]}", from=1-2, to=2-2]
      \arrow["{\lstr[A,B \tprod C]}", from=2-2, to=3-2]
      \arrow["{F\assoc[A,B,C]}"', from=3-1, to=3-2]
    \end{tikzcd}
    \hspace{1em}
    \begin{tikzcd}
      {\tunit \tprod FA} & {F(\tunit \tprod A)} \\
      & FA
      \arrow["{\lstr[\tunit,A]}", from=1-1, to=1-2]
      \arrow["{\lunit[FA]}"', from=1-1, to=2-2]
      \arrow["{F\lunit[A]}", from=1-2, to=2-2]
    \end{tikzcd}
  \]
  A \emph{(left) strong} functor on~$\cat{C}$ is
  an endofunctor~$F \colon \cat{C} \to \cat{C}$ equipped with a left strength~$\lstr$.
\end{definition}
We also have the notion of \emph{right strength}, which is defined symmetrically
and involves a natural transformation~$FA \tprod B \to F(A \tprod B)$.
In a symmetric monoidal category, the two notions are interdefinable using the braiding~$\braid$.

Commutative monads on symmetric monoidal closed categories were
originally defined by Kock~\cite{Koc:70} as strong monads satisfying a certain condition.
It was shown in~\cite{Koc:70,Koc:72} that commutative monads on
symmetric monoidal (closed) categories are equivalent to \emph{symmetric monoidal monads}.
In this paper, we mainly rely on the `double strength' of monoidal monads,
so we state the definition in monoidal form.
We chose to use the term `commutative' in the paper since
that is slightly more common in a theoretical computer science context.
\begin{definition}[\cite{Koc:70,Koc:72}]\label{def:comm-monad}
  Let~$\cat{C}$ be a monoidal category.
  A \emph{monoidal monad} on~$\cat{C}$ is a monad~$(T, \mu, \eta)$ equipped with
  a natural transformation~$\effmf[A,B] \colon TA \tprod TB \to T(A \tprod B)$
  making the following diagrams commute:
  \[
    \begin{array}{cc}
      \begin{tikzcd}[ampersand replacement=\&,column sep=4.5em]
        {(TA \tprod TB) \tprod TC} \& {T(A \tprod B) \tprod TC} \\
        {TA \tprod (TB \tprod TC)} \& {T((A \tprod B) \tprod C)} \\
        {TA \tprod T(B \tprod C)} \& {T(A \tprod (B \tprod C))}
        \arrow["{\effmf[A,B] \tprod \id[TC]}", from=1-1, to=1-2]
        \arrow["{\assoc[TA,TB,TC]}"', from=1-1, to=2-1]
        \arrow["{\effmf[A \tprod B,C]}", from=1-2, to=2-2]
        \arrow["{\id[TA] \tprod \effmf[B,C]}"', from=2-1, to=3-1]
        \arrow["{T\assoc[A,B,C]}", from=2-2, to=3-2]
        \arrow["{\effmf[A,B \tprod C]}"', from=3-1, to=3-2]
      \end{tikzcd}
      &
      \begin{gathered}
        \begin{tikzcd}[ampersand replacement=\&]
          {\tunit \tprod TA} \& TA \\
          {T\tunit \tprod TA} \& {T(\tunit \tprod A)}
          \arrow["{\lunit[TA]}", from=1-1, to=1-2]
          \arrow["{\eta_\tunit \tprod \id[TA]}"', from=1-1, to=2-1]
          \arrow["{\effmf[\tunit,A]}"', from=2-1, to=2-2]
          \arrow["{T\lunit[A]}"', from=2-2, to=1-2]
        \end{tikzcd}
        \\
        \begin{tikzcd}[ampersand replacement=\&]
          {TA \tprod \tunit} \& TA \\
          {TA \tprod T\tunit} \& {T(A \tprod \tunit)}
          \arrow["{\runit[TA]}", from=1-1, to=1-2]
          \arrow["{\id[TA] \tprod \eta_\tunit}"', from=1-1, to=2-1]
          \arrow["{\effmf[A,\tunit]}"', from=2-1, to=2-2]
          \arrow["{T\runit[A]}"', from=2-2, to=1-2]
        \end{tikzcd}
      \end{gathered}
    \end{array}
  \]
  \[
    \begin{tikzcd}[column sep=3.5em]
      {T^2A \tprod T^2B} & {T(TA \tprod TB)} & {T^2(A \tprod B)} \\
      {TA \tprod TB} && {T(A \tprod B)}
      \arrow["{\effmf[TA,TB]}", from=1-1, to=1-2]
      \arrow["{\mu_A \tprod \mu_B}"', from=1-1, to=2-1]
      \arrow["{T\effmf[A,B]}", from=1-2, to=1-3]
      \arrow["{\mu_{A \tprod B}}", from=1-3, to=2-3]
      \arrow["{\effmf[A,B]}"', from=2-1, to=2-3]
    \end{tikzcd}
    \hspace{0.5em}
    \begin{tikzcd}
      {A \tprod B} & \\
      {TA \tprod TB} & {T(A \tprod B)}
      \arrow["{\eta_A \tprod \eta_B}"', from=1-1, to=2-1]
      \arrow["{\eta_{A \tprod B}}", from=1-1, to=2-2]
      \arrow["{\effmf[A,B]}"', from=2-1, to=2-2]
    \end{tikzcd}
  \]
  If $\cat{C}$~is a symmetric monoidal category, then a \emph{symmetric monoidal monad} is
  a monoidal monad satisfying the following additional condition:
  \[\begin{tikzcd}
    {TA \tprod TB} & {T(A \tprod B)} \\
    {TB \tprod TA} & {T(B \tprod A)}
    \arrow["{\effmf[A,B]}", from=1-1, to=1-2]
    \arrow["{\braid[TA,TB]}"', from=1-1, to=2-1]
    \arrow["{T\braid[A,B]}", from=1-2, to=2-2]
    \arrow["{\effmf[B,A]}"', from=2-1, to=2-2]
  \end{tikzcd}\]
\end{definition}

Distributive laws between monads were defined by Beck~\cite{Bec:69} to study composite monads.
In~\cite{AM:75b}, they were generalised to distributive laws of functors over monads.
\begin{definition}[\cite{Bec:69,AM:75b}]\label{def:dist-law}
  Let~$F \colon \cat{C} \to \cat{C}$ be an endofunctor
  and let~$(T, \mu, \eta)$ be a monad on~$\cat{C}$.
  A \emph{distributive law} of~$F$ over~$T$ is
  a natural transformation~$\dlaw \colon FT \to TF$ making the following diagrams commute:
  \[
    \begin{tikzcd}
      {FT^2} & TFT & {T^2F} \\
      FT && TF
      \arrow["{\dlaw T}", from=1-1, to=1-2]
      \arrow["{F\mu}"', from=1-1, to=2-1]
      \arrow["{T\dlaw}", from=1-2, to=1-3]
      \arrow["{\mu F}", from=1-3, to=2-3]
      \arrow["\dlaw"', from=2-1, to=2-3]
    \end{tikzcd}
    \hspace{1em}
    \begin{tikzcd}
      F & \\
      FT & TF
      \arrow["{F\eta}"', from=1-1, to=2-1]
      \arrow["{\eta F}", from=1-1, to=2-2]
      \arrow["\dlaw"', from=2-1, to=2-2]
    \end{tikzcd}
  \]
\end{definition}

\section{Additional details for~\texorpdfstring{\cref{sec:applications}}{\autoref{sec:applications}}}\label[appendix]{sec:app-details}

\subsection{Nondeterministic automata}

We give the precise definition of the NFA~$\exNFA[n]$ for arbitrary~$n$ as follows.
The state space is the finite set~$Q \leteq \set{q_i}[0 \le i \le n+1]$,
with a single initial state~$Q_0 \leteq \set{q_0}$.
There is also a unique accepting state: $A \leteq \set{q_{n+1}}$.
The transition function~$\delta \colon Q \prodset \abcobj \to \powset Q$ is given by
\begin{alignat*}{2}
  \delta(q_0,     a) &\leteq \set{q_0, q_1}, \\
  \delta(q_0,     b) &\leteq \set{q_0}, \\
  \delta(q_i,     s) &\leteq \set{q_{i+1}}  &\hspace{3em}& \text{for~$1 \le i \le n$ and~$s \in \abcobj$}, \\
  \delta(q_{n+1}, s) &\leteq \emptyset      &            & \text{for~$s \in \abcobj$}.
\end{alignat*}

Note that the NFA~$\exNFA[n]$ has exactly $n+2$~states.
The determinisation~$\alglift{\klext}\exNFA[n]$ into~$\Set$ consists of
all subsets of~$Q$ containing the element~$q_1$ and so it has $2^{n+1}$~states.
The empty set does not need to be included, but it \emph{will} be included in the
determinisation~$\alglift{\klcomp}\exNFA[n]$ into~$\EM{\powset}$ (even after minimisation)
because $\powset$-algebras always contain a bottom element.

The size of the effectful state cover is not always linear in the size of the NFA.
The elements in the minimal state cover correspond to the join-irreducible elements
of the subalgebra generated by the reachable states,
which in turn correspond to states of the canonical RFSA.
Denis et al.~\cite{DLT:02} give an example of
a family of NFAs whose canonical RFSAs are exponentially large.
This example would also need exponentially many elements in any effectful state cover.

\paragraph*{Details for the test suite of~\texorpdfstring{$\exNFA[2]$}{the example NFA}}

\begingroup
\newcommand{\opt}[1]{{\color{lightgray}#1}\mathbin{\color{lightgray}\vee}}

Recall that the structure map~$\delta_C$ is given as follows:
\begin{align*}
  \empwd, a &\mapsto \opt{\empwd} a  &  a, a &\mapsto \opt{\empwd} a \vee ab  &  ab, a &\mapsto \opt{\empwd} a \vee abb  &  abb, a &\mapsto \opt{\empwd} a \\
  \empwd, b &\mapsto \empwd          &  a, b &\mapsto \opt{\empwd} ab         &  ab, b &\mapsto \opt{\empwd} abb         &  abb, b &\mapsto \empwd
\end{align*}
The map~$\delta_C$ is not uniquely determined.
For instance, the elements highlighted in grey are optional and can be left out.
In each case, the resulting map still satisfies the requirements of~\cref{def:effectful-st-cov}.
Since the set~$C$ is the same in all cases,
the corresponding effectful W-suite is independent of these choices.
For~$\exNFA[2]$, this test suite is
\begin{align*}
  \im{t} = \{
    & \empwd, a, aa, aaa, aaaa, aaaaa, ab, aba, abaa, abaaa, abaaaa, abb, abba \\
    & abbaa, abbaaa, abbaaaa, abbb, abbba, abbbaa, abbbaaa, b, ba, baa, baaa
  \}.
\end{align*}

\endgroup

\subsection{Weighted Automata}

\begingroup
\newcommand{\sring}{\mathbb{S}}
\newcommand{\field}{\mathbb{F}}

We generalise the monad~$\LinFunct$,
and we give the definition of the commutative monad~$\LinFunct[\sring]$ where we replace
integer coefficients with elements of an arbitrary commutative semiring~$\sring$.
The unit~$\effunit[X][\LinFunct[\sring]]$ of~$\LinFunct[\sring]$ sends~$x \in X$ to~$1 \cdot x$,
and the multiplication~$\effmult[X]$ of~$\LinFunct[\sring]$ is given by
\[
  \effmult[X]\mleft(\sum_{i=1}^m a_i \cdot v_i\mright)(x) \defeq \sum_{i=1}^m a_i v_i(x).
\]
Algebraically, $\Lin[\sring]{X}$~is the underlying set of the free $\sring$-semimodule on~$X$,
with pointwise addition and scalar multiplication.
If $\sring$~is a field, then we get a free vector space.
The monad~$\LinFunct[\sring]$ is commutative: the map~\(
  \effmf[X,Y] \colon \Lin[\sring]{X} \prodset \Lin[\sring]{Y} \to \Lin[\sring]{X \prodset Y}
\)
is given by
\[
  \effmf[X,Y]\mleft(\sum_{i=1}^m a_i \cdot x_i, \sum_{j=1}^n b_j \cdot y_j\mright) \defeq
    \sum_{i=1}^m \sum_{j=1}^n a_i b_j \cdot (x_i,y_j).
\]
Furthermore, we have a distributive law~\(
  \dlaw[X] \colon \Lin[\sring]{X} \prodset \abcobj \to \Lin[\sring]{X \prodset \abcobj}
\),
defined as
\[
  \dlaw[X]\mleft(\sum_{i=1}^m a_i \cdot x_i, s\mright) \defeq \sum_{i=1}^m a_i \cdot (x_i, s).
\]

For weighted automata over a field~$\field$, we can prove that the size of
the minimal state cover for an arbitrary WFA~$\spec$ is at most the number of states of~$\spec$.
To see this, write~$Q$ for the state space of~$\spec$
and let~$R \defeq \im{\rea{\alglift{\klext}\spec}}$.
Since $\Lin[\field]{Q}$~is a finite dimensional vector space,
the span of~$R$ in~$\Lin[\field]{Q}$ is also finite dimensional.
Hence, $R$~contains a finite basis~$B$ for~$\Span{R}$.
By choosing access sequences for the elements of~$B$, we thus obtain a minimal state cover
for~$\spec$ with size~$\card{B} = \dim\Span{R} \le \dim\Lin[\field]{Q} = \card{Q}$.

\endgroup

\paragraph*{Example WFA}

The state space of~$\exWFA[n]$ is~$Q \defeq \set{q_i}[0 \le i \le n+1]$,
with initial configuration~$i(*) \defeq 1 \cdot q_0$.
The final vector is given by $o(q_{n+1}) = 1$ and $o(q_i) = 0$ for $0 \le i \le n$.
Finally, the transition function~$\delta \colon Q \prodset \abcobj \to \Lin{Q}$ is defined as
\begin{alignat*}{2}
  \delta(q_0,     a) &\leteq 2 \cdot q_0 + 1 \cdot q_1, \\
  \delta(q_0,     b) &\leteq 1 \cdot q_0, \\
  \delta(q_i,     s) &\leteq 1 \cdot q_{i+1}  &\hspace{3em}& \text{for~$1 \le i \le n$ and~$s \in \abcobj$}, \\
  \delta(q_{n+1}, s) &\leteq 0                &            & \text{for~$s \in \abcobj$}.
\end{alignat*}
The WFA~$\exWFA[n]$ recognises the weighted language~$\exwlang[n]$ given by
\[
  \exwlang[n](v) \defeq
    \begin{cases}
      2^k & \text{if $v = uax$ for some~$u \in \words$ and~$x \in \abcobj^n$, and $|u|_a = k$\,;} \\
      0   & \text{otherwise.}
    \end{cases}
\]

\paragraph*{Details for the test suite of~\texorpdfstring{$\exWFA[2]$}{the example WFA}}

The map~$\delta_C$ for the state cover~$C$ is given as follows:
\begin{alignat*}{2}
  \empwd, a &\mapsto 1 \cdot a     &              \empwd, b &\mapsto 1 \cdot \empwd \\
  a,      a &\mapsto 1 \cdot aa    &              a,      b &\mapsto 2 \cdot \empwd - 2 \cdot a + 1 \cdot aa \\
  aa,     a &\mapsto 1 \cdot aaa   &              aa,     b &\mapsto 4 \cdot \empwd - 4 \cdot a + 1 \cdot aaa \\
  aaa,    a &\mapsto 2 \cdot aaa   &\hspace{5em}  aaa,    b &\mapsto 8 \cdot \empwd - 8 \cdot a + 2 \cdot aaa
\end{alignat*}
The resulting effectful W-suite for~$\exWFA[2]$ is
\begin{align*}
  \im{t} = \{
    & \empwd, a, aa, aaa, aaaa, aaaaa, aaaaaa, aaaaaaa, ab, aba, abaa, abaaa, \\
    & aab, aaba, aabaa, aabaaa, aaab, aaaba, aaabaa, aaabaaa, b, ba, baa, baaa
  \}.
\end{align*}

\section{Construction of~\texorpdfstring{$\flstr$}{σ*}}

For a varietor~$\dyn$ with strength~$\lstr$,  
the construction of a strength~$\flstr$ for~$\fdyn$ is somewhat folklore.
McDermott and Uustalu~\cite{MU:22} provide a nice characterisation
in terms of \emph{strongly free} algebras.
We summarise their results in~\cref{prop:flstr}.

\begin{definition}\label{def:strongly-free-alg}
  Let~$\bcat$ be a monoidal category,
  let~$\dyn \colon \bcat \to \bcat$ be a functor with strength~$\lstr$,
  and let~$A \in \bcat$.
  A \emph{strongly free $\dyn$-algebra over~$A$} consists of
  \an{\dyn} $\dyn$-algebra~$(\fdyn A, \con[A])$ and a map~$\emp[A] \colon A \to \fdyn A$
  \sth for all objects~$X \in \bcat$, $\dyn$-algebras~$(B, b)$,
  and maps~$f \colon X \tprod A \to B$, there exists a
  unique map~$\fext{f} \colon X \tprod \fdyn A \to B$ making the following diagrams commute:
  \[
    \begin{tikzcd}[baseline=(\tikzcdmatrixname.base),column sep=3.5em]
      {X \tprod \dyn\fdyn A} & {\dyn(X \tprod \fdyn A)} & {\dyn B} \\
      {X \tprod \fdyn A} && B
      \arrow["{\lstr[X,\fdyn A]}", from=1-1, to=1-2]
      \arrow["{\id[X] \tprod \con[A]}"', from=1-1, to=2-1]
      \arrow["{\dyn\fext{f}}", from=1-2, to=1-3]
      \arrow["b", from=1-3, to=2-3]
      \arrow["{\fext{f}}"', from=2-1, to=2-3]
    \end{tikzcd}
    \hspace{1.5em}
    \begin{tikzcd}[baseline=(\tikzcdmatrixname.base)]
      {X \tprod A} & \\
      {X \tprod \fdyn A} & B
      \arrow["{\id[X] \tprod \emp[A]}"', from=1-1, to=2-1]
      \arrow["f", from=1-1, to=2-2]
      \arrow["{\fext{f}}"', from=2-1, to=2-2]
    \end{tikzcd}
  \]
\end{definition}

\begin{proposition}\label{prop:flstr}
  Let~$\bcat$ be a monoidal closed category
  and let~$\dyn \colon \bcat \to \bcat$ be a functor with strength~$\lstr$.
  Then every free $\dyn$-algebra is strongly free.
  In particular, if $\dyn$~is a varietor, then $\fdyn$~has a strength~$\flstr$
  uniquely characterised by the fact that the following diagrams commute:
  \[
    \begin{tikzcd}[baseline=(\tikzcdmatrixname.base),column sep=3.5em]
      {A \tprod \dyn\fdyn B} & {\dyn(A \tprod \fdyn B)} & {\dyn\fdyn(A \tprod B)} \\
      {A \tprod \fdyn B} && {\fdyn(A \tprod B)}
      \arrow["{\lstr[A,\fdyn B]}", from=1-1, to=1-2]
      \arrow["{\id[A] \tprod \con[B]}"', from=1-1, to=2-1]
      \arrow["{\dyn\flstr[A,B]}", from=1-2, to=1-3]
      \arrow["{\con[A \tprod B]}", from=1-3, to=2-3]
      \arrow["{\flstr[A,B]}"', from=2-1, to=2-3]
    \end{tikzcd}
    \hspace{0.3em}
    \begin{tikzcd}[baseline=(\tikzcdmatrixname.base)]
      {A \tprod B} & \\
      {A \tprod \fdyn B} & {\fdyn(A \tprod B)}
      \arrow["{\id[A] \tprod \emp[B]}"', from=1-1, to=2-1]
      \arrow["{\emp[A \tprod B]}", from=1-1, to=2-2]
      \arrow["{\flstr[A,B]}"', from=2-1, to=2-2]
    \end{tikzcd}
  \]
\end{proposition}

\section{Proof of~\texorpdfstring{\cref{thm:w-comp}}{\autoref{thm:w-comp}}}

The proof presented here is a generalisation (and simplification)
of the one in~\cite[Appendix~C]{KR:25}.
We shall prove the following statement, which can be seen as
a simplified version of the W-method where $k = 0$.
\begin{theorem}\label{thm:w-k0-comp}
  Let $\spec$~and~$\impl$ be two automata in~$\Aut{\dyn}{\inobj}{\outobj}$,
  let~$(c, \delta_C)$ be a state cover for~$\impl$,
  and let~$w$ be a state characteriser for~$\spec$.
  Suppose $\spec$~is observable,
  and let~$t \leteq \trext{c} \tsconc w$.
  Then $\beh{\spec} \com t = \beh{\impl} \com t$ implies $\beh{\spec} = \beh{\impl}$.
\end{theorem}
The completeness of the W-method follows from this special case.
\begin{proof}[Proof of~\cref{thm:w-comp}]
  Writing~$t \leteq \trext{c}[k+1] \tsconc w$,
  let~$\impl \in \wfdom{c}{k}$, and suppose $\beh{\spec} \com t = \beh{\impl} \com t$.
  By the definition of~$\wfdom{c}{k}$ (\cref{def:w-fdom}),
  there exists~$\delta_C$ such that $(\trext{c}[k], \delta_C)$~is a state cover for~$\impl$.
  By assumption, we also have a state characteriser~$w$ for~$\spec$,
  and we know that $\spec$~is observable.
  Since $\trext*{\trext{c}[k]} = \trext{c}[k+1]$,
  \cref{thm:w-k0-comp} implies $\beh{\spec} = \beh{\impl}$.
\end{proof}

Before embarking on the proof of~\cref{thm:w-k0-comp}, we need some auxiliary results.

\begin{proposition}\label{prop:trconc-emp}
  The following triangle commutes:
  \[\begin{tikzcd}[column sep=1.5em]
    {\fdyn\inobj \tprod \tunit} & \\
    {\fdyn\inobj \tprod \fdyn\tunit} & {\fdyn\inobj}
    \arrow["{\id[\fdyn\inobj] \tprod \emp[\tunit]}"', from=1-1, to=2-1]
    \arrow["{\runit[\fdyn\inobj]}", from=1-1, to=2-2]
    \arrow["\trconc"', from=2-1, to=2-2]
  \end{tikzcd}\]
\end{proposition}
\begin{proof}
  The desired triangle is the outside of the commutative diagram below,
  where the left triangle commutes by~\cref{prop:flstr},
  the middle square commutes by the naturality of~$\emp$,
  and the right triangle commutes by the definition of~$\idext{\con[\inobj]}$.
  \[\begin{tikzcd}[baseline=(\tikzcdmatrixname.base)]
    &[-0.6em] {\fdyn\inobj \tprod \tunit} & {\fdyn\inobj} &[0.6em] \\
    {\fdyn\inobj \tprod \fdyn\tunit} & {\fdyn(\fdyn\inobj \tprod \tunit)} & {\fdyn\fdyn\inobj} & {\fdyn\inobj}
    \arrow["\runit", from=1-2, to=1-3]
    \arrow["{\id \tprod \emp}"', shift right, curve={height=3pt}, from=1-2, to=2-1]
    \arrow["\emp"', from=1-2, to=2-2]
    \arrow["\emp"', from=1-3, to=2-3]
    \arrow[shift left=1.5, equals, from=1-3, to=2-4]
    \arrow["\flstr", from=2-1, to=2-2]
    \arrow["\trconc"', shift right, curve={height=8pt}, from=2-1, to=2-4]
    \arrow["{\fdyn\runit}", from=2-2, to=2-3]
    \arrow["{\idext{\con[\inobj]}}"{pos=0.4}, from=2-3, to=2-4]
  \end{tikzcd}\qedhere\]
\end{proof}

\begin{proposition}\label{prop:rea-trconc}
  Let~$\aut{A} = (Q, \delta, i, o)$ be an automaton.
  Then the following commutes:
  \[\begin{tikzcd}[column sep=2em]
    {\fdyn\inobj \tprod \fdyn\tunit} & {\fdyn\inobj} \\
    {Q \tprod \fdyn\tunit} & Q
    \arrow["\trconc", from=1-1, to=1-2]
    \arrow["{\rea{\aut{A}} \tprod \id[\fdyn\tunit]}"', from=1-1, to=2-1]
    \arrow["{\rea{\aut{A}}}", from=1-2, to=2-2]
    \arrow["{\run{\aut{A}}}"', from=2-1, to=2-2]
  \end{tikzcd}\]
\end{proposition}
\begin{proof}
  The desired square is the outside of the following commuting diagram:
  
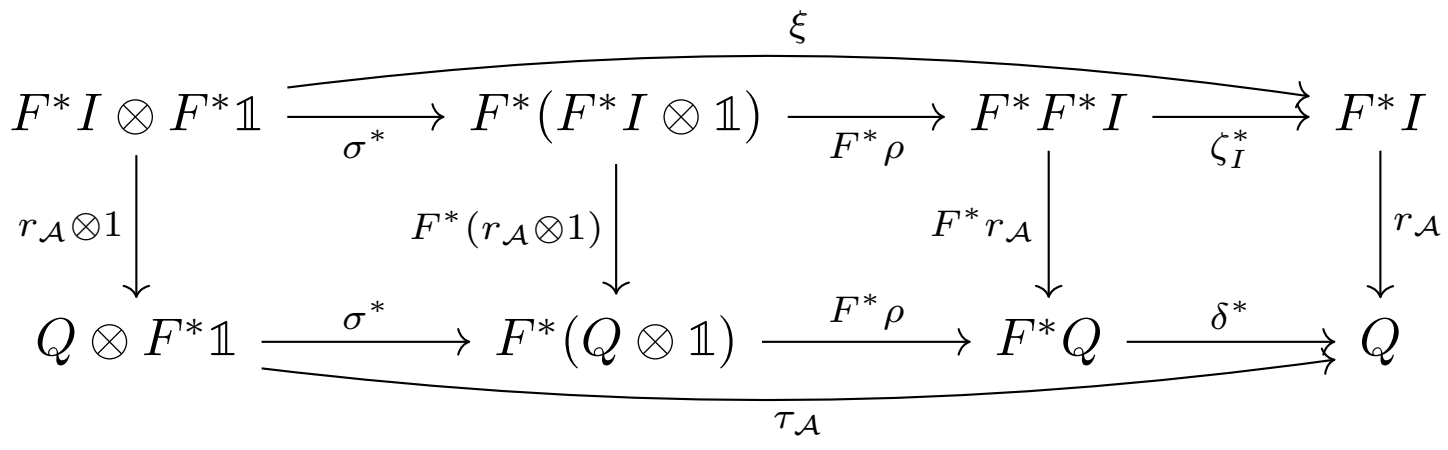
\[\begin{tikzcd}[column sep=3em,row sep=2.5em]
    {\fdyn\inobj \tprod \fdyn\tunit} & {\fdyn(\fdyn\inobj \tprod \tunit)} & {\fdyn\fdyn\inobj} & {\fdyn\inobj} \\
    {Q \tprod \fdyn\tunit} & {\fdyn(Q \tprod \tunit)} & {\fdyn Q} & Q
    \arrow["\flstr"', from=1-1, to=1-2]
    \arrow["\trconc", shift left, curve={height=-9pt}, from=1-1, to=1-4]
    \arrow["{\rea{\aut{A}} \tprod \id}"', from=1-1, to=2-1]
    \arrow["{\fdyn\runit}"', from=1-2, to=1-3]
    \arrow["{\fdyn(\rea{\aut{A}} \tprod \id)}"', from=1-2, to=2-2]
    \arrow["{\idext{\con[\inobj]}}"', from=1-3, to=1-4]
    \arrow["{\fdyn\rea{\aut{A}}}"', from=1-3, to=2-3]
    \arrow["{\rea{\aut{A}}}", from=1-4, to=2-4]
    \arrow["\flstr", from=2-1, to=2-2]
    \arrow["{\run{\aut{A}}}"', shift right, curve={height=9pt}, from=2-1, to=2-4]
    \arrow["{\fdyn\runit}", from=2-2, to=2-3]
    \arrow["{\idext{\delta}}", from=2-3, to=2-4]
  \end{tikzcd}\]
  The leftmost and the middle square commute
  by the naturality of~$\flstr$ and of~$\emp$, respectively.
  The rightmost square commutes since both sides are
  $\dyn$-algebra homomorphisms out of a free algebra such that
  precomposing them with~$\emp[\fdyn\inobj]$ gives~$\rea{\aut{A}}$.
\end{proof}

The remainder of this section is devoted to proving~\cref{thm:w-k0-comp}.
Hence, we fix a specification~$\spec = (\specname, \delta_\specname, i_\specname, o_\specname)$
and an implementation~$\impl = (\implname, \delta_\implname, i_\implname, o_\implname)$.
The strategy is to construct a `bisimulation' between $\spec$~and~$\impl$,
\ie a span~$\spec \xleftarrow{r_1} \aut{R} \xrightarrow{r_2} \impl$ in the category of automata.
Since automata connected by a homomorphism share the same behaviour, the theorem follows.
The bisimulation~$\aut{R}$ is constructed using the state cover,
and the morphisms $r_1$~and~$r_2$ are given by the reachability maps.
The next two lemmas cover parts of this proof.

\begin{lemma}\label{lem:w-k0-comp-1}
  Under the conditions of~\cref{thm:w-k0-comp}, the following diagram commutes:
  
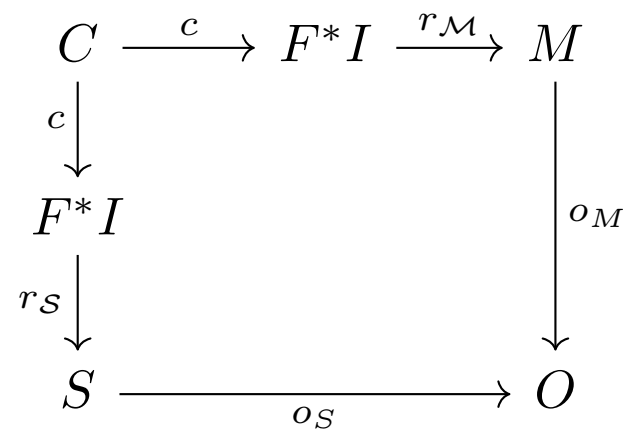
\[\begin{tikzcd}[column sep=2em]
    C & {\fdyn\inobj} & \implname \\
    {\fdyn\inobj} \\
    \specname && \outobj
    \arrow["c", from=1-1, to=1-2]
    \arrow["c"'{pos=0.4}, from=1-1, to=2-1]
    \arrow["{\rea{\impl}}", from=1-2, to=1-3]
    \arrow["{o_\implname}", from=1-3, to=3-3]
    \arrow["{\rea{\spec}}"', from=2-1, to=3-1]
    \arrow["{o_\specname}"', from=3-1, to=3-3]
  \end{tikzcd}\]
\end{lemma}
\begin{proof}
  By assumption, we have~\(
    o_\specname \com \rea{\spec} \com t = \beh{\spec} \circ t
      = \beh{\impl} \circ t = o_\implname \com \rea{\impl} \com t
  \).
  Thus, it suffices to show that $t \com f = c$
  for some morphism~$f \colon C \to \trext{C} \tprod W$.
  Since $w$~is a state characteriser, we have $\emp[\tunit] \le w$,
  \ie there is a morphism~$i_W \colon \tunit \to W$ with $w \com i_W = \emp[\tunit]$.
  Hence, we can take
  \[
    f \defeq
      C \xto{\inv{\runit[C]}}
      C \tprod \tunit \xto{\coprojl \tprod i_W}
      (C \catsum \dyn C) \tprod W =
      \trext{C} \tprod W.
  \]
  The equality~$t \com f = c$ follows from the following commuting diagram,
  using~\cref{prop:trconc-emp}:
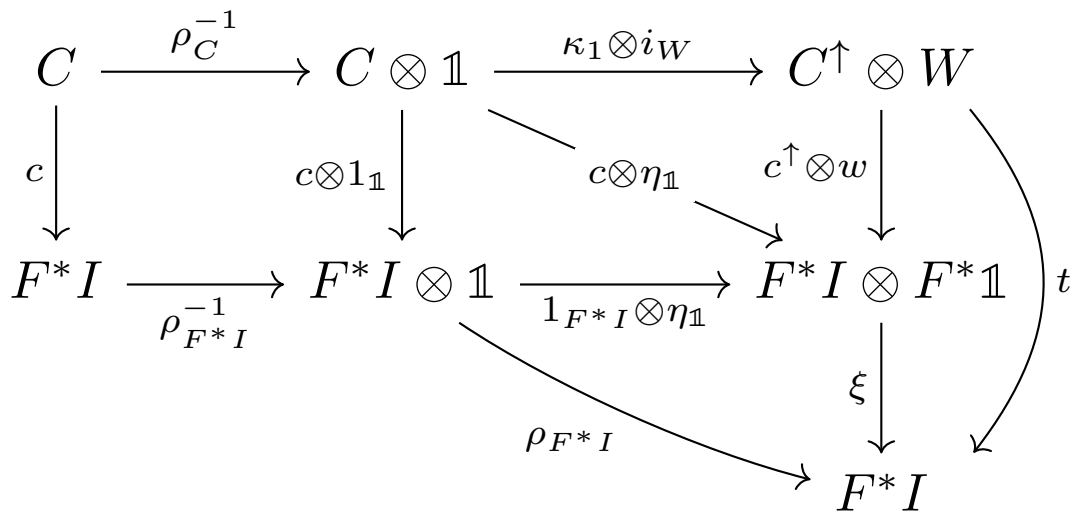
  \[\begin{tikzcd}[baseline=(\tikzcdmatrixname.base),column sep=3em,row sep=2.5em]
    C & {C \tprod \tunit} &[1em] {\trext{C} \tprod W} \\
    {\fdyn\inobj} & {\fdyn\inobj \tprod \tunit} & {\fdyn\inobj \tprod \fdyn\tunit} \\
    && {\fdyn\inobj}
    \arrow["{\inv{\runit[C]}}", from=1-1, to=1-2]
    \arrow["c"', from=1-1, to=2-1]
    \arrow["{\coprojl \tprod i_W}", from=1-2, to=1-3]
    \arrow["{c \tprod \id[\tunit]}"', from=1-2, to=2-2]
    \arrow["{c \tprod \emp[\tunit]}"{description}, from=1-2, to=2-3]
    \arrow["{\trext{c} \tprod w}"'{pos=0.4}, from=1-3, to=2-3]
    \arrow["t", shift left=5, curve={height=-18pt}, from=1-3, to=3-3]
    \arrow["{\inv{\runit[\fdyn\inobj]}}"', from=2-1, to=2-2]
    \arrow["{\id[\fdyn\inobj] \tprod \emp[\tunit]}"', from=2-2, to=2-3]
    \arrow["{\runit[\fdyn\inobj]}"', curve={height=6pt}, from=2-2, to=3-3]
    \arrow["\trconc"', from=2-3, to=3-3]
  \end{tikzcd}\qedhere\]
\end{proof}

\begin{lemma}\label{lem:w-k0-comp-2}
  Under the conditions of~\cref{thm:w-k0-comp}, the following diagram commutes:
  
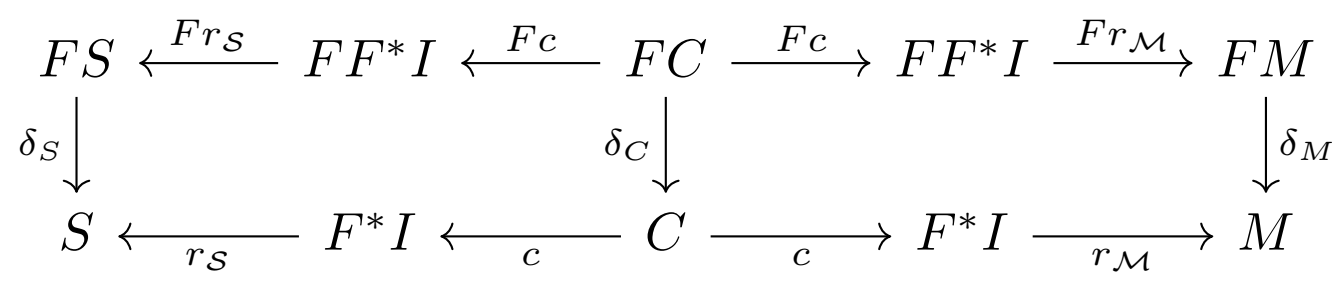
\[\begin{tikzcd}[column sep=2.6em]
    {\dyn\specname} & {\dyn\fdyn\inobj} & {\dyn C} & {\dyn\fdyn\inobj} & {\dyn\implname} \\
    \specname & {\fdyn\inobj} & C & {\fdyn\inobj} & \implname
    \arrow["{\delta_\specname}"', from=1-1, to=2-1]
    \arrow["{\dyn\rea{\spec}}"', from=1-2, to=1-1]
    \arrow["{\dyn c}"', from=1-3, to=1-2]
    \arrow["{\dyn c}", from=1-3, to=1-4]
    \arrow["{\delta_C}"', from=1-3, to=2-3]
    \arrow["{\dyn\rea{\impl}}", from=1-4, to=1-5]
    \arrow["{\delta_\implname}", from=1-5, to=2-5]
    \arrow["{\rea{\spec}}", from=2-2, to=2-1]
    \arrow["c", from=2-3, to=2-2]
    \arrow["c"', from=2-3, to=2-4]
    \arrow["{\rea{\impl}}"', from=2-4, to=2-5]
  \end{tikzcd}\]
\end{lemma}
\begin{proof}
  The right-hand rectangle commutes since $\rea{\impl} \com c$~is
  \an{\dyn} $\dyn$-algebra homomorphism by the definition of state cover.
  For the left-hand rectangle, we use that $\spec$~is observable
  and that $w$~is a state characteriser for~$\spec$.
  Therefore, it suffices to show that the two sides of the rectangle become equal
  after postcomposing with~$\inthom{w}{\id[\outobj]} \com \obs{\spec}$.
  This follows from the commuting diagram below,
  where we use again that $(c, \delta_C)$~is a state cover.
  
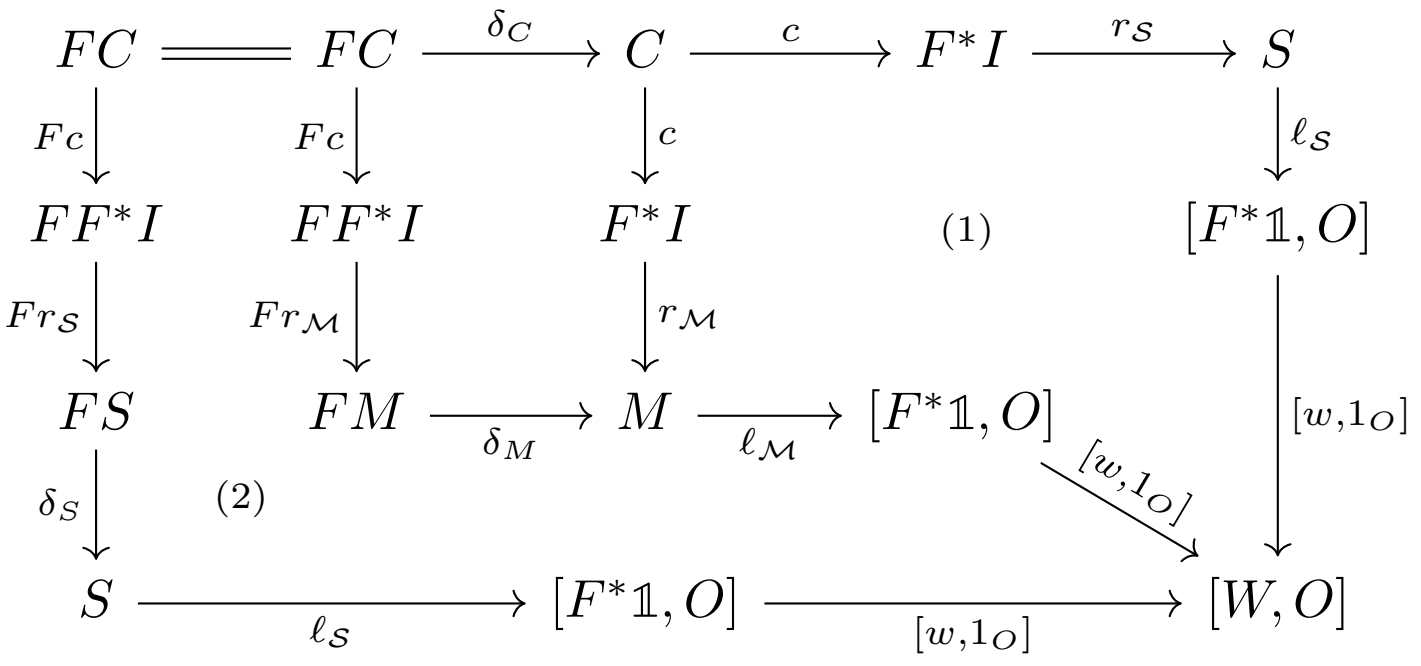
\[\begin{tikzcd}[column sep=1.5em]
    {\dyn C} & {\dyn C} & C & {\fdyn\inobj} & \specname \\
    {\dyn\fdyn\inobj} & {\dyn\fdyn\inobj} & {\fdyn\inobj} && {\inthom{\fdyn\tunit}{\outobj}} \\
    {\dyn\specname} & {\dyn\implname} & \implname & {\inthom{\fdyn\tunit}{\outobj}} \\
    \specname && {\inthom{\fdyn\tunit}{\outobj}} && {\inthom{W}{\outobj}}
    \arrow[equals, nfold, from=1-1, to=1-2]
    \arrow["{\dyn c}"', from=1-1, to=2-1]
    \arrow["{\delta_C}", from=1-2, to=1-3]
    \arrow["{\dyn c}"', from=1-2, to=2-2]
    \arrow["c", from=1-3, to=1-4]
    \arrow["c", from=1-3, to=2-3]
    \arrow["{\rea{\spec}}", from=1-4, to=1-5]
    \arrow["{\obs{\spec}}", from=1-5, to=2-5]
    \arrow["{\dyn\rea{\spec}}"', from=2-1, to=3-1]
    \arrow["{\dyn\rea{\impl}}"', from=2-2, to=3-2]
    \arrow["{\rea{\impl}}", from=2-3, to=3-3]
    \arrow["{\inthom{w}{\id[\outobj]}}", from=2-5, to=4-5]
    \arrow["{\delta_\specname}"', from=3-1, to=4-1]
    \arrow["{\delta_\implname}"', from=3-2, to=3-3]
    \arrow["{(1)}"{description}, draw=none, from=3-3, to=1-5]
    \arrow["{\obs{\impl}}"', from=3-3, to=3-4]
    \arrow["{\inthom{w}{\id[\outobj]}}"{sloped}, from=3-4, to=4-5]
    \arrow["{(2)}"{description, pos=0.6}, draw=none, from=4-1, to=3-2]
    \arrow["{\obs{\spec}}"', from=4-1, to=4-3]
    \arrow["{\inthom{w}{\id[\outobj]}}"', from=4-3, to=4-5]
  \end{tikzcd}\]

  It remains to show the commutativity of the areas labelled $(1)$~and~$(2)$.
  \begin{enumerate}[$(1)$]
    \item
    Taking adjuncts and writing~$\intev$ for the evaluation map, we need to show
    \[
      \intev \com (\id[\inthom{\fdyn\tunit}{\outobj}] \tprod w)
        \com ((\obs{\spec} \com \rea{\spec} \com c) \tprod \id[W]) =
      \intev \com (\id[\inthom{\fdyn\tunit}{\outobj}] \tprod w)
        \com ((\obs{\impl} \com \rea{\impl} \com c) \tprod \id[W]).
    \]
    Rearranging and using the definition of the observability map, this is equivalent to
    \[
      o_\specname \com \run{\spec} \com (\rea{\spec} \tprod \id[\fdyn\tunit]) \com (c \tprod w)
        = o_\implname \com \run{\impl} \com (\rea{\impl} \tprod \id[\fdyn\tunit]) \com (c \tprod w).
    \]
    This equality follows from the assumption~$\beh{\spec} \com t = \beh{\impl} \com t$
    by precomposing both sides with~\(
      (\coprojl \tprod \id[W]) \colon C \tprod W \to (C \catsum \dyn C) \tprod W = \trext{C} \tprod W
    \).
    Equality of the left sides is justified by the diagram below,
    using~\cref{prop:rea-trconc}; the right side is treated analogously.
    
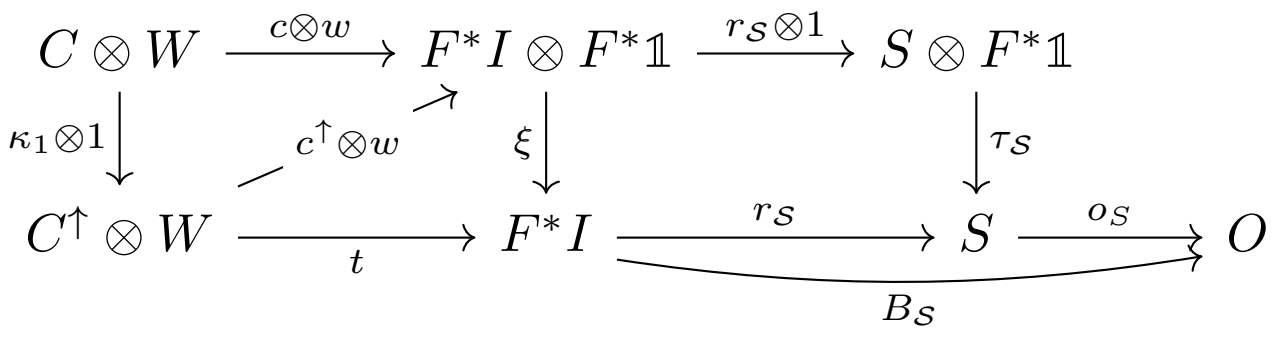
\[\begin{tikzcd}[column sep=3em]
      {C \tprod W} & {\fdyn\inobj \tprod \fdyn\tunit} & {\specname \tprod \fdyn\tunit} &[-1em] \\
      {\trext{C} \tprod W} & {\fdyn\inobj} & \specname & \outobj
      \arrow["{c \tprod w}", from=1-1, to=1-2]
      \arrow["{\coprojl \tprod \id}"', from=1-1, to=2-1]
      \arrow["{\rea{\spec} \tprod \id}", from=1-2, to=1-3]
      \arrow["\trconc"', from=1-2, to=2-2]
      \arrow["{\run{\spec}}", from=1-3, to=2-3]
      \arrow["{\trext{c} \tprod w}"{description}, from=2-1, to=1-2]
      \arrow["t"', from=2-1, to=2-2]
      \arrow["{\rea{\spec}}", from=2-2, to=2-3]
      \arrow["{\beh{\spec}}"', shift right, curve={height=6pt}, from=2-2, to=2-4]
      \arrow["{o_\specname}", from=2-3, to=2-4]
    \end{tikzcd}\]

    \item
    The proof is similar to the previous part.
    Taking adjuncts, we need to show
    \[
      \intev \com (\id \tprod w)
        \com ((\obs{\spec} \com \delta_\specname \com \dyn\rea{\spec} \com \dyn c) \tprod \id) =
      \intev \com (\id \tprod w)
        \com ((\obs{\impl} \com \delta_\implname \com \dyn\rea{\impl} \com \dyn c) \tprod \id).
    \]
    Rearranging and using the definition of the observability map, this is equivalent to
    \[
      o_\specname \com \run{\spec} \com (\delta_\specname \tprod \id)
        \com (\dyn\rea{\spec} \tprod \id) \com (\dyn c \tprod w) =
      o_\implname \com \run{\impl} \com (\delta_\implname \tprod \id)
        \com (\dyn\rea{\impl} \tprod \id) \com (\dyn c \tprod w).
    \]
    This equality follows from the assumption~$\beh{\spec} \com t = \beh{\impl} \com t$
    by precomposing both sides with~\(
      (\coprojr \tprod \id[W]) \colon \dyn C \tprod W \to (C \catsum \dyn C) \tprod W = \trext{C} \tprod W
    \).
    Equality of the left sides is justified by the diagram below,
    using~\cref{prop:rea-trconc} and that $\rea{\spec}$~is \an{\dyn} $\dyn$-algebra homomorphism;
    the right side is treated analogously.
    
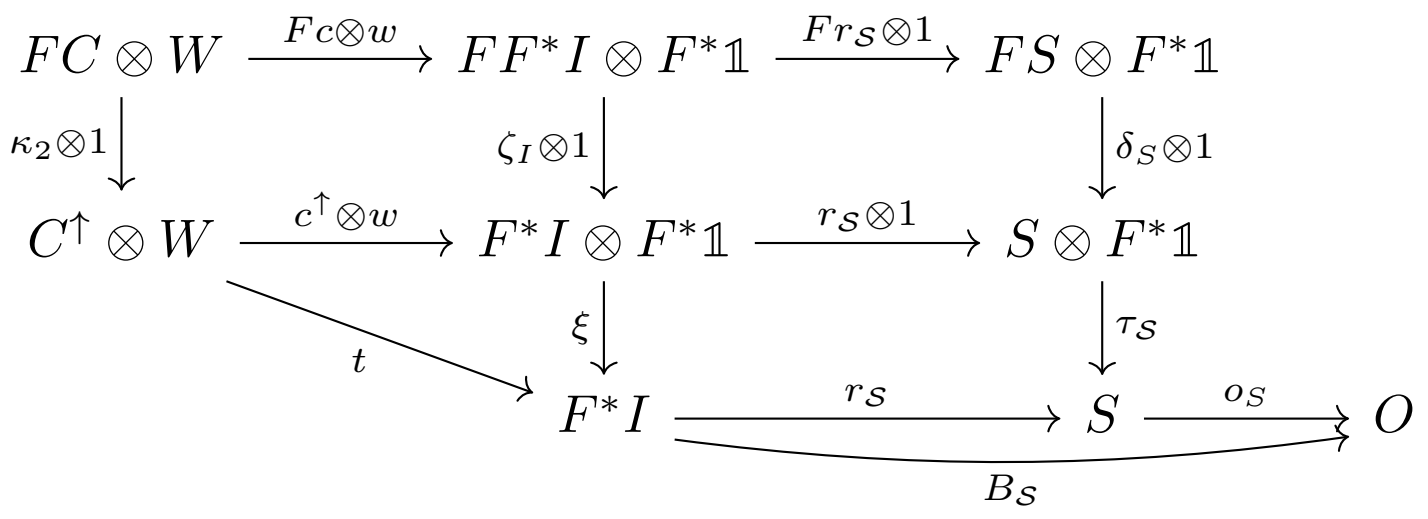
\[\begin{tikzcd}[baseline=(\tikzcdmatrixname.south),column sep=3.5em]
      {\dyn C \tprod W} & {\dyn\fdyn\inobj \tprod \fdyn\tunit} & {\dyn\specname \tprod \fdyn\tunit} &[-1.5em] \\
      {\trext{C} \tprod W} & {\fdyn\inobj \tprod \fdyn\tunit} & {\specname \tprod \fdyn\tunit} \\
      & {\fdyn\inobj} & \specname & \outobj
      \arrow["{\dyn c \tprod w}", from=1-1, to=1-2]
      \arrow["{\coprojr \tprod \id}"', from=1-1, to=2-1]
      \arrow["{\dyn\rea{\spec} \tprod \id}", from=1-2, to=1-3]
      \arrow["{\con[\inobj] \tprod \id}"', from=1-2, to=2-2]
      \arrow["{\delta_\specname \tprod \id}", from=1-3, to=2-3]
      \arrow["{\trext{c} \tprod w}", from=2-1, to=2-2]
      \arrow["t"', from=2-1, to=3-2]
      \arrow["{\rea{\spec} \tprod \id}", from=2-2, to=2-3]
      \arrow["\trconc"', from=2-2, to=3-2]
      \arrow["{\run{\spec}}", from=2-3, to=3-3]
      \arrow["{\rea{\spec}}", from=3-2, to=3-3]
      \arrow["{\beh{\spec}}"', shift right, curve={height=6pt}, from=3-2, to=3-4]
      \arrow["{o_\specname}", from=3-3, to=3-4]
    \end{tikzcd}\qedhere\]
  \end{enumerate}
\end{proof}

\begin{proof}[Proof of~\cref{thm:w-k0-comp}]
  Since $(c, \delta_C)$~is a state cover, we have $\emp[\inobj] \le c$,
  \ie there exists a map~$i_C \colon \inobj \to C$ with $c \com i_C = \emp[\inobj]$.
  Thus, we can define an automaton~$\aut{R} \defeq (C, \delta_C, i_C, \beh{\spec} \com c)$.
  Since $\rea{\spec} \com c \com i_C = \rea{\spec} \com \emp[\inobj] = i_\specname$,
  the map~$\rea{\spec} \com c$ preserves initial states; similarly for~$\rea{\impl} \com c$.
  By~\cref{lem:w-k0-comp-1,lem:w-k0-comp-2}, these maps also preserve transitions and outputs;
  hence, they are automata homomorphisms.
  Therefore, we obtain a span of automata~\(
    \spec \xleftarrow{\rea{\spec} \com c} \aut{R} \xrightarrow{\rea{\impl} \com c} \impl
  \).
  Since the existence of a homomorphism between two automata
  implies that they have the same behaviour~\cite{AM:75b},
  it follows that $\beh{\spec} = \beh{\impl}$, as desired.
\end{proof}

\section{Proof of~\texorpdfstring{\cref{thm:dlaw-emdyn-equiv}}{\autoref{thm:dlaw-emdyn-equiv}}}

We first recall Mulry's construction\footnotemark\ of~$\emdyn$
and~$\emfreedyn$~\cite[Lemma~2.20]{Mul:02} (which is slightly more general).
We then verify properties~\labelcref{prop:emdyn-pres-can,prop:emfreedyn-effmult}
and conclude with the uniqueness argument.
\footnotetext{A nearly identical construction can also be found in~\cite[Theorem~2]{Joh:75}.}

\begingroup
\newcommand{\q}[1][]{q_{#1}}

\proofsubparagraph{Construction of~$\emdyn$.}

For any $\eff$-algebra~$\malg{A} = (A, a)$, the parallel pair~\(
  \begin{tikzcd}[cramped,column sep=4em]
    {\emfree\dyn\eff A} & {\emfree\dyn A}
    \arrow["{\effmult[\dyn A] \com \emfree\dlaw[A]}"{pos=0.48}, shift left, from=1-1, to=1-2]
    \arrow["{\emfree\dyn a}"', shift right, from=1-1, to=1-2]
  \end{tikzcd}
\)
is reflexive, with common section~$\emfree\dyn\effunit[A]$.
Hence, we can define~$\emdyn\malg{A}$ as the coequaliser
\[\begin{tikzcd}
  {\emfree\dyn\eff A} &[2em] {\emfree\dyn A} & {\emdyn\malg{A}}
  \arrow["{\effmult[\dyn A] \com \emfree\dlaw[A]}", shift left, from=1-1, to=1-2]
  \arrow["{\emfree\dyn a}"', shift right, from=1-1, to=1-2]
  \arrow["{\q[\malg{A}]}", from=1-2, to=1-3]
\end{tikzcd}\]
in~$\emcat$, which exists by~\cref{ass:monad}.
For a map of algebras~$f \colon \malg{A} \to \malg{B}$, where $\malg{B} = (B, b)$,
we define the map~$\emdyn f$ using the universal property of the coequaliser~$\emdyn\malg{A}$:
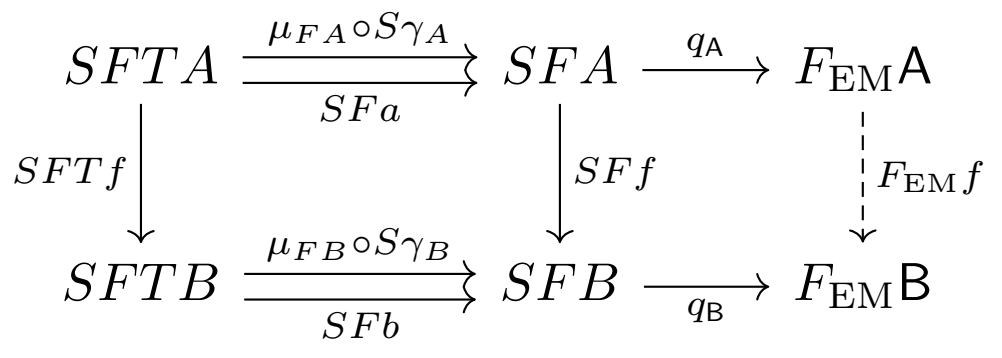
\[\begin{tikzcd}[row sep=2.5em]
	{\emfree\dyn\eff A} &[2em] {\emfree\dyn A} & {\emdyn\malg{A}} \\
	{\emfree\dyn\eff B} & {\emfree\dyn B} & {\emdyn\malg{B}}
	\arrow["{\effmult[\dyn A] \com \emfree\dlaw[A]}", shift left, from=1-1, to=1-2]
	\arrow["{\emfree\dyn a}"', shift right, from=1-1, to=1-2]
	\arrow["{\emfree\dyn\eff f}"', from=1-1, to=2-1]
	\arrow["{\q[\malg{A}]}", from=1-2, to=1-3]
	\arrow["{\emfree\dyn f}", from=1-2, to=2-2]
	\arrow["{\emdyn f}", dashed, from=1-3, to=2-3]
	\arrow["{\effmult[\dyn B] \com \emfree\dlaw[B]}", shift left, from=2-1, to=2-2]
	\arrow["{\emfree\dyn b}"', shift right, from=2-1, to=2-2]
	\arrow["{\q[\malg{B}]}"', from=2-2, to=2-3]
\end{tikzcd}\]
Here, the left square commutes `serially': we have
\(
  \emfree\dyn f \com \effmult[\dyn A] \com \emfree\dlaw[A]
    = \effmult[\dyn B] \com \emfree\dlaw[B] \com \emfree\dyn\eff f
\)
by the naturality of $\effmult$~and~$\dlaw$, and
$\emfree\dyn f \com \emfree\dyn a = \emfree\dyn b \com \emfree\dyn\eff f$
since $f$~preserves the $\eff$-algebra structure.
This implies that the morphism~$\q[\malg{B}] \com \emfree\dyn f$
coequalises the upper parallel pair;
thus, the coequaliser~$\emdyn\malg{A}$ induces
a map~$\emdyn f \colon \emdyn\malg{A} \to \emdyn\malg{B}$
with $\emdyn f \com \q[\malg{A}] = \q[\malg{B}] \com \emfree\dyn f$.
One can check that $\emdyn$~preserves composition and identities
using that $\q[\malg{A}]$~is an epi.

\proofsubparagraph{Construction of~$\emfreedyn$.}

Consider the following diagam:
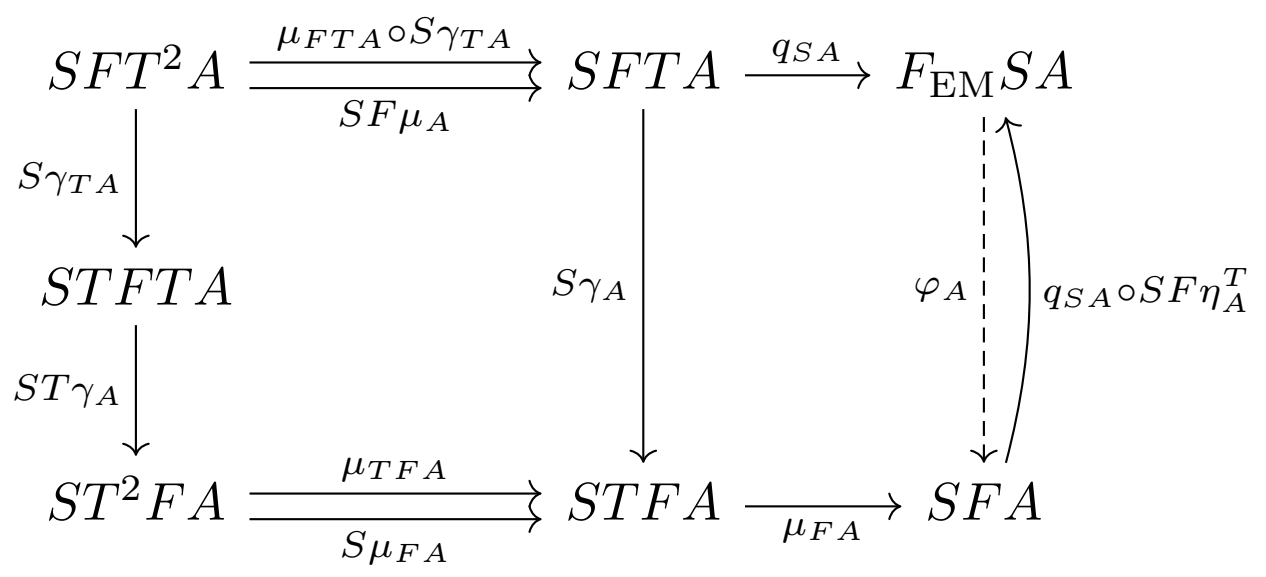
\[\begin{tikzcd}[row sep=2.5em]
	{\emfree\dyn\eff^2 A} &[3em] {\emfree\dyn\eff A} & {\emdyn\emfree A} \\
	{\emfree\eff\dyn\eff A} \\
	{\emfree\eff^2\dyn A} & {\emfree\eff\dyn A} & {\emfree\dyn A}
	\arrow["{\effmult[\dyn\eff A] \com \emfree\dlaw[\eff A]}", shift left, from=1-1, to=1-2]
	\arrow["{\emfree\dyn\effmult[A]}"', shift right, from=1-1, to=1-2]
	\arrow["{\emfree\dlaw[\eff A]}"', from=1-1, to=2-1]
	\arrow["{\q[\emfree A]}", from=1-2, to=1-3]
	\arrow["{\emfree\dlaw[A]}"', from=1-2, to=3-2]
	\arrow["{\emfreedyn[A]}"', dashed, from=1-3, to=3-3]
	\arrow["{\emfree\eff\dlaw[A]}"', from=2-1, to=3-1]
	\arrow["{\emfree\effmult[\dyn A]}"', shift right, from=3-1, to=3-2]
	\arrow["{\effmult[\eff\dyn A]}", shift left, from=3-1, to=3-2]
	\arrow["{\effmult[\dyn A]}"', from=3-2, to=3-3]
	\arrow["{\q[\emfree A] \com \emfree\dyn\effunit[A]}"', shift right, curve={height=6pt}, from=3-3, to=1-3]
\end{tikzcd}\]
The left square commutes serially:
the upper square commutes by the naturality of~$\effmult$,
and the lower square commutes by the multiplication axiom of distributive laws.
Since the bottom row is a fork, this implies that
the morphism~$\effmult[\dyn A] \com \emfree\dlaw[A]$ coequalises the upper parallel pair.
Hence, we get an induced morphism~$\emfreedyn[A] \colon \emdyn\emfree A \to \emfree\dyn A$
with $\emfreedyn[A] \com \q[\emfree A] = \effmult[\dyn A] \com \emfree\dlaw[A]$.

It remains to check that $\emfreedyn$~is natural and an isomorphism.
Naturality of~$\emfreedyn$ follows from the following calculation using that
$\q[\emfree A]$~is epic, where $f \colon A \to B$:
\begin{align*}
  \emfreedyn[B] \com \emdyn\emfree f \com \q[\emfree A]
   &= \emfreedyn[B] \com \q[\emfree B] \com \emfree\dyn\eff f
    = \effmult[\dyn B] \com \emfree\dlaw[B] \com \emfree\dyn\eff f \\
   &= \effmult[\dyn B] \com \emfree\eff\dyn f \com \emfree\dlaw[A]
    = \emfree\dyn f \com \effmult[\dyn A] \com \emfree\dlaw[A]
    = \emfree\dyn f \com \emfreedyn[A] \com \q[\emfree A].
\end{align*}
The inverse of~$\emfreedyn[A]$ is
$\inv{\emfreedyn[A]} = \q[\emfree A] \com \emfree\dyn\effunit[A]$.
We check that both roundtrips are the identity:
\[
  \emfreedyn[A] \com \inv{\emfreedyn[A]}
    = \emfreedyn[A] \com \q[\emfree A] \com \emfree\dyn\effunit[A]
    = \effmult[\dyn A] \com \emfree\dlaw[A] \com \emfree\dyn\effunit[A]
    = \effmult[\dyn A] \com \emfree\effunit[\dyn A]
    = \id,
\]
and, using the definition of~$\q[\emfree A]$ as the coequaliser of
$\effmult[\dyn\eff A] \com \emfree\dlaw[\eff A]$~and~$\emfree\dyn\effmult[A]$:
\begin{align*}
  \inv{\emfreedyn[A]} \com \emfreedyn[A] \com \q[\emfree A]
   &= \q[\emfree A] \com \emfree\dyn\effunit[A] \com \effmult[\dyn A] \com \emfree\dlaw[A]
    = \q[\emfree A] \com \effmult[\dyn\eff A] \com \emfree\eff\dyn\effunit[A] \com \emfree\dlaw[A] \\
   &= \q[\emfree A] \com \effmult[\dyn\eff A] \com \emfree\dlaw[\eff A] \com \emfree\dyn\eff\effunit[A]
    = \q[\emfree A] \com \emfree\dyn\effmult[A] \com \emfree\dyn\eff\effunit[A]
    = \q[\emfree A],
\end{align*}
which implies $\inv{\emfreedyn[A]} \com \emfreedyn[A] = \id$ since $\q[\emfree A]$~is epic.

\proofsubparagraph{Properties of~$\emdyn$.}

We check that $\emdyn$~satisfies
properties~\labelcref{prop:emdyn-pres-can,prop:emfreedyn-effmult} in the theorem.
\begin{enumerate}[(1)]
  \setcounter{enumi}{\numexpr\getrefnumber{prop:emfreedyn-effmult}-1\relax}
  \item
  We use that $\q[\emfree\eff A]$~is an epi and calculate as follows:
  \begin{align*}
    \emfreedyn[A] \com \emdyn\effmult[A] &\com \q[\emfree\eff A]
      = \emfreedyn[A] \com \q[\emfree A] \com \emfree\dyn\effmult[A]
      = \effmult[\dyn A] \com \emfree\dlaw[A] \com \emfree\dyn\effmult[A] \\
     &= \effmult[\dyn A] \com \emfree\effmult[\dyn A] \com \emfree\eff\dlaw[A] \com \emfree\dlaw[\eff A]
      = \effmult[\dyn A] \com \effmult[\eff\dyn A] \com \emfree\eff\dlaw[A] \com \emfree\dlaw[\eff A] \\
     &= \effmult[\dyn A] \com \emfree\dlaw[A] \com \effmult[\dyn\eff A] \com \emfree\dlaw[\eff A]
      = \effmult[\dyn A] \com \emfree\dlaw[A] \com \emfreedyn[\eff A] \com \q[\emfree\eff A].
  \end{align*}

  \setcounter{enumi}{\numexpr\getrefnumber{prop:emdyn-pres-can}-1\relax}
  \item
  Let~$\malg{A} = (A, a)$ be \an{\eff} $\eff$-algebra with canonical presentation~\(
    \begin{tikzcd}[cramped,column sep=1.8em]
      {\emfree\eff A} & {\emfree A} &[-0.4em] {\malg{A}}
      \arrow["{\emfree a}"', shift right, from=1-1, to=1-2]
      \arrow["{\effmult[A]}", shift left, from=1-1, to=1-2]
      \arrow["a", from=1-2, to=1-3]
    \end{tikzcd}
  \).
  We need to show that the top row in the following diagram is a coequaliser:
  \begin{equation}\label{eq:emdyn-pres-can}
    \begin{tikzcd}[column sep=3em,row sep=2.5em]
      {\emdyn\emfree\eff A} &[1em] {\emdyn\emfree A} & {\emdyn\malg{A}} \\
      {\emfree\dyn\eff A} & {\emfree\dyn A} & {\emdyn\malg{A}}
      \arrow["{\emdyn\effmult[A]}", shift left, from=1-1, to=1-2]
      \arrow["{\emdyn\emfree a}"', shift right, from=1-1, to=1-2]
      \arrow["{\emfreedyn[\eff A]}"', "\cong"{sloped}, from=1-1, to=2-1]
      \arrow["{\emdyn a}", from=1-2, to=1-3]
      \arrow["{\emfreedyn[A]}"', "\cong"{sloped}, from=1-2, to=2-2]
      \arrow[equals, nfold, from=1-3, to=2-3]
      \arrow["{\effmult[\dyn A] \com \emfree\dlaw[A]}", shift left, from=2-1, to=2-2]
      \arrow["{\emfree\dyn a}"', shift right, from=2-1, to=2-2]
      \arrow["{\q[\malg{A}]}"', from=2-2, to=2-3]
    \end{tikzcd}
  \end{equation}
  The left square commutes serially:
  the upper square commutes by property~\labelcref{prop:emfreedyn-effmult},
  and the lower square commutes by the naturality of~$\emfreedyn$.
  The right square commutes since
  \[
    \q[\malg{A}] \com \emfreedyn[A] \com \q[\emfree A]
      = \q[\malg{A}] \com \effmult[\dyn A] \com \emfree\dlaw[A]
      = \q[\malg{A}] \com \emfree\dyn a
      = \emdyn a \com \q[\emfree A]
  \]
  and $\q[\emfree A]$~is an epi.
  Since all vertical arrows are isomorphisms and the bottom row is a coequaliser by definition,
  it follows that the top row is also a coequaliser.
\end{enumerate}

\proofsubparagraph{Uniqueness.}

Let~$\emdyn' \colon \emcat \to \emcat$ be a functor
and~$\emfreedyn' \colon \emdyn'\emfree \ito \emfree\dyn$ a natural isomorphism
satisfying~\labelcref{prop:emdyn-pres-can,prop:emfreedyn-effmult}.
Then we get a diagram similar to~\labelcref{eq:emdyn-pres-can}:
\[\begin{tikzcd}[column sep=3em,row sep=2.5em]
	{\emdyn'\emfree\eff A} &[1em] {\emdyn'\emfree A} & {\emdyn'\malg{A}} \\
	{\emfree\dyn\eff A} & {\emfree\dyn A} & {\emdyn\malg{A}}
	\arrow["{\emdyn'\effmult[A]}", shift left, from=1-1, to=1-2]
	\arrow["{\emdyn'\emfree a}"', shift right, from=1-1, to=1-2]
	\arrow["{\emfreedyn[\eff A]'}"', "\cong"{sloped}, from=1-1, to=2-1]
	\arrow["{\emdyn'a}", from=1-2, to=1-3]
	\arrow["{\emfreedyn[A]'}"', "\cong"{sloped}, from=1-2, to=2-2]
	\arrow["{\iota_{\malg{A}}}"', "\cong"{sloped}, dashed, from=1-3, to=2-3]
	\arrow["{\effmult[\dyn A] \com \emfree\dlaw[A]}", shift left, from=2-1, to=2-2]
	\arrow["{\emfree\dyn a}"', shift right, from=2-1, to=2-2]
	\arrow["{\q[\malg{A}]}"', from=2-2, to=2-3]
\end{tikzcd}\]
Again, the left squares commute by assumption of
property~\labelcref{prop:emfreedyn-effmult} and naturality of~$\emfreedyn'$.
Moreover, the top row is a coequaliser by assumption of property~\labelcref{prop:emdyn-pres-can},
and the bottom row is a coequaliser by definition.
Since the left and middle vertical arrows are isomorphisms,
we get an induced isomorphism~$\iota_{\malg{A}} \colon \emdyn'\malg{A} \ito \emdyn\malg{A}$
with $\iota_{\malg{A}} \com \emdyn'a = \q[\malg{A}] \com \emfreedyn[A]'$.

To see that $\iota$~is natural, let~$\malg{B} = (B, b)$ be \an{\eff} $\eff$-algebra
and~$f \colon \malg{A} \to \malg{B}$ \an{\eff} $\eff$-algebra map.
We use that $\emdyn'a$~is an epimorphism (since it is a coequaliser):
\begin{align*}
  \iota_{\malg{B}} \com \emdyn'f \com \emdyn'a
   &= \iota_{\malg{B}} \com \emdyn'b \com \emdyn'\emfree f
    = \q[\malg{B}] \com \emfreedyn[B]' \com \emdyn'\emfree f \\
   &= \q[\malg{B}] \com \emfree\dyn f \com \emfreedyn[A]'
    = \emdyn f \com \q[\malg{A}] \com \emfreedyn[A]'
    = \emdyn f \com \iota_{\malg{A}} \com \emdyn'a.
\end{align*}
Similarly, we have $\emfreedyn[A] \com \iota_{\emfree A} = \emfreedyn[A]'$
since $\emdyn'\effmult[A]$~is an epimorphism and
\[
  \emfreedyn[A] \com \iota_{\emfree A} \com \emdyn'\effmult[A]
    = \emfreedyn[A] \com \q[\emfree A] \com \emfreedyn[\eff A]'
    = \effmult[\dyn A] \com \emfree\dlaw[A] \com \emfreedyn[\eff A]'
    = \emfreedyn[A]' \com \emdyn'\effmult[A].
\]

Finally, if $\iota' \colon \emdyn' \to \emdyn$~is a natural transformation
with $\emfreedyn \com \iota'\emfree = \emfreedyn'$, then the diagram below commutes
(where the triangle was shown to be commutative in diagram~\labelcref{eq:emdyn-pres-can}):
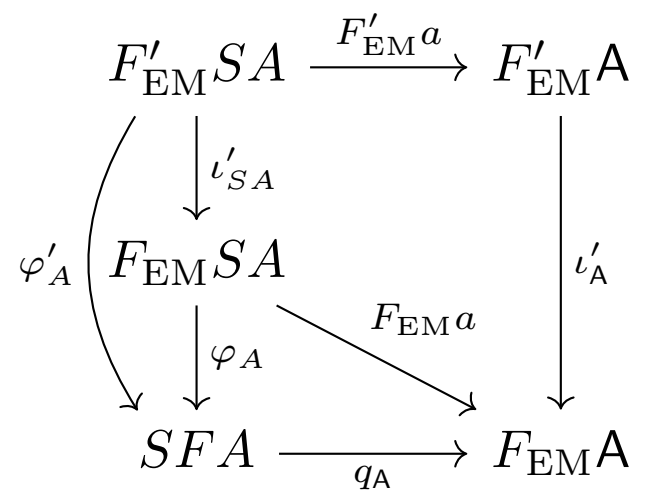
\[\begin{tikzcd}[column sep=3em,row sep=2em]
	{\emdyn'\emfree A} & {\emdyn'\malg{A}} \\
	{\emdyn\emfree A} \\
	{\emfree\dyn A} & {\emdyn\malg{A}}
	\arrow["{\emdyn'a}", from=1-1, to=1-2]
	\arrow["{\iota'_{\emfree A}}", from=1-1, to=2-1]
	\arrow["{\emfreedyn[A]'}"', shift right=3, curve={height=12pt}, from=1-1, to=3-1]
	\arrow["{\iota'_{\malg{A}}}", from=1-2, to=3-2]
	\arrow["{\emfreedyn[A]}", from=2-1, to=3-1]
	\arrow["{\emdyn a}"{pos=0.4}, from=2-1, to=3-2]
	\arrow["{\q[\malg{A}]}"', from=3-1, to=3-2]
\end{tikzcd}\]
Since $\emdyn'a$~is an epimorphism, we conclude $\iota_{\malg{A}} = \iota'_{\malg{A}}$,
showing the uniqueness of~$\iota$. \qed

\endgroup

\section{Proof of~\texorpdfstring{\cref{lem:klcompdyn}}{\autoref{lem:klcompdyn}}}

The map~$\klcompdyn[A]$ is defined to be identical to~$\emfreedyn[A]$, which is an isomorphism.
To see naturality, let~$f \colon A \to \eff B$ be a Kleisli map~$A \klto B$.
By~\cite[Theorem~2.2]{Mul:94}, the map~$\kldyn f \colon \dyn A \klto \dyn B$
is defined as~$\dlaw[B] \com \dyn f$.
Using this, the naturality of~$\emfreedyn$, and diagram~\labelcref{eq:emfreedyn-effmult},
we calculate as~follows:
\begin{align*}
  \klcompdyn[B] \com \emdyn\klcomp f
   &= \emfreedyn[B] \com \emdyn\effmult[B] \com \emdyn\emfree f
    = \effmult[\dyn B] \com \emfree\dlaw[B] \com \emfreedyn[\eff B] \com \emdyn\emfree f \\
   &= \effmult[\dyn B] \com \emfree\dlaw[B] \com \emfree\dyn f \com \emfreedyn[A]
    = \effmult[\dyn B] \com \emfree\kldyn f \com \emfreedyn[A]
    = \klcomp\kldyn f \com \klcompdyn[A]. \qed
\end{align*}

\section{Proof of~\texorpdfstring{\cref{thm:em-sat-ass}}{\autoref{thm:em-sat-ass}}}

In this section,
we work under~\cref{ass:effectful-base,ass:monad,ass:dist-law,ass:dlaw-effmf-lstr}
and verify the conditions in~\cref{ass:w-method}
for $\bcat \leteq \emcat$~and~$\dyn \leteq \emdyn$.
We begin with the base category.

\begin{lemma}\label{lem:em-sat-ass-bcat}
  The category~$\emcat$ is symmetric monoidal closed and cocomplete.
\end{lemma}
\begin{proof}
  By~\cref{ass:effectful-base}, $\bcat$~is symmetric monoidal closed and has equalisers.
  Furthermore, by~\cref{ass:monad}, $\eff$~is commutative (hence monoidal)
  and $\emcat$~has reflexive coequalisers (hence tensors~\cite[Section~2.2]{Sea:13}).
  Thus, applying~\cite[Corollary~2.5.7]{Sea:13}, we obtain that $\emcat$~is monoidal.
  In fact, examining the proof of~\cite[Corollary~2.5.7]{Sea:13},
  we conclude at once that $\emcat$~is symmetric monoidal closed.
  Finally, since $\bcat$~is cocomplete by~\cref{ass:effectful-base},
  we deduce by~\cite[Corollary~2]{Lin:69} that $\emcat$~is cocomplete.
\end{proof}

Before proceeding with the other conditions, we introduce some notation and results
in connection with the monoidal closed structure of~$\emcat$
that will be used in the rest of this proof and also in later proofs.
We write~$\emprod$ for the monoidal product in~$\emcat$.
The tensor unit in~$\emcat$ is~$\emfree\tunit$,
and we write~$\emrunit$ for the right unitor in~$\emcat$.

The tensor product~$\emprod$ in~$\emcat$ is constructed as a certain reflexive coequaliser,
similarly to the construction of~$\emdyn$.
As shown in~\cite[Proposition~2.3.4]{Sea:13}, this tensor product represents \emph{bimorphisms}.
Recall~\cite[Section~2.1]{Sea:13} that for $\eff$-algebras $(A,a)$, $(B,b)$, and~$(C,c)$,
a \emph{bimorphism}~$(A,a), (B,b) \to (C,c)$ is a
morphism~$f \colon A \tprod B \to C$ in~$\bcat$ such that the diagram
\[\begin{tikzcd}
	{\eff A \tprod \eff B} & {\eff(A \tprod B)} & {\eff C} \\
	{A \tprod B} && C
	\arrow["{\effmf[A,B]}", from=1-1, to=1-2]
	\arrow["{a \tprod b}"', from=1-1, to=2-1]
	\arrow["{\eff f}", from=1-2, to=1-3]
	\arrow["c", from=1-3, to=2-3]
	\arrow["f"', from=2-1, to=2-3]
\end{tikzcd}\]
commutes.
Then for any pair of algebras~$\malg{A},\malg{B}$, we have a
bimorphism~$\emforgetmf[\malg{A},\malg{B}] \colon \malg{A}, \malg{B} \to \malg{A} \emprod \malg{B}$
satisfying the following universal property:
for all algebras~$\malg{C}$ and bimorphisms~$f \colon \malg{A}, \malg{B} \to \malg{C}$,
there exists a unique algebra morphism~$\bar{f} \colon \malg{A} \emprod \malg{B} \to \malg{C}$
such that $\emforget\bar{f} \com \emforgetmf[\malg{A},\malg{B}] = f$.
Furthermore, the morphisms~\(
  \emforgetmf[\malg{A},\malg{B}] \colon
    \emforget\malg{A} \tprod \emforget\malg{B} \to \emforget(\malg{A} \emprod \malg{B})
\)
form a natural transformation.

\begin{lemma}\label{lem:emfreemf}
  We have isomorphisms~$\emfreemf[A,B] \colon \emfree A \emprod \emfree B \to \emfree(A \tprod B)$,
  natural in~$A,B \in \bcat$, making the following diagram commute:
  \[\begin{tikzcd}[column sep=3.5em]
    {\eff A \tprod \eff B} && {\eff(A \tprod B)} \\
    {\emforget\emfree A \tprod \emforget\emfree B} & {\emforget(\emfree A \emprod \emfree B)} & {\emforget\emfree(A \tprod B)}
    \arrow["{\effmf[A,B]}", from=1-1, to=1-3]
    \arrow[equals, nfold, from=1-1, to=2-1]
    \arrow["{\emforgetmf[\emfree A,\emfree B]}"', from=2-1, to=2-2]
    \arrow["{\emforget\emfreemf[A,B]}"', from=2-2, to=2-3]
    \arrow[equals, nfold, from=2-3, to=1-3]
  \end{tikzcd}\]
\end{lemma}
\begin{proof}
  The morphism~$\emfreemf[A,B]$ is obtained by applying the universal property of~$\emprod$
  to~$\effmf[A,B]$, which is a bimorphism~$\emfree A, \emfree B \to \emfree(A \tprod B)$
  by the multiplication axiom for~$\effmf$.
  Then the diagram above commutes by definition.
  The naturality of~$\emfreemf$ follows from uniqueness for~$\emprod$:
  \begin{multline*}
    \emforget\emfreemf[A',B'] \com \emforget(\emfree f \emprod \emfree g) \com \emforgetmf[\emfree A,\emfree B]
      = \emforget\emfreemf[A',B'] \com \emforgetmf[\emfree A',\emfree B'] \com (\emforget\emfree f \tprod \emforget\emfree g) \\
      = \effmf[A',B'] \com (\eff f \tprod \eff g)
      = \eff(f \tprod g) \com \effmf[A,B]
      = \emforget\emfree(f \tprod g) \com \emforget\emfreemf[A,B] \com \emforgetmf[\emfree A,\emfree B].
  \end{multline*}

  \noindent The inverse of~$\emfreemf[A,B]$ is given by the free $\eff$-extension of the composite
  \[
    A \tprod B
      \xto{\effunit[A] \tprod \effunit[B]} \eff A \tprod \eff B
      = \emforget\emfree A \tprod \emforget\emfree B
      \xto{\emforgetmf[\emfree A,\emfree B]} \emforget(\emfree A \emprod \emfree B).
  \]
  We check that the two roundtrips are the identity.
  We have
  \[
    \emforget\emfreemf[A,B] \com \emforget\inv{\emfreemf[A,B]} \com \effunit[A \tprod B]
      = \emforget\emfreemf[A,B] \com \emforgetmf[\emfree A,\emfree B] \com (\effunit[A] \tprod \effunit[B])
      = \effmf[A,B] \com (\effunit[A] \tprod \effunit[B])
      = \effunit[A \tprod B],
  \]
  which implies $\emfreemf[A,B] \com \inv{\emfreemf[A,B]} = \id$
  using the adjunction~$\emfree \dashv \emforget$.
  Furthermore, writing~$\algmap$ for the algebra structure of~$\emfree A \emprod \emfree B$
  and using that $\emforgetmf[\emfree A,\emfree B]$~is a bimorphism, we have
  \begin{align*}
    \emforget\inv{\emfreemf[A,B]} \com \emforget\emfreemf[A,B] \com \emforgetmf[\emfree A,\emfree B]
      &= \algmap \com \eff\emforgetmf[\emfree A,\emfree B] \com \eff(\effunit[A] \tprod \effunit[B]) \com \effmf[A,B] \\
      &= \algmap \com \eff\emforgetmf[\emfree A,\emfree B] \com \effmf[\eff A,\eff B] \com (\eff\effunit[A] \tprod \eff\effunit[B]) \\
      &= \emforgetmf[\emfree A,\emfree B] \com (\effmult[A] \tprod \effmult[B]) \com (\eff\effunit[A] \tprod \eff\effunit[B])
       = \emforgetmf[\emfree A,\emfree B],
  \end{align*}
  implying $\inv{\emfreemf[A,B]} \com \emfreemf[A,B] = \id$
  by the uniqueness part of the universal property of~$\emprod$.
\end{proof}

\begin{lemma}\label{lem:emmfprod-effmult}
  The following diagram commutes:
  \[\begin{tikzcd}[column sep=3.5em]
    {\emfree\eff A \emprod \emfree\eff B} & {\emfree(\eff A \tprod \eff B)} & {\emfree\eff(A \tprod B)} \\
    {\emfree A \emprod \emfree B} && {\emfree(A \tprod B)}
    \arrow["{\emfreemf[\eff A,\eff B]}", from=1-1, to=1-2]
    \arrow["{\effmult[A] \emprod \effmult[B]}"', from=1-1, to=2-1]
    \arrow["{\emfree\effmf[A,B]}", from=1-2, to=1-3]
    \arrow["{\effmult[A \tprod B]}", from=1-3, to=2-3]
    \arrow["{\emfreemf[A,B]}"', from=2-1, to=2-3]
  \end{tikzcd}\]
\end{lemma}
\begin{proof}
  It suffices to show that the two sides become equal after precomposition with~$\emforgetmf$.
  This holds by the naturality of~$\emforgetmf$, \cref{lem:emfreemf},
  and the multiplication axiom for~$\effmf$:
  \begin{align*}
    \emforget\emfreemf[A,B] &\com \emforget(\effmult[A] \emprod \effmult[B])
      \com \emforgetmf[\emfree\eff A,\emfree\eff B]
        = \emforget\emfreemf[A,B] \com \emforgetmf[\emfree A,\emfree B]
            \com (\effmult[A] \tprod \effmult[B])
        = \effmf[A,B] \com (\effmult[A] \tprod \effmult[B]) \\
       &= \effmult[A \tprod B] \com \eff\effmf[A,B] \com \effmf[\eff A,\eff B]
        = \effmult[A \tprod B] \com \emforget\emfree\effmf[A,B]
            \com \emforget\emfreemf[\eff A,\eff B] \com \emforgetmf[\emfree\eff A,\emfree\eff B].
        \qedhere
  \end{align*}
\end{proof}

\begingroup
\newcommand{\enr}[1][]{\chi_{#1}}

As for the closed structure of~$\emcat$,
we write~$\emhom{-}{-}$ for the internal hom in~$\emcat$.
Recall from~\cite{Koc:71} that for algebras $(A,a)$~and~$(B,b)$,
(the carrier of) the algebra~$\emhom{(A,a)}{(B,b)}$ is given by
the equaliser of~$\inthom{A}{B} \xto{\inthom{a}{\id}} \inthom{\eff A}{B}$ and~\(
  \inthom{A}{B} \xto{\enr[A,B]}
  \inthom{\eff A}{\eff B} \xto{\inthom{\id}{b}}
  \inthom{\eff A}{B}
\),
where $\enr$~is the enrichment of~$\eff$, which can be defined
for any commutative monad on a monoidal closed category (see \eg \cite{Koc:72}).
Hence, we have a family~\(
  \emhominc[\malg{A},\malg{B}] \colon
    \emforget\emhom{\malg{A}}{\malg{B}} \to \inthom{\emforget\malg{A}}{\emforget\malg{B}}
\)
for algebras $\malg{A}$~and~$\malg{B}$, and this family is natural in $\malg{A}$~and~$\malg{B}$.
We write~$\emcurry{f} \colon \malg{A} \to \emhom{\malg{B}}{\malg{C}}$ for the
currying of \an{\eff} $\eff$-algebra map~$f \colon \malg{A} \emprod \malg{B} \to \malg{C}$.
The evaluation map~\(
  \emev[\malg{A},\malg{B}] \colon \emhom{\malg{A}}{\malg{B}} \emprod \malg{A} \to \malg{B}
\)
in~$\emcat$ is defined using the universal property of tensor products as the unique morphism
making the following diagram commute, where $\intev$~is the evaluation map in~$\bcat$:
\begin{equation}\label{eq:emev-def}
  \begin{tikzcd}[column sep=3.5em]
    {\emforget\emhom{\malg{A}}{\malg{B}} \tprod \emforget\malg{A}} & {\inthom{\emforget\malg{A}}{\emforget\malg{B}} \tprod \emforget\malg{A}} \\
    {\emforget(\emhom{\malg{A}}{\malg{B}} \emprod \malg{A})} & {\emforget\malg{B}}
    \arrow["{\emhominc[\malg{A},\malg{B}] \tprod \id}", from=1-1, to=1-2]
    \arrow["{\emforgetmf[\emhom{\malg{A}}{\malg{B}},\malg{A}]}"', from=1-1, to=2-1]
    \arrow["{\intev[\emforget\malg{A},\emforget\malg{B}]}", from=1-2, to=2-2]
    \arrow["{\emforget\emev[\malg{A},\malg{B}]}"', from=2-1, to=2-2]
  \end{tikzcd}
\end{equation}

\begin{proposition}
  There is a natural family of isomorphisms~\(
    \freehomiso[A,\malg{B}] \colon
      \emforget\emhom{\emfree A}{\malg{B}} \ito \inthom{A}{\emforget\malg{B}}
  \)
  for~$A \in \bcat$ and~$\malg{B} \in \emcat$,
  where $\freehomiso[A,\malg{B}]$~is given by the composite
  \[
    \emforget\emhom{\emfree A}{\malg{B}} \xto{\emhominc[\emfree A,\malg{B}]}
    \inthom{\emforget\emfree A}{\emforget\malg{B}} =
    \inthom{\eff A}{\emforget\malg{B}} \xto{\inthom{\effunit[A]}{\id}}
    \inthom{A}{\emforget\malg{B}}.
  \]
\end{proposition}
The family~$\freehomiso[A,\malg{B}]$ internalises
the natural bijection~$\Hom[\emcat]{\emfree A}{\malg{B}} \cong \Hom[\bcat]{A}{\emforget\malg{B}}$
arising from the adjunction~$\emfree \dashv \emforget$.
\begin{proof}
  The naturality of~$\freehomiso$ is clear since it is a composite of natural transformations.
  Writing~$\malg{B} = (B,b)$,
  the inverse of~$\freehomiso[A,\malg{B}]$ is obtained by factoring the composite
  \[
    \inthom{A}{\emforget\malg{B}} =
    \inthom{A}{B} \xto{\enr[A,B]}
    \inthom{\eff A}{\eff B} \xto{\inthom{\id}{b}}
    \inthom{\eff A}{B} =
    \inthom{\emforget\emfree A}{\emforget\malg{B}}
  \]
  through the equaliser~$\emhominc[\emfree A,\malg{B}]$
  of $\inthom{\effmult[A]}{\id}$~and~$\inthom{\id}{b} \com \enr[\eff A,B]$.
  This is possible since
  \begin{multline*}
    \inthom{\id}{b} \com \enr[\eff A,B] \com \inthom{\id}{b} \com \enr[A,B]
                      = \inthom{\id}{b} \com \inthom{\id}{\eff b} \com \enr[\eff A,\eff B] \com \enr[A,B] \\
                      = \inthom{\id}{b} \com \inthom{\id}{\effmult[B]} \com \enr[\eff A,\eff B] \com \enr[A,B]
      \overset{(\ast)}= \inthom{\id}{b} \com \inthom{\effmult[A]}{\id} \com \enr[A,B]
                      = \inthom{\effmult[A]}{\id} \com \inthom{\id}{b} \com \enr[A,B],
  \end{multline*}
  where in~$(\ast)$ we used that $\effmult$~is an enriched natural transformation.
  We check that this indeed defines an inverse.
  Since $\effunit$~is an enriched natural transformation, we have
  \begin{align*}
    \freehomiso[A,\malg{B}] \com \inv{\freehomiso[A,\malg{B}]}
      &= \inthom{\effunit[A]}{\id} \com \emhominc[\emfree A,\malg{B}] \com \inv{\freehomiso[A,\malg{B}]}
       = \inthom{\effunit[A]}{\id} \com \inthom{\id}{b} \com \enr[A,B] \\
      &= \inthom{\id}{b} \com \inthom{\effunit[A]}{\id} \com \enr[A,B]
       = \inthom{\id}{b} \com \inthom{\id}{\effunit[B]}
       = \id.
  \end{align*}
  Finally, using the definition of~$\emhominc[\emfree A,\malg{B}]$ as an equaliser, we have
  \begin{multline*}
    \emhominc[\emfree A,\malg{B}] \com \inv{\freehomiso[A,\malg{B}]} \com \freehomiso[A,\malg{B}]
      = \inthom{\id}{b} \com \enr[A,B] \com \inthom{\effunit[A]}{\id} \com \emhominc[\emfree A,\malg{B}]
      = \inthom{\id}{b} \com \inthom{\eff\effunit[A]}{\id} \com \enr[\eff A,B] \com \emhominc[\emfree A,\malg{B}] \\
      = \inthom{\eff\effunit[A]}{\id} \com \inthom{\id}{b} \com \enr[\eff A,B] \com \emhominc[\emfree A,\malg{B}]
      = \inthom{\eff\effunit[A]}{\id} \com \inthom{\effmult[A]}{\id} \com \emhominc[\emfree A,\malg{B}]
      = \emhominc[\emfree A,\malg{B}],
  \end{multline*}
  which implies $\inv{\freehomiso[A,\malg{B}]} \com \freehomiso[A,\malg{B}] = \id$
  since $\emhominc[\emfree A,\malg{B}]$~is monic.
\end{proof}

\endgroup

In the remaining proofs, we will frequently omit the forgetful functor~$\emforget$.
Thus, if $\malg{A} = (A, a)$~and~$\malg{B} = (B, b)$ are $\eff$-algebras, then
we will write \eg $\emforgetmf[\malg{A},\malg{B}] \colon A \tprod B \to \malg{A} \emprod \malg{B}$
and~$\emhominc[\malg{A},\malg{B}] \colon \emhom{\malg{A}}{\malg{B}} \to \inthom{A}{B}$.

We now move on to showing that $\emdyn$~admits a strength.
\begin{lemma}\label{lem:emlstr}
  There is a family of maps~\(
    \emlstr[\malg{A},\malg{B}] \colon
      \malg{A} \emprod \emdyn\malg{B} \to \emdyn(\malg{A} \emprod \malg{B})
  \),
  natural in $\malg{A}$~and~$\malg{B}$, that makes the following diagram commute
  (where $\emfreedyn$~and~$\emfreemf$~are the isomorphisms
  from~\cref{thm:dlaw-emdyn-equiv,lem:emfreemf}, respectively):
  \begin{equation}\label{eq:emfreedyn-emfreemf-emlstr}
    \begin{tikzcd}[column sep=4em]
      {\emfree A \emprod \emdyn\emfree B} &[1.2em] {\emdyn(\emfree A \emprod \emfree B)} & {\emdyn\emfree(A \tprod B)} \\
      {\emfree A \emprod \emfree\dyn B} & {\emfree(A \tprod \dyn B)} & {\emfree\dyn(A \tprod B)}
      \arrow["{\emlstr[\emfree A,\emfree B]}", from=1-1, to=1-2]
      \arrow["{\id[\emfree A] \emprod \emfreedyn[B]}"', from=1-1, to=2-1]
      \arrow["{\emdyn\emfreemf[A,B]}", from=1-2, to=1-3]
      \arrow["{\emfreedyn[A \tprod B]}", from=1-3, to=2-3]
      \arrow["{\emfreemf[A,\dyn B]}"', from=2-1, to=2-2]
      \arrow["{\emfree\lstr[A,B]}"', from=2-2, to=2-3]
    \end{tikzcd}
  \end{equation}
\end{lemma}
The diagram above is reminiscent of the one in~\cref{ass:dlaw-effmf-lstr}.
It expresses that on free algebras,
$\emlstr$~coincides with~$\emfree\lstr$ (up to $\emfreedyn$~and~$\emfreemf$).
\begin{proof}
\newcommand{\emfreelstr}[1][]{\chi_{#1}}
  To construct~$\emlstr[\malg{A},\malg{B}]$, we use the observation that
  the domain of~$\emlstr[\malg{A},\malg{B}]$ can be described as a coequaliser.
  To see this, note that since
  $\emcat$~is a symmetric monoidal closed category by~\cref{lem:em-sat-ass-bcat},
  all functors of the form $\malg{A} \emprod (-)$~or~$(-) \emprod \malg{B}$
  for~$\malg{A}, \malg{B} \in \emcat$ preserve reflexive coequalisers.
  Hence, by the dual of~\cite[Corollary~1.2.12]{Joh:02},
  the tensor product~$\emprod$ preserves reflexive coequalisers in both variables together
  (meaning that it maps pairs of reflexive coequaliser diagrams to coequaliser diagrams).
  Furthermore, we know by~\cref{thm:dlaw-emdyn-equiv} that
  $\emdyn$~preserves canonical presentations.
  Applying these facts to the
  canonical presentations of $\malg{A} = (A, a)$~and~$\malg{B} = (B, b)$,
  we get that the left column in the diagram below is a coequaliser:
  \begin{equation}\label{eq:emlstr-def}
    \begin{tikzcd}[column sep=4em]
      {\emfree\eff A \emprod \emdyn\emfree\eff B} & {\emdyn(\emfree\eff A \emprod \emfree\eff B)} \\
      {\emfree A \emprod \emdyn\emfree B} & {\emdyn(\emfree A \emprod \emfree B)} \\
      {\malg{A} \emprod \emdyn\malg{B}} & {\emdyn(\malg{A} \emprod \malg{B})}
      \arrow["{\emfreelstr[\eff A,\eff B]}", from=1-1, to=1-2]
      \arrow["{\emfree a \emprod \emdyn\emfree b}", shift left, from=1-1, to=2-1]
      \arrow["{\effmult[A] \emprod \emdyn\effmult[B]}"', shift right, from=1-1, to=2-1]
      \arrow["{\emdyn(\emfree a \emprod \emfree b)}", shift left, from=1-2, to=2-2]
      \arrow["{\emdyn(\effmult[A] \emprod \effmult[B])}"', shift right, from=1-2, to=2-2]
      \arrow["{\emfreelstr[A,B]}"', from=2-1, to=2-2]
      \arrow["{a \emprod \emdyn b}"', from=2-1, to=3-1]
      \arrow["{\emdyn(a \emprod b)}", from=2-2, to=3-2]
      \arrow["{\emlstr[\malg{A},\malg{B}]}"', dashed, from=3-1, to=3-2]
    \end{tikzcd}
  \end{equation}
  Hence, if we can construct a family~\(
    \emfreelstr[A,B] \colon \emfree A \emprod \emdyn\emfree B \to \emdyn(\emfree A \emprod \emfree B)
  \)
  making the top square in the diagram above serially commutative,
  then we get an induced morphism~$\emlstr[\malg{A},\malg{B}]$
  making the lower square commute.

  To make sure that the resulting map is natural,
  $\emfreelstr[A,B]$~should be equal to~$\emlstr[\emfree A,\emfree B]$,
  the value of which is forced by diagram~\labelcref{eq:emfreedyn-emfreemf-emlstr}.
  Thus, we define~$\emfreelstr[A,B]$ as the composite
  \[
    \emfree A \emprod \emdyn\emfree B \cong
    \emfree(A \tprod \dyn B) \xto{\emfree\lstr[A,B]}
    \emfree\dyn(A \tprod B) \cong
    \emdyn(\emfree A \emprod \emfree B),
  \]
  where the first isomorphism is the composite
  \[
    \emfree A \emprod \emdyn\emfree B \xto[\cong]{\id \emprod \emfreedyn[B]}
    \emfree A \emprod \emfree\dyn B \xto[\cong]{\emfreemf[A,\dyn B]}
    \emfree(A \tprod \dyn B),
  \]
  and the second isomorphism is (the inverse of) the composite
  \[
    \emdyn(\emfree A \emprod \emfree B) \xto[\cong]{\emdyn\emfreemf[A,B]}
    \emdyn\emfree(A \tprod B) \xto[\cong]{\emfreedyn[A \tprod B]}
    \emfree\dyn(A \tprod B).
  \]
  Since $\emfreelstr$~is a composite of natural transformations, it is itself natural.
  This implies that the upper square in~\labelcref{eq:emlstr-def}
  involving the algebra maps $a$~and~$b$ is commutative.
  The other upper square is obtained by composing the following three diagrams horizontally:
  \[\begin{tikzcd}
    {\emfree\eff A \emprod \emdyn\emfree\eff B} & {\emfree\eff A \emprod \emfree\dyn\eff B} & {\emfree(\eff A \tprod \dyn\eff B)} \\
    {\emfree A \emprod \emdyn\emfree B} & {\emfree A \emprod \emfree\dyn B} & {\emfree(A \tprod \dyn B)}
    \arrow["{\id \emprod \emfreedyn}", "\cong"', from=1-1, to=1-2]
    \arrow["{\effmult \emprod \emdyn\effmult}"', from=1-1, to=2-1]
    \arrow["\emfreemf", "\cong"', from=1-2, to=1-3]
    \arrow["{\effmult \com \emfree\effmf \com \emfree(\id \tprod \dlaw)}", from=1-3, to=2-3]
    \arrow["{\id \emprod \emfreedyn}"', "\cong", from=2-1, to=2-2]
    \arrow["\emfreemf"', "\cong", from=2-2, to=2-3]
  \end{tikzcd}\]
  \[\begin{tikzcd}
    {\emfree(\eff A \tprod \dyn\eff B)} & {\emfree\dyn(\eff A \tprod \eff B)} \\
    {\emfree(A \tprod \dyn B)} & {\emfree\dyn(A \tprod B)}
    \arrow["{\emfree\lstr}", from=1-1, to=1-2]
    \arrow["{\effmult \com \emfree\effmf \com \emfree(\id \tprod \dlaw)}"', from=1-1, to=2-1]
    \arrow["{\effmult \com \emfree\dlaw \com \emfree\dyn\effmf}", from=1-2, to=2-2]
    \arrow["{\emfree\lstr}"', from=2-1, to=2-2]
  \end{tikzcd}\]
  \[\begin{tikzcd}[column sep=3em]
    {\emfree\dyn(\eff A \tprod \eff B)} & {\emdyn\emfree(\eff A \tprod \eff B)} & {\emdyn(\emfree\eff A \emprod \emfree\eff B)} \\
    {\emfree\dyn(A \tprod B)} & {\emdyn\emfree(A \tprod B)} & {\emdyn(\emfree A \emprod \emfree B)}
    \arrow["{\effmult \com \emfree\dlaw \com \emfree\dyn\effmf}"', from=1-1, to=2-1]
    \arrow["\emfreedyn"', "\cong", from=1-2, to=1-1]
    \arrow["{\emdyn\emfreemf}"', "\cong", from=1-3, to=1-2]
    \arrow["{\emdyn(\effmult \emprod \effmult)}", from=1-3, to=2-3]
    \arrow["\emfreedyn", "\cong"', from=2-2, to=2-1]
    \arrow["{\emdyn\emfreemf}", "\cong"', from=2-3, to=2-2]
  \end{tikzcd}\]

  We now verify the commutativity of each of the three diagrams above in order.
  \begin{enumerate}[(1)]
    \item
    The first diagram commutes by diagram~\labelcref{eq:emfreedyn-effmult},
    \cref{lem:emmfprod-effmult}, and the naturality of~$\emfreemf$:
    \[\begin{tikzcd}[column sep=3em]
      {\emfree\eff A \emprod \emdyn\emfree\eff B} & {\emfree\eff A \emprod \emfree\dyn\eff B} & {\emfree(\eff A \tprod \dyn\eff B)} \\
      & {\emfree\eff A \emprod \emfree\eff\dyn B} & {\emfree(\eff A \tprod \eff\dyn B)} \\
      && {\emfree\eff(A \tprod \dyn B)} \\
      {\emfree A \emprod \emdyn\emfree B} & {\emfree A \emprod \emfree\dyn B} & {\emfree(A \tprod \dyn B)}
      \arrow["{\id \emprod \emfreedyn}", from=1-1, to=1-2]
      \arrow["{\effmult \emprod \emdyn\effmult}"', from=1-1, to=4-1]
      \arrow["\emfreemf", from=1-2, to=1-3]
      \arrow["{\id \emprod \emfree\dlaw}"', from=1-2, to=2-2]
      \arrow["{\emfree(\id \tprod \dlaw)}", from=1-3, to=2-3]
      \arrow["\emfreemf", from=2-2, to=2-3]
      \arrow["{\effmult \emprod \effmult}"', from=2-2, to=4-2]
      \arrow["{\emfree\effmf}", from=2-3, to=3-3]
      \arrow["\effmult", from=3-3, to=4-3]
      \arrow["{\id \emprod \emfreedyn}"', from=4-1, to=4-2]
      \arrow["\emfreemf"', from=4-2, to=4-3]
    \end{tikzcd}\]

    \item
    The second diagram commutes by~\cref{ass:dlaw-effmf-lstr} and the naturality of~$\effmult$:
    \[\begin{tikzcd}[column sep=3em]
      {\emfree(\eff A \tprod \dyn\eff B)} & {\emfree\dyn(\eff A \tprod \eff B)} \\
      {\emfree(\eff A \tprod \eff\dyn B)} & {\emfree\dyn\eff(A \tprod B)} \\
      {\emfree\eff(A \tprod \dyn B)} & {\emfree\eff\dyn(A \tprod B)} \\
      {\emfree(A \tprod \dyn B)} & {\emfree\dyn(A \tprod B)}
      \arrow["{\emfree\lstr}", from=1-1, to=1-2]
      \arrow["{\emfree(\id \tprod \dlaw)}"', from=1-1, to=2-1]
      \arrow["{\emfree\dyn\effmf}", from=1-2, to=2-2]
      \arrow["{\emfree\effmf}"', from=2-1, to=3-1]
      \arrow["{\emfree\dlaw}", from=2-2, to=3-2]
      \arrow["{\emfree\eff\lstr}", from=3-1, to=3-2]
      \arrow["\effmult"', from=3-1, to=4-1]
      \arrow["\effmult", from=3-2, to=4-2]
      \arrow["{\emfree\lstr}"', from=4-1, to=4-2]
    \end{tikzcd}\]

    \item
    The third diagram commutes by~\cref{lem:emmfprod-effmult},
    diagram~\labelcref{eq:emfreedyn-effmult}, and the naturality of~$\emfreedyn$:
    \[\begin{tikzcd}[column sep=3em]
      {\emdyn(\emfree\eff A \emprod \emfree\eff B)} & {\emdyn\emfree(\eff A \tprod \eff B)} & {\emfree\dyn(\eff A \tprod \eff B)} \\
      & {\emdyn\emfree\eff(A \tprod B)} & {\emfree\dyn\eff(A \tprod B)} \\
      && {\emfree\eff\dyn(A \tprod B)} \\
      {\emdyn(\emfree A \emprod \emfree B)} & {\emdyn\emfree(A \tprod B)} & {\emfree\dyn(A \tprod B)}
      \arrow["{\emdyn\emfreemf}", from=1-1, to=1-2]
      \arrow["{\emdyn(\effmult \emprod \effmult)}"', from=1-1, to=4-1]
      \arrow["\emfreedyn", from=1-2, to=1-3]
      \arrow["{\emdyn\emfree\effmf}"', from=1-2, to=2-2]
      \arrow["{\emfree\dyn\effmf}", from=1-3, to=2-3]
      \arrow["\emfreedyn", from=2-2, to=2-3]
      \arrow["{\emdyn\effmult}"', from=2-2, to=4-2]
      \arrow["{\emfree\dlaw}", from=2-3, to=3-3]
      \arrow["\effmult", from=3-3, to=4-3]
      \arrow["{\emdyn\emfreemf}"', from=4-1, to=4-2]
      \arrow["\emfreedyn"', from=4-2, to=4-3]
    \end{tikzcd}\]
  \end{enumerate}

  \noindent It remains to show that $\emlstr$~is natural
  and that diagram~\labelcref{eq:emfreedyn-emfreemf-emlstr} commutes.
  The naturality of~$\emlstr$ is a consequence of the naturality of~$\emfreelstr$.
  The commutativity of diagram~\labelcref{eq:emfreedyn-emfreemf-emlstr} can be shown
  using the definition of~$\emlstr$, the upper square in~\labelcref{eq:emlstr-def},
  and that $\effmult \emprod \emdyn\effmult$~is an epi.
\end{proof}

\begin{lemma}\label{lem:emlstr-is-str}
  The natural transformation~$\emlstr$ from~\cref{lem:emlstr} is a strength for~$\emdyn$.
\end{lemma}
\begin{proof}
  We need to show that $\emlstr$~is compatible with the left unitor and the associator of~$\emcat$.
  For the left unitor axiom, we need to show
  \(
    \emdyn\emlunit[\malg{A}] \com \emlstr[\emfree\tunit,\malg{A}]
      = \emlunit[\emdyn\malg{A}]
  \),
  where $\malg{A} = (A,a)$.
  Since $\emprod$~preserves reflexive coequalisers and $\emdyn$~preserves canonical presentations,
  $\effmult[\tunit] \emprod \emdyn a$~is an epimorphism,
  so it suffices to show the above equality after precomposing with this morphism.
  This follows via routine but elaborate diagram chasing using the naturality of~$\emlstr$.
  The equation involving the associator~$\emassoc$ can be proved similarly.
\end{proof}

Finally, we show that $\emdyn$~is a varietor.
We do this in two steps.

\begin{proposition}\label{prop:emdyn-pres-colim}
  The functor~$\emdyn$ preserves colimits of all shapes that $\dyn$~and~$\eff$ preserve.
\end{proposition}
\begin{proof}
  Let~$D = (\malg{A}_i)_{i \in \cat{I}}$ be a diagram~$\cat{I} \to \emcat$
  of arbitrary shape~$\cat{I}$ with $\malg{A}_i = (A_i, a_i)$,
  and let~$\malg{A} = (A, a)$ be a colimit of~$D$ 
  with coprojections~$\coproj{i} \colon \malg{A}_i \to \malg{A}$.
  We need to show that $\emdyn\coproj{i} \colon \emdyn\malg{A}_i \to \emdyn\malg{A}$
  is a colimiting cocone.
  That is, for any cocone~$(f_i)_{i \in \cat{I}}$ with $f_i \colon \emdyn\malg{A}_i \to \malg{B}$,
  we have to find a unique $\eff$-algebra map~$f \colon \emdyn\malg{A} \to \malg{B}$
  such that $f \com \emdyn\coproj{i} = f_i$ for all~$i \in \cat{I}$,
  as indicated in the diagram below.
  \[\begin{tikzcd}[column sep=3em]
    {\emdyn\emfree\eff A_i} & {\emfree\dyn\eff A_i} & {\emfree\dyn\eff A} & {\emdyn\emfree\eff A} \\
    {\emdyn\emfree A_i} & {\emfree\dyn A_i} & {\emfree\dyn A} & {\emdyn\emfree A} \\
    {\emdyn\malg{A}_i} &&& {\emdyn\malg{A}} \\
    &&& {\malg{B}}
    \arrow["\emfreedyn", "\cong"', from=1-1, to=1-2]
    \arrow["{\emdyn\effmult}"', shift right, from=1-1, to=2-1]
    \arrow["{\emdyn\emfree a_i}", shift left, from=1-1, to=2-1]
    \arrow["{\emfree\dyn\eff\coproj{i}}", from=1-2, to=1-3]
    \arrow["{\emfree\dyn a_i}", shift left, from=1-2, to=2-2]
    \arrow["{\effmult \com \emfree\dlaw}"', shift right, from=1-2, to=2-2]
    \arrow["{\emfree\dyn a}", shift left, from=1-3, to=2-3]
    \arrow["{\effmult \com \emfree\dlaw}"', shift right, from=1-3, to=2-3]
    \arrow["\emfreedyn"', "\cong", from=1-4, to=1-3]
    \arrow["{\emdyn\emfree a}", shift left, from=1-4, to=2-4]
    \arrow["{\emdyn\effmult}"', shift right, from=1-4, to=2-4]
    \arrow["\emfreedyn"', "\cong", from=2-1, to=2-2]
    \arrow["{\emdyn a_i}"', from=2-1, to=3-1]
    \arrow["{\emfree\dyn\coproj{i}}"', from=2-2, to=2-3]
    \arrow["\emfreedyn", "\cong"', from=2-4, to=2-3]
    \arrow["{\emdyn a}", from=2-4, to=3-4]
    \arrow["{\emdyn\coproj{i}}", from=3-1, to=3-4]
    \arrow["{f_i}"', from=3-1, to=4-4]
    \arrow["f", dashed, from=3-4, to=4-4]
  \end{tikzcd}\]

  To construct~$f$, we use the colimit preservation properties of the functors involved.
  Firstly, since $\emdyn$ preserves canonical presentations and $\emfreefdyn$ is an isomorphism,
  it suffices to find a map~$\bar{f} \colon \emfree\dyn A \to \malg{B}$
  such that $\bar{f} \com \emfree\dyn\coproj{i} \com \emfreefdyn = f_i \com \emdyn a_i$
  for all~$i \in \cat{I}$.
  Next, we note that $\dyn$~and~$\eff$ (and hence~$\emforget$)
  preserve colimits of shape~$\cat{I}$ by assumption,
  and $\emfree$~preserves all colimits since it is is a left adjoint.
  Hence, the cocones
  $(\emfree\dyn\coproj{i})_{i \in \cat{I}}$~and~$(\emfree\dyn\eff\coproj{i})_{i \in \cat{I}}$
  are both colimiting.
  Thus, to construct~$\bar{f}$, it suffices to find
  a cocone over~$\emfree \circ \dyn \circ \emforget \circ D$,
  and we can choose~$\mleft(f_i \com \emdyn a_i \com \inv*{\emfreefdyn}\mright)_{i \in \cat{I}}$.
  The verification that this is a cocone and that the universal property holds
  now follows from elementary calculations; we omit the details.
\end{proof}

\begin{lemma}\label{lem:emdyn-varietor}
  The functor~$\emdyn$ is a varietor.
\end{lemma}
\begin{proof}
  We have seen in~\cref{lem:em-sat-ass-bcat} that $\emcat$~is cocomplete.
  Hence, by~\cref{lem:varietor},
  it suffices to show that $\emdyn$~preserves colimits of $\omega$-chains.
  Since both $\dyn$~and~$\eff$ preserve colimits of $\omega$-chains
  by~\cref{ass:effectful-base,ass:monad}, this follows from~\cref{prop:emdyn-pres-colim}.
\end{proof}

\begin{proof}[Proof of~\cref{thm:em-sat-ass}]
  Consequence of~\cref{lem:em-sat-ass-bcat,lem:emlstr-is-str,lem:emdyn-varietor}.
\end{proof}

There is a generalisation of diagram~\labelcref{eq:emfreedyn-emfreemf-emlstr}
to free algebras, which we shall need later.
\begin{lemma}\label{lem:emfreefdyn-emfreemf-emflstr}
  The following diagram commutes:
  \[\begin{tikzcd}[column sep=4em]
    {\emfree A \emprod \emfdyn\emfree B} &[1.2em] {\emfdyn(\emfree A \emprod \emfree B)} & {\emfdyn\emfree(A \tprod B)} \\
    {\emfree A \emprod \emfree\fdyn B} & {\emfree(A \tprod \fdyn B)} & {\emfree\fdyn(A \tprod B)}
    \arrow["{\emflstr[\emfree A,\emfree B]}", from=1-1, to=1-2]
    \arrow["{\id[\emfree A] \emprod \emfreefdyn[B]}"', from=1-1, to=2-1]
    \arrow["{\emfdyn\emfreemf[A,B]}", from=1-2, to=1-3]
    \arrow["{\emfreefdyn[A \tprod B]}", from=1-3, to=2-3]
    \arrow["{\emfreemf[A,\fdyn B]}"', from=2-1, to=2-2]
    \arrow["{\emfree\flstr[A,B]}"', from=2-2, to=2-3]
  \end{tikzcd}\]
\end{lemma}
\begin{proof}
  We use that $\emfdyn\emfree B$~is a strongly free $\emdyn$-algebra
  by~\cref{thm:em-sat-ass,prop:flstr}; thus it suffices to show that
  both composites are `parametrised pointed algebra maps',
  \ie that the diagrams in~\cref{def:strongly-free-alg} commute.
  This latter claim follows from the definition of~$\emflstr$ (see~\cref{prop:flstr})
  and diagram~\labelcref{eq:emfreedyn-emfreemf-emlstr}.
\end{proof}

\section{Lifting functors and adjunctions to algebras and automata}

We recall the result of Hermida and Jacobs~\cite{HJ:98}
and generalise it slightly to automata.
If $L \dashv R$ is an adjunction,
we write~$\ladj{f} \colon LA \to B$ for the adjunct of a morphism~$A \to RB$,
and similarly, we write~$\radj{g} \colon A \to RB$ for the adjunct of~$g \colon LA \to B$.
\begin{theorem}\label{thm:lift-adj-alg}
  Let~$\nu \colon HL \to LG$ be a natural transformation as in the following situation:
  \[\begin{tikzcd}
    {\cat{A}} & {\cat{B}} \\
    {\cat{A}} & {\cat{B}}
    \arrow["L", from=1-1, to=1-2]
    \arrow["G"', from=1-1, to=2-1]
    \arrow["\nu", between={0.2}{0.8}, Rightarrow, from=1-2, to=2-1]
    \arrow["H", from=1-2, to=2-2]
    \arrow["L"', from=2-1, to=2-2]
  \end{tikzcd}\]
  Then there is an induced functor~$\alglift{L} \colon \Alg{G} \to \Alg{H}$
  given by $\alglift{L}(A, a) \defeq (LA, La \com \nu_A)$ and $\alglift{L}f \defeq Lf$.
  If $\nu$~is an isomorphism, then any adjunction~$L \dashv R$
  with unit~$\iota$ and counit~$\delta$ lifts to an adjunction~$\alglift{L} \dashv \alglift{R}$
  where the functor~$\alglift{R}$ is induced by
  the natural transformation\footnotemark~$\tilde{\nu}$ given by
  $\tilde{\nu} \defeq GR \xto{\iota GR} RLGR \xto{R\inv{\nu}R} RHLR \xto{RH\delta} RH$.
  \footnotetext{The natural transformation~$\tilde{\nu}$ is called the \emph{mate} of~$\inv{\nu}$.}
  \[
    \begin{tikzcd}[column sep=3em]
      {\cat{A}} & {\cat{B}}
      \arrow[""{name=0, anchor=center, inner sep=0}, "L", shift left, curve={height=-4pt}, from=1-1, to=1-2]
      \arrow[""{name=1, anchor=center, inner sep=0}, "R", shift left, curve={height=-4pt}, from=1-2, to=1-1]
      \arrow["\dashv"{anchor=center, rotate=-90}, draw=none, from=0, to=1]
    \end{tikzcd}
    \quad\implies\quad
    \begin{tikzcd}[column sep=3em]
      {\Alg{G}} & {\Alg{H}}
      \arrow[""{name=0, anchor=center, inner sep=0}, "{\alglift{L}}", shift left, curve={height=-3pt}, from=1-1, to=1-2]
      \arrow[""{name=1, anchor=center, inner sep=0}, "{\alglift{R}}", shift left, curve={height=-3pt}, from=1-2, to=1-1]
      \arrow["\dashv"{anchor=center, rotate=-90}, draw=none, from=0, to=1]
    \end{tikzcd}
  \]
  Lifting means that the natural bijection~\(
    \Hom[\Alg{G}]{(A, a)}{\alglift{R}(B, b)} \cong \Hom[\Alg{H}]{\alglift{L}(A, a)}{(B, b)}
  \)
  is the restriction of the bijection~$\Hom[\cat{A}]{A}{RB} \cong \Hom[\cat{B}]{LA}{B}$
  to algebra homomorphisms.
\end{theorem}

We state a variant of~\cref{thm:lift-adj-alg} for pointed algebras.
For a fixed~$I$, an \emph{($I$-)pointed $G$-algebra}
is a $G$-algebra~$(A, a)$ equipped with a point~$I \to A$,
and a \emph{pointed homomorphism} is a $G$-algebra homomorphism that preserves the point.
These form a category~$\Alg[I]{G}$.
\begin{theorem}\label{thm:lift-adj-pt-alg}
  In the situation of~\cref{thm:lift-adj-alg} and for a fixed object~$I \in \cat{A}$,
  we have an induced functor~$\alglift{L} \colon \Alg[I]{G} \to \Alg[LI]{H}$
  given by $\alglift{L}(A, a, i) \defeq (LA, La \com \nu_A, Li)$ and $\alglift{L}f \defeq Lf$.
  If $\nu$~is an isomorphism, then an adjunction~$L \dashv R$ lifts to pointed algebras:
  \[
    \begin{tikzcd}[column sep=3em]
      {\cat{A}} & {\cat{B}}
      \arrow[""{name=0, anchor=center, inner sep=0}, "L", shift left, curve={height=-4pt}, from=1-1, to=1-2]
      \arrow[""{name=1, anchor=center, inner sep=0}, "R", shift left, curve={height=-4pt}, from=1-2, to=1-1]
      \arrow["\dashv"{anchor=center, rotate=-90}, draw=none, from=0, to=1]
    \end{tikzcd}
    \quad\implies\quad
    \begin{tikzcd}[column sep=3em]
      {\Alg[I]{G}} & {\Alg[LI]{H}}
      \arrow[""{name=0, anchor=center, inner sep=0}, "{\alglift{L}}", shift left, curve={height=-3pt}, from=1-1, to=1-2]
      \arrow[""{name=1, anchor=center, inner sep=0}, "{\alglift{R}}", shift left, curve={height=-3pt}, from=1-2, to=1-1]
      \arrow["\dashv"{anchor=center, rotate=-90}, draw=none, from=0, to=1]
    \end{tikzcd}
  \]
  Here, $\alglift{L}$~is defined as above, and $\alglift{R}$ is given by
  $\alglift{R}(B, b, j) \defeq (RB, Rb \com \tilde{\nu}_B, \radj{j})$ and $\alglift{R}g \defeq Rg$.
  Again, the natural bijection~\(
    \Hom[\Alg[I]{G}]{(A, a, i)}{\alglift{R}(B, b, j)}
      \cong \Hom[\Alg[LI]{H}]{\alglift{L}(A, a, i)}{(B, b, j)}
  \)
  is the restriction of the bijection~$\Hom[\cat{A}]{A}{RB} \cong \Hom[\cat{B}]{LA}{B}$
  to pointed algebra homomorphisms.
\end{theorem}
\begin{proof}
  The proof is analogous to~\cref{thm:lift-adj-alg}.
  One has to check additionally that
  $\alglift{L}f$ (respectively~$\alglift{R}g$)~is pointed if $f$ (respectively~$g$)~is pointed,
  and that $f \colon LA \to B$~is pointed iff $\radj{f}$~is pointed.
  These are all routine calculations.
\end{proof}
If $\cat{A}$~and~$\cat{B}$ have coproducts,
then \cref{thm:lift-adj-pt-alg} also follows directly from~\cref{thm:lift-adj-alg}
if we take~$G \leteq I \catsum G(-)$, $H \leteq LI \catsum H(-)$, and~\(
  \nu \leteq LI \catsum HLA \xto{\id \catsum \nu}
             LI \catsum LGA \xto{\casem{\coprojl}{\coprojr}}
             L(I \catsum GA)
\)
(the composite is an isomorphism if $\nu$~is since $L$, being a left adjoint, preserves colimits).
Then $\Alg{I \catsum G(-)}$~and~$\Alg{LI \catsum H(-)}$ are isomorphic to
$\Alg[I]{G}$~and~$\Alg[LI]{H}$, respectively.
One can then verify that $\alglift{L}$~and~$\alglift{R}$ are as given above.

\begin{proof}[Proof of~\cref{thm:triangle-lift}]
  By entirely analogous considerations as in the proof of~\cref{thm:lift-adj-pt-alg},
  we get that for a natural isomorphism~$\nu \colon HL \ito LG$ and a fixed~$O \in \cat{B}$,
  an adjunction~$L \dashv R$ lifts to an adjunction
  \[\begin{tikzcd}[column sep=3em]
    {\Aut{G}{I}{RO}} & {\Aut{H}{LI}{O}}
    \arrow[""{name=0, anchor=center, inner sep=0}, "{\alglift{L}}", shift left, curve={height=-3pt}, from=1-1, to=1-2]
    \arrow[""{name=1, anchor=center, inner sep=0}, "{\alglift{R}}", shift left, curve={height=-3pt}, from=1-2, to=1-1]
    \arrow["\dashv"{anchor=center, rotate=-90}, draw=none, from=0, to=1]
  \end{tikzcd}\]
  where $\alglift{L}$~and~$\alglift{R}$ are given by
  \begin{alignat*}{2}
    \alglift{L}(Q, \delta, i, o) &\defeq (LQ, L\delta \com \nu, Li, \ladj{o}),         \qquad& \alglift{L}f &\defeq Lf, \\
    \alglift{R}(Q, \delta, i, o) &\defeq (RQ, R\delta \com \tilde{\nu}, \radj{i}, Ro),       & \alglift{R}g &\defeq Rg.
  \end{alignat*}
  Hence, we obtain the adjunctions in the right triangle of~\labelcref{eq:aut-triangle},
  where the adjunction~$\alglift{\klemb} \dashv \alglift{\klext}$ is
  induced by the identity natural transformation on~$\kldyn\klemb = \klemb\dyn$
  and the adjunction~$\alglift{\emfree} \dashv \alglift{\emforget}$ is induced by~$\emfreedyn$.
  The algebra part of~$\alglift{\klcomp}$
  is induced by~$\klcompdyn$ (see~\cref{lem:klcompdyn}) as in~\cref{thm:lift-adj-alg};
  additionally, $\alglift{\klcomp}$~sends the input and outputs maps $i$~and~$o$
  to $\klcomp i$~and~$\klcomp o$, respectively.
\end{proof}

\begin{proof}[Proof of~\cref{prop:kldyn-free-alg}]
  The result follows from~\cref{thm:lift-adj-pt-alg}
  applied to the adjunction~$\klemb \dashv \klext$,
  using the fact that a free algebra over~$A$ is the same thing
  as an initial $A$-pointed algebra, and that left adjoints preserve initial objects.
\end{proof}

Let us write~$(\emfdyn\malg{A}, \emcon[\malg{A}], \ememp[\malg{A}])$
for the free $\emdyn$-algebra over~$\malg{A} \in \emcat$.
\begin{proof}[Proof of~\cref{prop:emdyn-free-alg}]
  Similarly to~\cref{prop:kldyn-free-alg}, we apply~\cref{thm:lift-adj-pt-alg}
  to the adjunction~$\emfree \dashv \emforget$ to deduce the first part of the statement.
  Since initial (or free) objects are unique up to unique isomorphism,
  we obtain, for each~$A \in \bcat$, a unique isomorphism of $\emdyn$-algebras~\(
    \emfreefdyn[A] \colon
      (\emfdyn\emfree A, \emcon[A]) \ito (\emfree\fdyn A, \emfree\con[A] \com \emfreedyn[A])
  \)
  \sth $\emfreedyn[A] \com \ememp[\emfree A] = \emfree\emp[A]$.
\end{proof}

\section{Proof of~\texorpdfstring{\cref{prop:effectful-st-cov-set}}{\autoref{prop:effectful-st-cov-set}}}

Let us write~$c \colon C \to \fdyn\inobj$
for the inclusion of the subset~$C \subseteq \fdyn\inobj$.
Note that the condition $\emp[\inobj](a) \in C$ ($a \in \inobj$)
is equivalent to the statement that $\emp[\inobj]$~factors through~$c$,
which is equivalent to the existence of a morphism~$i_C \colon \emp[\inobj] \to c$.
Hence, it remains to show that a map~$\delta_C$ making diagram~\labelcref{eq:effectful-st-cov}
commute exists if and only if the second condition holds.

For the `if' direction, let us denote the image factorisation
of~$\rea{\alglift{\emforget}\spec}$ by~$\fdyn\inobj \xto{e} R \xto{m} Q$.
First choose for every~$q \in R$ an element~$z \in \eff C$
such that $\mext{\rea{\alglift{\emforget}\spec}}(\eff c(z)) = q$ (such a~$z$ exists by assumption).
This defines a function~$s \colon R \to \eff C$ satisfying
$\mext{\rea{\alglift{\emforget}\spec}} \com \eff c \com s = m$.
Then we can define~$\delta_C$ as the composite~\(
  \dyn C \xto{\dyn c} \dyn\fdyn\inobj \xto{\con[\inobj]} \fdyn\inobj \xto{e} R \xto{s} \eff C
\).
This makes diagram~\labelcref{eq:effectful-st-cov} commute since
\(
  \mext{\rea{\alglift{\emforget}\spec}} \com \eff c \com \delta_C
    = \mext{\rea{\alglift{\emforget}\spec}} \com \eff c \com s \com e \com \con[\inobj] \com \dyn c
    = m \com e \com \con[\inobj] \com \dyn c
    = \rea{\alglift{\emforget}\spec} \com \con[\inobj] \com \dyn c
\).

Conversely, let~$\delta_C \colon \dyn C \to \eff C$ be a morphism
making~\labelcref{eq:effectful-st-cov} commute and let~$i_C \colon \emp[\inobj] \to c$.
Define~$Z \subseteq \fdyn\inobj$ as~\(
  Z \defeq
    \set{v \in \fdyn\inobj}
        [\exists z \in \eff C.\,
          \mext{\rea{\alglift{\emforget}\spec}}(\eff c(z)) = \rea{\alglift{\emforget}\spec}(v)]
\).
By the definition of~$R$, we have for all~$q \in R$ some~$v \in \fdyn\inobj$
such that $\rea{\alglift{\emforget}\spec}(v) = q$.
Hence, it suffices to show that $Z = \fdyn\inobj$, which is equivalent to
the inclusion map~$m \colon Z \to \fdyn\inobj$ being an isomorphism.
Our strategy is to use `induction', \ie to show that $Z$~carries a pointed $\dyn$-algebra structure
that makes~$m$ a pointed homomorphism from~$Z$ into the free algebra~$\fdyn\inobj$.
Then $m$~is an isomorphism since it is a subobject of the initial object
in the category of pointed $\dyn$-algebras.

To construct the pointed $\dyn$-algebra structure on~$Z$,
we use that $Z$~can be characterised categorically as the image
of the projection~$r$ in the following pullback diagram:
\[\begin{tikzcd}
	P && {\fdyn\inobj} \\
	{\eff C} & {\eff\fdyn\inobj} & Q
	\arrow["r", from=1-1, to=1-3]
	\arrow["p"', from=1-1, to=2-1]
	\arrow["{\rea{\alglift{\emforget}\spec}}", from=1-3, to=2-3]
	\arrow["{\eff c}"', from=2-1, to=2-2]
	\arrow["{\mext{\rea{\alglift{\emforget}\spec}}}"', from=2-2, to=2-3]
\end{tikzcd}\]
Hence, to show that a morphism~$v \colon A \to \fdyn\inobj$
factors through~$m \colon Z \to \fdyn\inobj$,
it suffices to give a morphism~$u \colon A \to \eff C$ such that
$\mext{\rea{\alglift{\emforget}\spec}} \com \eff c \com u = \rea{\alglift{\emforget}\spec} \com v$.
In particular, $\emp[\inobj] \colon \inobj \to \fdyn\inobj$~factors through~$m$
since
\(
  \mext{\rea{\alglift{\emforget}\spec}} \com \eff c \com \effunit[C] \com i_C
    = \mext{\rea{\alglift{\emforget}\spec}} \com \effunit[\fdyn\inobj] \com c \com i_C
    = \rea{\alglift{\emforget}\spec} \com \emp[\inobj]
\).
This gives us a point~$i_Z \colon \inobj \to Z$ satisfying $m \com i_Z = \emp[\inobj]$.

For the algebra structure on~$Z$, we need a lemma:
\begin{lemma}\label{lem:emdyn-alg}
  Let~$(\malg{A},\alpha)$ be \an{\emdyn} $\emdyn$-algebra, where $\malg{A} = (A,a)$,
  and let us denote the structure map of~$\alglift{\emforget}(\malg{A},\alpha)$
  by~$\bar{\alpha} \colon \dyn A \to A$.
  Then the following diagram commutes:
  \[\begin{tikzcd}
    {\dyn\eff A} & {\eff\dyn A} & {\eff A} \\
    {\dyn A} && A
    \arrow["{\dlaw[A]}", from=1-1, to=1-2]
    \arrow["{\dyn a}"', from=1-1, to=2-1]
    \arrow["{\eff\bar{\alpha}}", from=1-2, to=1-3]
    \arrow["a", from=1-3, to=2-3]
    \arrow["{\bar{\alpha}}"', from=2-1, to=2-3]
  \end{tikzcd}\]
\end{lemma}
\begin{proof}
  Since the adjunction~$\alglift{\emfree} \dashv \alglift{\emforget}$ is induced by~$\emfreedyn$,
  the functor~$\alglift{\emforget}$ is induced by
  the natural transformation~$\emforgetdyn \colon \dyn\emforget \to \emforget\emdyn$
  given by $\emforgetdyn[(B,b)] = \emdyn b \com \inv{\emfreedyn[B]} \com \effunit[\dyn B]$
  (see~\cref{thm:lift-adj-alg}).
  Hence, we have $\bar{\alpha} = \alpha \com \emforgetdyn[\malg{A}]$.
  Writing~$\algmap$ for the $\eff$-algebra structure of~$\emdyn\malg{A}$,
  the desired result follows from the following commuting diagram:
  \[\begin{tikzcd}
    {\dyn\eff A} & {\eff\dyn A} & {\eff\emdyn\malg{A}} & {\eff A} \\
    {\dyn\eff A} & {\emdyn\emfree A} \\
    {\dyn A} && {\emdyn\malg{A}} & A
    \arrow[""{name=0, anchor=center, inner sep=0}, "{\dlaw[A]}", from=1-1, to=1-2]
    \arrow[equals, nfold, from=1-1, to=2-1]
    \arrow["{\eff\emforgetdyn[\malg{A}]}", from=1-2, to=1-3]
    \arrow["{\eff\alpha}", from=1-3, to=1-4]
    \arrow["\algmap", from=1-3, to=3-3]
    \arrow["a", from=1-4, to=3-4]
    \arrow[""{name=1, anchor=center, inner sep=0}, "{\emforgetdyn[\emfree A]}"', from=2-1, to=2-2]
    \arrow["{\dyn a}"', from=2-1, to=3-1]
    \arrow["{\emfreedyn[A]}", from=2-2, to=1-2]
    \arrow["{(\ast)}"'{pos=0.4}, draw=none, from=2-2, to=1-3]
    \arrow["{\emdyn a}"'{pos=0.4}, from=2-2, to=3-3]
    \arrow["{\emforgetdyn[\malg{A}]}"', from=3-1, to=3-3]
    \arrow["\alpha"', from=3-3, to=3-4]
    \arrow["{(\times)}"{description}, draw=none, from=0, to=1]
  \end{tikzcd}\]
  The lower trapezoid commutes by the naturality of~$\emforgetdyn$,
  and the right-hand square commutes since
  $\alpha \colon \emdyn\malg{A} \to \malg{A}$~is \an{\eff} $\eff$-algebra morphism.
  Furthermore, $(\times)$~commutes since
  \[
    \emfreedyn[A] \com \emforgetdyn[\emfree A]
      = \emfreedyn[A] \com \emdyn\effmult[A] \com \inv{\emfreedyn[\eff A]} \com \effunit[\dyn\eff A]
      = \effmult[\dyn A] \com \eff\dlaw[A] \com \effunit[\dyn\eff A]
      = \effmult[\dyn A] \com \effunit[\eff\dyn A] \com \dlaw[A]
      = \dlaw[A],
  \]
  using diagram~\labelcref{eq:emfreedyn-effmult}.
  Finally, $(\ast)$~commutes since
  \[
    \algmap \com \eff\emforgetdyn[\malg{A}] \com \emfreedyn[A]
      = \algmap \com \eff(\emdyn a \com \inv{\emfreedyn[A]}) \com \eff\effunit[\dyn A] \com \emfreedyn[A]
      = \emdyn a \com \inv{\emfreedyn[A]} \com \effmult[\dyn A] \com \eff\effunit[\dyn A] \com \emfreedyn[A]
      = \emdyn a,
  \]
  using that $\emdyn a \com \inv{\emfreedyn[A]}$~is
  \an{\eff} $\eff$-algebra morphism~$\emfree\dyn A \to \emdyn\malg{A}$.
\end{proof}

Now choose for each~$v \in Z$ some~$z \in \eff C$ such that
$\mext{\rea{\alglift{\emforget}\spec}}(\eff c(z)) = \rea{\alglift{\emforget}\spec}(v)$
(such a~$z$ exists by the definition of~$Z$).
This defines a map~$s \colon Z \to \eff C$ that satisfies
$\mext{\rea{\alglift{\emforget}\spec}} \com \eff c \com s = \rea{\alglift{\emforget}\spec} \com m$.
Let us write~$q \colon \eff Q \to Q$ for the $\eff$-algebra structure of~$Q$
and $\bar{\delta}$~for the transition map of~$\alglift{\emforget}\spec$,
and consider the following diagram:
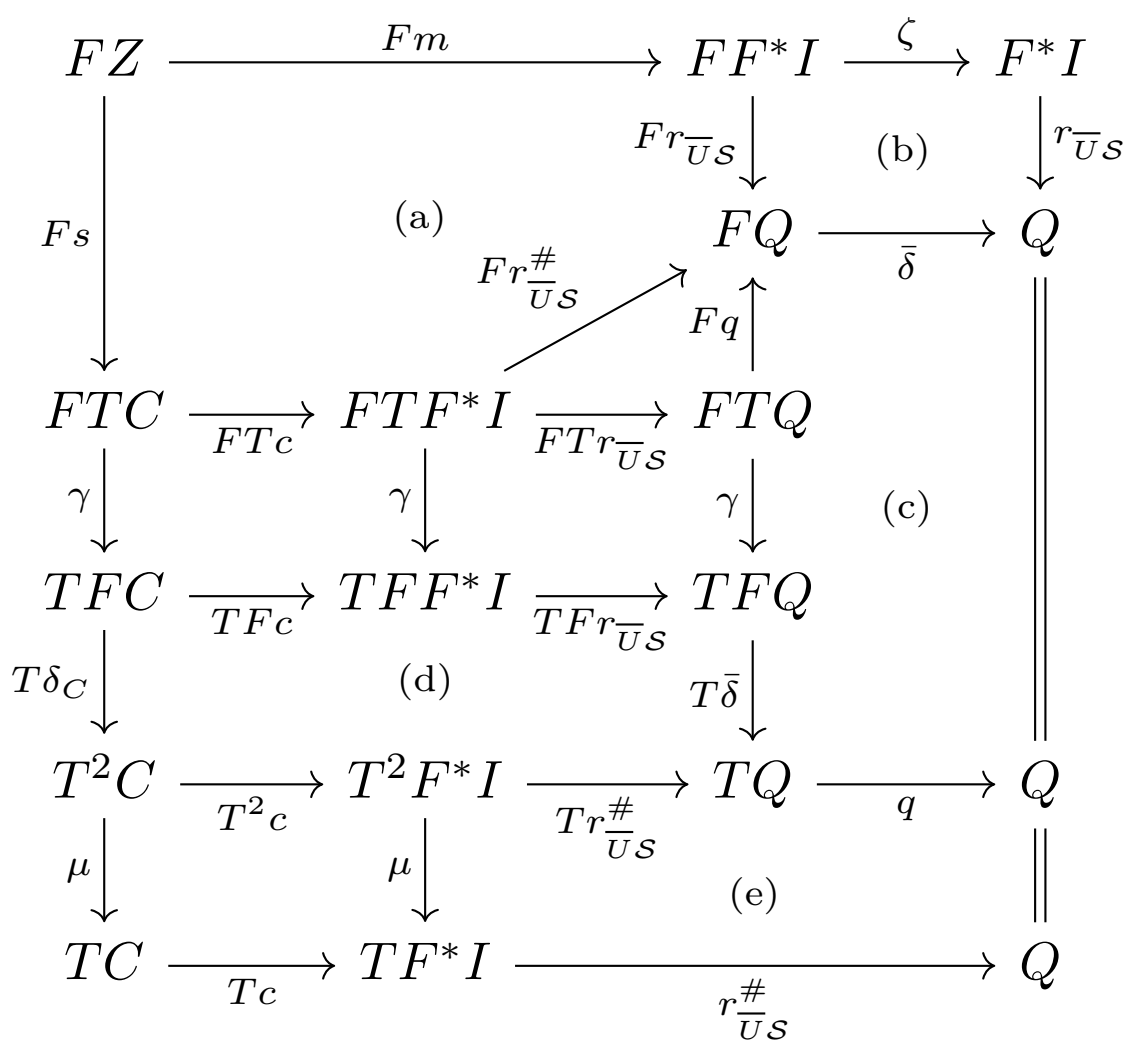
\[\begin{tikzcd}
	{\dyn Z} && {\dyn\fdyn\inobj} & {\fdyn\inobj} \\
	&& {\dyn Q} & Q \\
	{\dyn\eff C} & {\dyn\eff\fdyn\inobj} & {\dyn\eff Q} \\
	{\eff\dyn C} & {\eff\dyn\fdyn\inobj} & {\eff\dyn Q} \\
	{\eff^2 C} & {\eff^2\fdyn\inobj} & {\eff Q} & Q \\
	{\eff C} & {\eff\fdyn\inobj} && Q
	\arrow[""{name=0, anchor=center, inner sep=0}, "{\dyn m}", from=1-1, to=1-3]
	\arrow["{\dyn s}"', from=1-1, to=3-1]
	\arrow["\con", from=1-3, to=1-4]
	\arrow["{\dyn\rea{\alglift{\emforget}\spec}}"', from=1-3, to=2-3]
	\arrow["{(\text{b})}"{description}, draw=none, from=1-3, to=2-4]
	\arrow["{\rea{\alglift{\emforget}\spec}}", from=1-4, to=2-4]
	\arrow[""{name=1, anchor=center, inner sep=0}, "{\bar{\delta}}"', from=2-3, to=2-4]
	\arrow[equals, nfold, from=2-4, to=5-4]
	\arrow["{\dyn\eff c}"', from=3-1, to=3-2]
	\arrow["\dlaw"', from=3-1, to=4-1]
	\arrow["{\dyn\mext{\rea{\alglift{\emforget}\spec}}}", from=3-2, to=2-3]
	\arrow["{\dyn\eff\rea{\alglift{\emforget}\spec}}"', from=3-2, to=3-3]
	\arrow["\dlaw"', from=3-2, to=4-2]
	\arrow["{\dyn q}", from=3-3, to=2-3]
	\arrow["\dlaw"', from=3-3, to=4-3]
	\arrow["{\eff\dyn c}"', from=4-1, to=4-2]
	\arrow["{\eff\delta_C}"', from=4-1, to=5-1]
	\arrow["{\eff\dyn\rea{\alglift{\emforget}\spec}}"', from=4-2, to=4-3]
	\arrow["{(\text{d})}"{description}, draw=none, from=4-2, to=5-2]
	\arrow["{\eff\bar{\delta}}"', from=4-3, to=5-3]
	\arrow["{\eff^2 c}"', from=5-1, to=5-2]
	\arrow["\effmult"', from=5-1, to=6-1]
	\arrow["{\eff\mext{\rea{\alglift{\emforget}\spec}}}"', from=5-2, to=5-3]
	\arrow["\effmult"', from=5-2, to=6-2]
	\arrow[""{name=2, anchor=center, inner sep=0}, "q"', from=5-3, to=5-4]
	\arrow[equals, nfold, from=5-4, to=6-4]
	\arrow["{\eff c}"', from=6-1, to=6-2]
	\arrow[""{name=3, anchor=center, inner sep=0}, "{\mext{\rea{\alglift{\emforget}\spec}}}"', from=6-2, to=6-4]
	\arrow["{(\text{a})}"{description}, draw=none, from=0, to=3-2]
	\arrow["{(\text{c})}"{description}, draw=none, from=1, to=2]
	\arrow["{(\text{e})}"{description}, draw=none, from=5-3, to=3]
\end{tikzcd}\]
Here, (a)~commutes by the definition of~$s$,
(b)~commutes since $\rea{\alglift{\emforget}\spec}$~is
\an{\dyn} $\dyn$-algebra map~$(\fdyn\inobj,\con) \to (Q,\bar{\delta})$,
(c)~commutes by~\cref{lem:emdyn-alg},
(d)~commutes by diagram~\labelcref{eq:effectful-st-cov} and the equality
$\rea{\alglift{\emforget}\spec} \com \con = \bar{\delta} \com \dyn\rea{\alglift{\emforget}\spec}$,
(e)~commutes since $\mext{\rea{\alglift{\emforget}\spec}}$~is
\an{\eff} $\eff$-algebra map~$\emfree\fdyn\inobj \to (Q,q)$,
the remaining squares commute by naturality, and the triangle commutes by definition.

Hence, we obtain a map~$z \colon \dyn Z \to Z$ such that $m \com z = \con[\inobj] \com \dyn m$.
This, together with the equality $m \com i_Z = \emp[\inobj]$,
shows that $m$~is a pointed homomorphism~$(Z,z,i_Z) \to (\fdyn\inobj,\con[\inobj],\emp[\inobj])$,
which is what needed to show. \qed

\section{Proof of~\texorpdfstring{\cref{thm:effectful-iff-free}}{\autoref{thm:effectful-iff-free}}}

\begin{lemma}\label{lem:emfreets-cont-ememp}
  Let~$a \colon B \to \fdyn A$ be a morphism in~$\bcat$.
  Then a map~$h \colon A \to B$ in~$\bcat$ is a morphism~$\emp[A] \to a$ if and only if
  $\emfree h \colon \emfree A \to \emfree B$~is a morphism~$\ememp[\emfree A] \to \emfreets{a}$.
\end{lemma}
\begin{proof}
  Follows from the definition of~$\emfreets$
  and the equality~$\emfreedyn[A] \com \ememp[\emfree A] = \emfree\emp[A]$.
\end{proof}

\paragraph*{Part~\labelcref{prop:effectful-iff-free:st-cov}}

Let us write~$\spec = (\malg{Q}, \delta, i, o)$,
where $\malg{Q} = (Q, q)$~is \an{\eff} $\eff$-algebra.
By~\cref{lem:emfreets-cont-ememp}, we have
$i_C \colon \emp[\inobj] \to c$ if and only if $\emfree i_C \colon \ememp[\emfree\inobj] \to \emfreets{c}$.
Furthermore, $\rea{\spec} \com \emfreets{c}$~is \an{\emdyn} $\emdyn$-algebra
morphism~$(\emfree C, \klcomp\delta_C \com \emfreedyn[C]) \to (\malg{Q}, \delta)$
iff the outer square in the following diagram commutes:
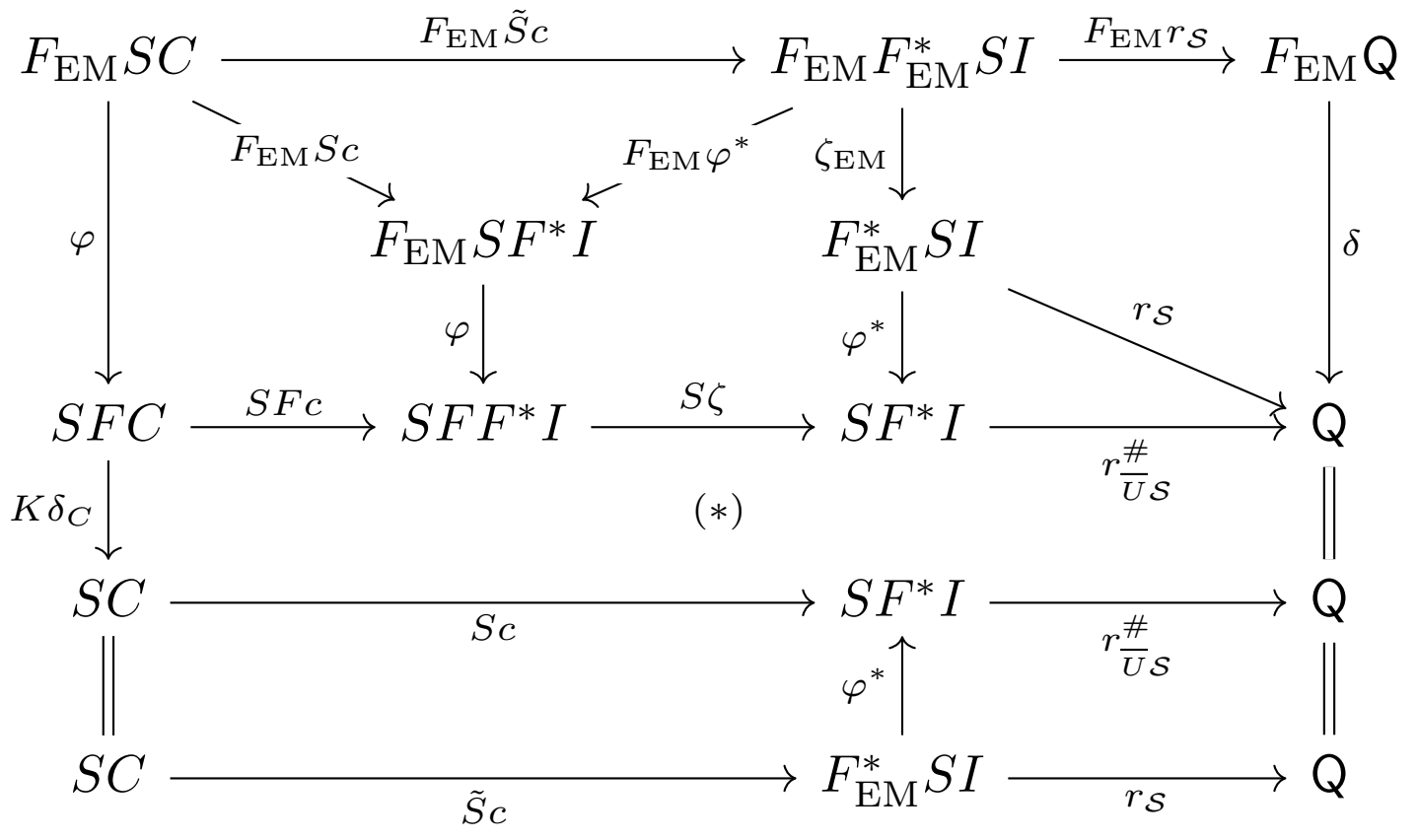
\[\begin{tikzcd}
	{\emdyn\emfree C} && {\emdyn\emfdyn\emfree\inobj} &[1em] {\emdyn\malg{Q}} \\
	& {\emdyn\emfree\fdyn\inobj} & {\emfdyn\emfree\inobj} \\
	{\emfree\dyn C} & {\emfree\dyn\fdyn\inobj} & {\emfree\fdyn\inobj} & {\malg{Q}} \\
	{\emfree C} && {\emfree\fdyn\inobj} & {\malg{Q}} \\
	{\emfree C} && {\emfdyn\emfree\inobj} & {\malg{Q}}
	\arrow["{\emdyn\emfreets{c}}", from=1-1, to=1-3]
	\arrow["{\emdyn\emfree c}"{description}, from=1-1, to=2-2]
	\arrow["\emfreedyn"', from=1-1, to=3-1]
	\arrow["{\emdyn\rea{\spec}}", from=1-3, to=1-4]
	\arrow["{\emdyn\emfreefdyn}"{description}, from=1-3, to=2-2]
	\arrow["\emcon"', from=1-3, to=2-3]
	\arrow["\delta", from=1-4, to=3-4]
	\arrow["\emfreedyn"', from=2-2, to=3-2]
	\arrow["\emfreefdyn"', from=2-3, to=3-3]
	\arrow["{\rea{\spec}}"{pos=0.4}, from=2-3, to=3-4]
	\arrow["{\emfree\dyn c}", from=3-1, to=3-2]
	\arrow[""{name=0, anchor=center, inner sep=0}, "{\klcomp\delta_C}"', from=3-1, to=4-1]
	\arrow["{\emfree\con}", from=3-2, to=3-3]
	\arrow["{\mext{\rea{\alglift{\emforget}\spec}}}"', from=3-3, to=3-4]
	\arrow[""{name=1, anchor=center, inner sep=0}, equals, from=3-4, to=4-4]
	\arrow["{\emfree c}"', from=4-1, to=4-3]
	\arrow[equals, nfold, from=4-1, to=5-1]
	\arrow["{\mext{\rea{\alglift{\emforget}\spec}}}"', from=4-3, to=4-4]
	\arrow[equals, nfold, from=4-4, to=5-4]
	\arrow["{\emfreets{c}}"', from=5-1, to=5-3]
	\arrow["\emfreefdyn", from=5-3, to=4-3]
	\arrow["{\rea{\spec}}"', from=5-3, to=5-4]
	\arrow["{(\ast)}"{description}, draw=none, from=0, to=1]
\end{tikzcd}\]
In the diagram above, all parts other than the one labelled~$(\ast)$ commute
by the definition of~$\emfreets$, the naturality of~$\emfreefdyn$,
\cref{prop:emdyn-free-alg}, and the fact that
the maps~$\rea{\spec} \colon (\emfdyn\emfree\inobj, \emcon) \to (\malg{Q}, \delta)$ and~\(
  \emfreefdyn[\inobj] \colon
    (\emfdyn\emfree\inobj, \emcon[\inobj]) \to
    (\emfree\fdyn\inobj, \emfree\con[\inobj] \com \emfreefdyn[\inobj])
\)
are homomorphisms.
Hence, the outer square commutes iff $(\ast)$ commutes.
Taking adjuncts along~$\emfree \dashv \emforget$,
this holds iff \labelcref{eq:effectful-st-cov} commutes. \qed

\paragraph*{Part~\labelcref{prop:effectful-iff-free:st-char}}

The proof of the second claim is much more involved.
The key result is the following:
\begin{proposition}\label{prop:base-eq-iff-em-eq}
  Let~$\aut{A} = (\malg{Q}, \delta, i, o)$ be \an{\emdyn} $\emdyn$-automaton
  with $\malg{Q} = (Q,q)$ \an{\eff} $\eff$-algebra.
  Let~$A \in \bcat$, $f,g \colon A \to Q$, and~$w \colon W \to \fdyn\tunit$ be arbitrary.
  Then the composite
  \[
    Q \xto{\obs{\aut{A}}} \emhom{\emfdyn\emfree\tunit}{\emfree\outobj}
      \xto{\emhom{\emfreets{w}}{\id}} \emhom{\emfree W}{\emfree \outobj}
  \]
  equalises $f$~and~$g$ iff the composite
  \[
    A \xto{\obs{\alglift{\emforget}\aut{A}}} \inthom{\fdyn\tunit}{\eff\outobj}
      \xto{\inthom{w}{\id}} \inthom{W}{\eff\outobj}
  \]
  does so.
\end{proposition}
With~\cref{prop:base-eq-iff-em-eq} at hand,
the proof of~\itemref{thm:effectful-iff-free}{prop:effectful-iff-free:st-char}
is relatively straightforward.
\begin{proof}[Proof of~\itemref{thm:effectful-iff-free}{prop:effectful-iff-free:st-char}]
  Let us write~$\spec = (\malg{Q}, \delta, i, o)$,
  where $\malg{Q} = (Q, q)$~is \an{\eff} $\eff$-algebra.
  By~\cref{lem:emfreets-cont-ememp}, we have
  $i_W \colon \emp[\tunit] \to w$ iff $\emfree i_W \colon \ememp[\emfree\tunit] \to \emfreets{w}$.
  We prove the remaining implications in both directions.
  \begin{description}
    \item[($\Rightarrow$)]
    Let~$f,g \colon \malg{A} \to \malg{Q}$ such that
    \(
      \emhom{\emfreets{w}}{\id} \com \obs{\spec} \com f
        = \emhom{\emfreets{w}}{\id} \com \obs{\spec} \com g
    \).
    By~\cref{prop:base-eq-iff-em-eq}, we have
    \(
      \inthom{w}{\id} \com \obs{\alglift{\emforget}\spec} \com f
        = \inthom{w}{\id} \com \obs{\alglift{\emforget}\spec} \com g
    \).
    Since $w$~is an effectful state characteriser, we get
    $\obs{\alglift{\emforget}\spec} \com f = \obs{\alglift{\emforget}\spec} \com g$.
    Applying~\cref{prop:base-eq-iff-em-eq} again with~$w \leteq \id$
    and using that $\emfreets{\id}$ is an isomorphism,
    we conclude that $\obs{\spec} \com f = \obs{\spec} \com g$.

    \item[($\Leftarrow$)]
    Let~$f,g \colon A \to Q$ such that
    \(
      \inthom{w}{\id} \com \obs{\alglift{\emforget}\spec} \com f
        = \inthom{w}{\id} \com \obs{\alglift{\emforget}\spec} \com g
    \).
    By~\cref{prop:base-eq-iff-em-eq}, we have
    \(
      \emhom{\emfreets{w}}{\id} \com \obs{\spec} \com f
        = \emhom{\emfreets{w}}{\id} \com \obs{\spec} \com g
    \).
    Let us write~$\mext{f} \defeq q \com \eff f$ for the free $\eff$-extension of~$f$,
    and similary for~$g$; these are $\eff$-algebra maps~$\emfree A \to \malg{Q}$.
    Then, writing~$\algmap$ for the algebra map of~$\emhom{\emfree W}{\emfree\outobj}$, we have
    \begin{align*}
      \emhom{\emfreets{w}}{\id} &\com \obs{\spec} \com \mext{f}
        = \emhom{\emfreets{w}}{\id} \com \obs{\spec} \com q \com \eff f
        = \algmap \com \eff(\emhom{\emfreets{w}}{\id} \com \obs{\spec} \com f) \\
       &= \algmap \com \eff(\emhom{\emfreets{w}}{\id} \com \obs{\spec} \com g)
        = \emhom{\emfreets{w}}{\id} \com \obs{\spec} \com q \com \eff g
        = \emhom{\emfreets{w}}{\id} \com \obs{\spec} \com \mext{g}.
    \end{align*}
    Hence, since $\emfreets{w}$~is a state characteriser,
    we get $\obs{\spec} \com \mext{f} = \obs{\spec} \com \mext{g}$,
    implying $\obs{\spec} \com f = \obs{\spec} \com g$.
    Applying~\cref{prop:base-eq-iff-em-eq} again with~$w \leteq \id$,
    we conclude that $\obs{\alglift{\emforget}\spec} \com f = \obs{\alglift{\emforget}\spec} \com g$.
    \qedhere
  \end{description}
\end{proof}

\noindent To prove~\cref{prop:base-eq-iff-em-eq}, we need some lemmas.
\begin{lemma}\label{lem:freehomiso-emcurry}
  Let~$\malg{A} = (A, a)$ and~$\malg{C} = (C, c)$ be $\eff$-algebras,
  and let~$f \colon \malg{A} \emprod \emfree B \to \malg{C}$ be \an{\eff} $\eff$-algebra map.
  Then the composite
  \[
    \malg{A} \xto{\emcurry{f}}
    \emhom{\emfree B}{\malg{C}} \xto{\freehomiso}
    \inthom{B}{C}
  \]
  is equal to the currying (in~$\bcat$) of the composite
  \[
    A \tprod B \xto{\id \tprod \effunit}
    A \tprod \eff B \xto{\emforgetmf}
    \malg{A} \emprod \emfree B \xto{f}
    C.
  \]
\end{lemma}
\begin{proof}
  The two morphisms are equal iff their adjuncts are equal.
  Hence, we need to prove
  \(
    \intev \com (\freehomiso \tprod \id) \com (\emcurry{f} \tprod \id)
      = f \com \emforgetmf \com (\id \tprod \effunit)
  \).
  This follows from the following diagram, which commutes by the definition of~$\freehomiso$,
  the naturality of~$\emforgetmf$, and diagram~\labelcref{eq:emev-def}:
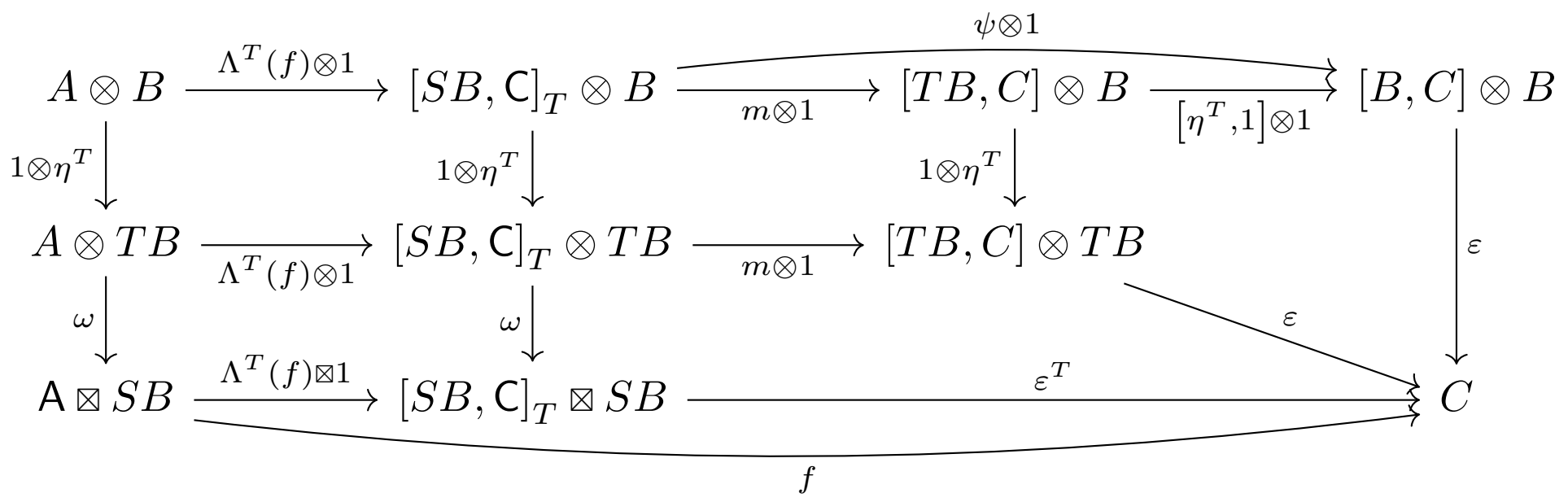
  \[\begin{tikzcd}[baseline=(\tikzcdmatrixname.south),column sep=4em]
    {A \tprod B} & {\emhom{\emfree B}{\malg{C}} \tprod B} & {\inthom{\eff B}{C} \tprod B} & {\inthom{B}{C} \tprod B} \\
    {A \tprod \eff B} & {\emhom{\emfree B}{\malg{C}} \tprod \eff B} & {\inthom{\eff B}{C} \tprod \eff B} \\
    {\malg{A} \emprod \emfree B} & {\emhom{\emfree B}{\malg{C}} \emprod \emfree B} && C
    \arrow["{\emcurry{f} \tprod \id}", from=1-1, to=1-2]
    \arrow["{\id \tprod \effunit}"', from=1-1, to=2-1]
    \arrow["{\emhominc \tprod \id}"', from=1-2, to=1-3]
    \arrow["{\freehomiso \tprod \id}", shift left, curve={height=-6pt}, from=1-2, to=1-4]
    \arrow["{\id \tprod \effunit}"', from=1-2, to=2-2]
    \arrow["{\inthom{\effunit}{\id} \tprod \id}"', from=1-3, to=1-4]
    \arrow["{\id \tprod \effunit}"', from=1-3, to=2-3]
    \arrow["\intev", from=1-4, to=3-4]
    \arrow["{\emcurry{f} \tprod \id}"', from=2-1, to=2-2]
    \arrow["\emforgetmf"', from=2-1, to=3-1]
    \arrow["{\emhominc \tprod \id}"', from=2-2, to=2-3]
    \arrow["\emforgetmf"', from=2-2, to=3-2]
    \arrow["\intev", from=2-3, to=3-4]
    \arrow["{\emcurry{f} \emprod \id}", from=3-1, to=3-2]
    \arrow["f"', shift right, curve={height=12pt}, from=3-1, to=3-4]
    \arrow["\emev", from=3-2, to=3-4]
  \end{tikzcd}\qedhere\]
\end{proof}

\begin{proposition}\label{prop:run-emforget}
  Let~$\aut{A} = (\malg{Q}, \delta, i, o)$ be \an{\emdyn} $\emdyn$-automaton
  with $\malg{Q} = (Q, q)$ \an{\eff} $\eff$-algebra.
  Then the run map~$\run{\alglift{\emforget}\aut{A}}$ of
  the automaton~$\alglift{\emforget}\aut{A}$ is the composite
  \[
    Q \tprod \fdyn\tunit \xto{\id \tprod \effunit}
    Q \tprod \eff\fdyn\tunit \xto{\emforgetmf}
    \malg{Q} \emprod \emfree\fdyn\tunit \xto{\id \emprod \inv*{\emfreefdyn}}
    \malg{Q} \emprod \emfdyn\emfree\tunit \xto{\run{\aut{A}}}
    Q.
  \]
\end{proposition}
\begin{proof}
  Let us write~$\bar{\delta} \colon \dyn Q \to Q$
  for the $\dyn$-algebra structure of~$\alglift{\emforget}\aut{A}$.
  Since $\emforgetmf$~is natural and $q \com \effunit = \id$, we have
  \begin{align*}
    \mleft(\id \emprod \inv*{\emfreefdyn}\mright)
      \com \emforgetmf \com \mleft(\id \tprod \effunit\mright)
     &= \emforgetmf \com \mleft(\id \tprod \inv*{\emfreefdyn}\mright)
          \com \mleft(\id \tprod \effunit\mright) \\
     &= \emforgetmf \com (q \tprod \id) \com \mleft(\id \tprod \inv*{\emfreefdyn}\mright)
          \com \mleft(\effunit \tprod \effunit\mright).
  \end{align*}
  Hence, it suffices to show that the outside of the following diagram commutes:
  
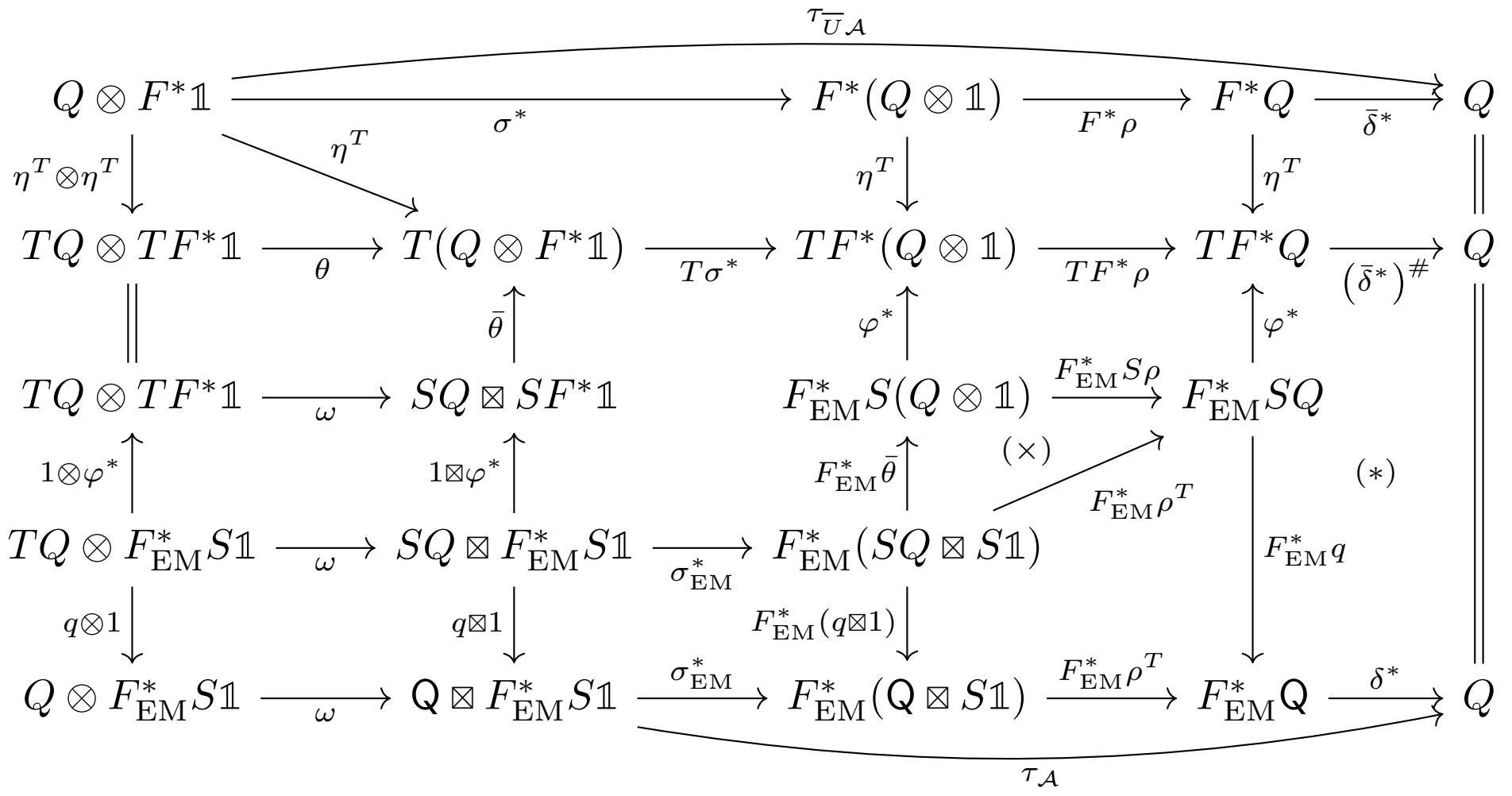
\[\begin{tikzcd}
    {Q \tprod \fdyn\tunit} && {\fdyn(Q \tprod \tunit)} & {\fdyn Q} & Q \\
    {\eff Q \tprod \eff\fdyn\tunit} & {\eff(Q \tprod \fdyn\tunit)} & {\eff\fdyn(Q \tprod \tunit)} & {\eff\fdyn Q} & Q \\
    {\eff Q \tprod \eff\fdyn\tunit} & {\emfree Q \emprod \emfree\fdyn\tunit} & {\emfdyn\emfree(Q \tprod \tunit)} & {\emfdyn\emfree Q} \\
    {\eff Q \tprod \emfdyn\emfree\tunit} & {\emfree Q \emprod \emfdyn\emfree\tunit} & {\emfdyn(\emfree Q \emprod \emfree\tunit)} \\
    {Q \tprod \emfdyn\emfree\tunit} & {\malg{Q} \emprod \emfdyn\emfree\tunit} & {\emfdyn(\malg{Q} \emprod \emfree\tunit)} & {\emfdyn\malg{Q}} & Q
    \arrow["\flstr"', from=1-1, to=1-3]
    \arrow["{\run{\alglift{\emforget}\aut{A}}}", shift left, curve={height=-12pt}, from=1-1, to=1-5]
    \arrow["{\effunit \tprod \effunit}"', from=1-1, to=2-1]
    \arrow["\effunit", from=1-1, to=2-2]
    \arrow["{\fdyn\runit}"', from=1-3, to=1-4]
    \arrow["\effunit"', from=1-3, to=2-3]
    \arrow["{\idext{\bar{\delta}}}"', from=1-4, to=1-5]
    \arrow["\effunit", from=1-4, to=2-4]
    \arrow[equals, nfold, from=1-5, to=2-5]
    \arrow["\effmf"', from=2-1, to=2-2]
    \arrow[equals, nfold, from=2-1, to=3-1]
    \arrow["{\eff\flstr}"', from=2-2, to=2-3]
    \arrow["{\eff\fdyn\runit}"', from=2-3, to=2-4]
    \arrow[""{name=0, anchor=center, inner sep=0}, "{\mext*{\idext{\bar{\delta}}}}"', from=2-4, to=2-5]
    \arrow[equals, nfold, from=2-5, to=5-5]
    \arrow["\emforgetmf"', from=3-1, to=3-2]
    \arrow["\emfreemf", from=3-2, to=2-2]
    \arrow["\emfreefdyn", from=3-3, to=2-3]
    \arrow["{\emfdyn\emfree\runit}", from=3-3, to=3-4]
    \arrow["\emfreefdyn"', from=3-4, to=2-4]
    \arrow["{\emfdyn q}", from=3-4, to=5-4]
    \arrow["{\id \tprod \emfreefdyn}", from=4-1, to=3-1]
    \arrow["\emforgetmf"', from=4-1, to=4-2]
    \arrow["{q \tprod \id}"', from=4-1, to=5-1]
    \arrow["{\id \emprod \emfreefdyn}", from=4-2, to=3-2]
    \arrow["\emflstr"', from=4-2, to=4-3]
    \arrow["{q \emprod \id}"', from=4-2, to=5-2]
    \arrow["{\emfdyn\emfreemf}", from=4-3, to=3-3]
    \arrow[""{name=1, anchor=center, inner sep=0}, "{\emfdyn\emrunit}"', from=4-3, to=3-4]
    \arrow["{\emfdyn(q \emprod \id)}"', from=4-3, to=5-3]
    \arrow["\emforgetmf"', from=5-1, to=5-2]
    \arrow["\emflstr", from=5-2, to=5-3]
    \arrow["{\run{\aut{A}}}"', shift right, curve={height=12pt}, from=5-2, to=5-5]
    \arrow["{\emfdyn\emrunit}", from=5-3, to=5-4]
    \arrow[""{name=2, anchor=center, inner sep=0}, "{\idext{\delta}}", from=5-4, to=5-5]
    \arrow["{(\ast)}"{description}, shift right, draw=none, from=0, to=2]
    \arrow["{(\times)}"{description, pos=0.4}, draw=none, from=3-3, to=1]
  \end{tikzcd}\]

  In the diagram above, all unlabelled areas commute either by definition, naturality properties,
  the commutativity of~$\eff$, \cref{lem:emfreemf}, or \cref{lem:emfreefdyn-emfreemf-emflstr}.
  The area labelled~$(\times)$ commutes since
  $(\emfree, \emfreemf, \id)$~is a monoidal functor~\cite[Remark~2.7.3]{Sea:13}.
  To show that $(\ast)$~commutes, we use the freeness of~$\emfdyn\emfree Q$
  and show that both sides are $\emdyn$-algebra homomorphisms
  such that precomposition with~$\ememp$ results in the same morphism.
  By definition, the bottom side is a composite homomorphism~\(
    (\emfdyn\emfree Q, \emcon) \xto{\emfdyn q}
    (\emfdyn\malg{Q}, \emcon) \xto{\idext{\delta}}
    (\malg{Q}, \delta)
  \)
  and $\emfreefdyn[Q]$~is a
  homomorphism~$(\emfdyn\emfree Q, \emcon) \to \alglift{\emfree}(\fdyn Q, \con)$.
  Furthermore, $\idext{\bar{\delta}}$~is \an{\dyn} $\dyn$-algebra
  homomorphism~$(\fdyn Q, \con) \to (Q, \bar{\delta}) = \alglift{\emforget}(\malg{Q}, \delta)$
  by definition, so \cref{thm:lift-adj-alg} implies that $\mext*{\idext{\bar{\delta}}}$~is
  a homomorphism~$\alglift{\emfree}(\fdyn Q, \con) \to (\malg{Q}, \delta)$.
  Finally, we have
  \[
    \idext{\delta} \com \emfdyn q \com \ememp
      = \idext{\delta} \com \ememp \com q
      = q
      = q \com \emfree\idext{\bar{\delta}} \com \emfree\emp
      = \mext*{\idext{\bar{\delta}}} \com \emfreefdyn \com \ememp,
  \]
  which concludes the proof.
\end{proof}

\begin{proof}[Proof of~\cref{prop:base-eq-iff-em-eq}]
  Consider the following diagram:
  
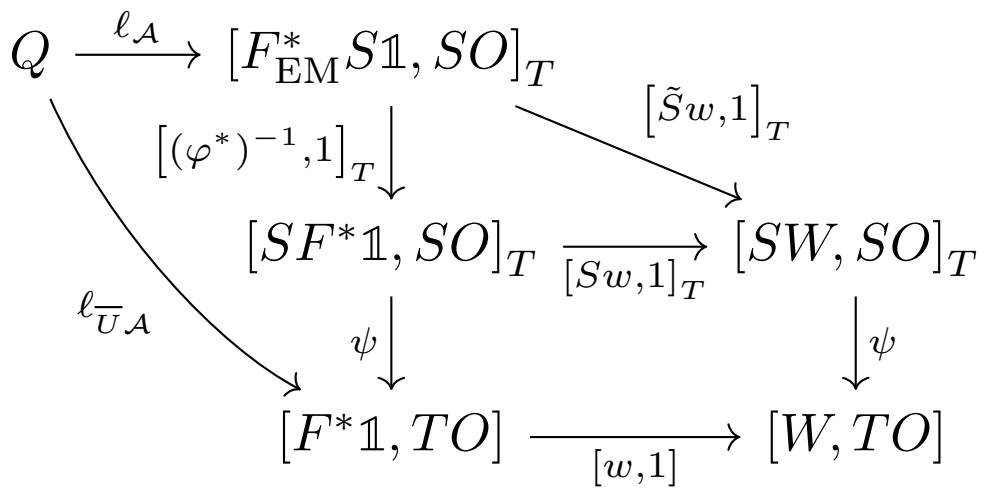
\[\begin{tikzcd}
    Q & {\emhom{\emfdyn\emfree\tunit}{\emfree\outobj}} & \\
    & {\emhom{\emfree\fdyn\tunit}{\emfree\outobj}} & {\emhom{\emfree W}{\emfree\outobj}} \\
    & {\inthom{\fdyn\tunit}{\eff\outobj}} & {\inthom{W}{\eff\outobj}}
    \arrow["{\obs{\aut{A}}}", from=1-1, to=1-2]
    \arrow["{\obs{\alglift{\emforget}\aut{A}}}"', curve={height=12pt}, from=1-1, to=3-2]
    \arrow["{\emhom{\inv*{\emfreefdyn}}{\id}}"', from=1-2, to=2-2]
    \arrow["{\emhom{\emfreets{w}}{\id}}", from=1-2, to=2-3]
    \arrow["{\emhom{\emfree w}{\id}}"', from=2-2, to=2-3]
    \arrow["\freehomiso"', from=2-2, to=3-2]
    \arrow["\freehomiso", from=2-3, to=3-3]
    \arrow["{\inthom{w}{\id}}"', from=3-2, to=3-3]
  \end{tikzcd}\]
  The square commutes by the naturality of~$\freehomiso$,
  the triangle above it commutes by the definition of~$\emfreets$,
  and the triangle on the left commutes by~\cref{lem:freehomiso-emcurry,prop:run-emforget} since
  \(
    \emhom{\inv*{\emfreefdyn}}{\id} \com \obs{\aut{A}}
      = \emcurry{o \com \run{\aut{A}} \com \mleft(\id \emprod \inv*{\emfreefdyn}\mright)}
  \).
  Hence, the top composite equalises $f$~and~$g$ if and only if the bottom composite does.
  Since $\freehomiso$~is an isomorphism, the claim follows.
\end{proof}

\section{Proof of~\texorpdfstring{\cref{lem:emfree-w-suite}}{\autoref{lem:emfree-w-suite}}}

Recall the concatenation map~$\trconc$ from~\cref{def:word-extension}.
Let us write~$\emtrconc$ for the morphism~$\trconc$ interpreted in~$\emcat$.
\begin{lemma}\label{lem:emtrconc}
  The following diagram commutes:
  \[\begin{tikzcd}
    {\emfdyn\emfree\inobj \emprod \emfdyn\emfree\tunit} && {\emfdyn\emfree\inobj} \\
    {\emfree\fdyn\inobj \emprod \emfree\fdyn\tunit} & {\emfree(\fdyn\inobj \tprod \fdyn\tunit)} & {\emfree\fdyn\inobj}
    \arrow["\emtrconc", from=1-1, to=1-3]
    \arrow["{\emfreefdyn \emprod \emfreefdyn}"', from=1-1, to=2-1]
    \arrow["\emfreefdyn", from=1-3, to=2-3]
    \arrow["\emfreemf"', from=2-1, to=2-2]
    \arrow["{\emfree\trconc}"', from=2-2, to=2-3]
  \end{tikzcd}\]
\end{lemma}
\begin{proof}
  Diagram chase using the naturality of~$\emfreefdyn$, $\emflstr$, $\runit$, and~$\emrunit$,
  that $\emfreefdyn$~is \an{\emdyn} $\emdyn$-algebra morphism~\(
    (\emfdyn\emfree\inobj, \emcon[\emfree\inobj]) \to \alglift{\emfree}(\fdyn\inobj, \con)
  \),
  that $\eff$~is commutative, and \cref{lem:emfreefdyn-emfreemf-emflstr}.
\end{proof}

\begin{lemma}\label{lem:emtrext+emtsconc}
  For all~$k \in \N$ and~$a \colon A \to \fdyn\inobj$,
  there is an isomorphism making the diagram on the left commute.
  For $a \colon A \to \fdyn\inobj$~and~$b \colon B \to \fdyn\tunit$,
  the diagram on the right commutes.
  \[
    \begin{tikzcd}[baseline=(\tikzcdmatrixname.base),column sep=-1.6em,row sep=2.5em]
      {\trext*{\emfree A}[k]} & \cong & {\emfree(\trext{A}[k])} \\
      & {\emfdyn\emfree\inobj}
      \arrow["{\trext*{\emfreets{a}}[k]}"'{pos=0.3}, from=1-1, to=2-2]
      \arrow["{\emfreets*{\trext{a}[k]}}"{pos=0.3}, from=1-3, to=2-2]
    \end{tikzcd}
    \hspace{2em}
    \begin{tikzcd}[baseline=(\tikzcdmatrixname.base),column sep=-1.2em,row sep=2.5em]
      {\emfree A \emprod \emfree B} && {\emfree(A \tprod B)} \\
      & {\emfdyn\emfree\inobj}
      \arrow["\emfreemf", from=1-1, to=1-3]
      \arrow["{\emfreets{a} \tsconc \emfreets{b}}"'{pos=0.3}, from=1-1, to=2-2]
      \arrow["{\emfreets*{a \tsconc b}}"{pos=0.3}, from=1-3, to=2-2]
    \end{tikzcd}
  \]
\end{lemma}
\begin{proof}
  The first diagram is proved commutative by induction on~$k$.
  The base case~$k = 0$ is trivial since $\trext{a}[0] = a$.
  The inductive case follows from the special case~$k = 1$,
  where we use the isomorphism $\emfreefdyn$~and~$\emfreemf$,
  that $\emfree$~preserves coproducts (being a left adjoint),
  and the definition of~$\emfreets$.
  The second diagram commutes by the definition of~$\emfreets$ and~\cref{lem:emtrconc}.
\end{proof}

\begin{proof}[Proof of~\cref{lem:emfree-w-suite}]
  Immediate from~\cref{lem:emtrext+emtsconc}.
\end{proof}

\section{Remaining proofs}

\begin{proof}[Proof of~\cref{thm:effectful-w-comp}]
  Follows from~\cref{thm:em-sat-ass,thm:w-comp,thm:effectful-iff-free,lem:emfree-w-suite},
  noting that $\emwfdom{c}{k} \subseteq \wfdom{\emfreets{c}}{k}$ in~$\emcat$
  by~\itemref{thm:effectful-iff-free}{prop:effectful-iff-free:st-cov}.
\end{proof}

\begin{proof}[Proof of~\cref{cor:effectful-w-comp}]
  Immediate consequence of~\cref{thm:effectful-w-comp},
  using that $\klemb t$~is complete for~$\spec$ \wrt $\klwfdom{c}{k}$
  iff $\emfreets{t}$~is complete for~$\alglift{\klcomp}\spec$ \wrt $\emwfdom{c}{k}$.
  This latter fact relies
  on the equality~$\beh{\alglift{\klcomp}\spec} = \klcomp\beh{\spec} \com \emfreefdyn$,
  which follows from~\cref{prop:kldyn-free-alg,prop:emdyn-free-alg}.
\end{proof}

\end{document}